\documentclass[aps,prb,twocolumn,showpacs,10pt,floatfix,superscriptaddress,floatfix]{revtex4-2}
\usepackage{epsf}
\usepackage{bm}
\usepackage{hyperref}
\usepackage{amsfonts}
\usepackage{amssymb}
\usepackage{amsmath,mathtools}
\usepackage{array}
\usepackage{enumerate,dsfont}
\usepackage{dcolumn,multirow}
\usepackage[utf8]{inputenc}
\usepackage{latexsym}
\usepackage{xcolor}

\newcommand{\ve}{\vec e}
\newcommand{\vb}{\vec b}

\newcommand{\va}{\vec a}

\newcommand{\vd}{\vec d}

\newcommand{\vR}{\vec R}

\newcommand{\vv}{\vec v}

\newcommand{\RR}{\mathbb{R}}

\usepackage{ bbold }
\newcommand{\id}{\textbf{1}}

\renewcommand{\vec}[1]{\mathbf{#1}}

\newcommand{\refeqWC}{\ref{eq:WC1}}
\newcommand{\refeqWtriangle}{\ref{eq:Wtriangle}}
\newcommand{\refeqvortho}{\ref{eq:vortho}}
\newcommand{\refeqwform}{\ref{eq:wform}}
\newcommand{\refeqvxy}{\ref{eq:v_xy}}
\newcommand{\refeqHxy}{\ref{eq:Hxy}}
\begin{document}

\newcommand{\refeqGammaW}{\ref{eq:GammaW}}
\newcommand{\refeqQf}{\ref{eq:Qf3_def}}
\newcommand{\refeqHsimple}{\ref{eq:Hsimple}}

\newcommand{\reffigone}{\ref{fig:1}}
\newcommand{\reffigthree}{\ref{fig:3}}

\title{Non-Abelian holonomy of Majorana zero modes coupled to a chaotic quantum dot}

\author{Max Geier}
\affiliation{Center for Quantum Devices, Niels Bohr Institute, University of Copenhagen, DK-2100 Copenhagen, Denmark}
\author{Svend Kr{\o}jer}
\affiliation{Center for Quantum Devices, Niels Bohr Institute, University of Copenhagen, DK-2100 Copenhagen, Denmark}
\author{Felix von Oppen}
\affiliation{Dahlem Center for Complex Quantum Systems and Physics Department, Freie Universit\"at Berlin, Arnimallee 14, 14195 Berlin, Germany}
\author{Charles M. Marcus}
\affiliation{Center for Quantum Devices, Niels Bohr Institute, University of Copenhagen, DK-2100 Copenhagen, Denmark}
\date{\today}
\author{Karsten Flensberg}
\affiliation{Center for Quantum Devices, Niels Bohr Institute, University of Copenhagen, DK-2100 Copenhagen, Denmark}
\date{\today}
\author{Piet W. Brouwer}
\affiliation{Dahlem Center for Complex Quantum Systems and Physics Department, Freie Universit\"at Berlin, Arnimallee 14, 14195 Berlin, Germany}
\begin{abstract}
If a quantum dot is coupled to a topological superconductor via tunneling contacts, each contact hosts a Majorana zero mode in the limit of zero
transmission. Close to a resonance and at a finite contact transparency, the resonant level in the quantum dot couples the Majorana modes,
but a ground state degeneracy per fermion parity subspace 
remains if the number of Majorana modes coupled to the dot is five or larger.
Upon varying shape-defining gate voltages while remaining close to
resonance, a nontrivial evolution within the degenerate ground-state
manifold is achieved. We characterize the corresponding non-Abelian
holonomy for a quantum dot with chaotic classical dynamics
using random matrix theory and discuss measurable
signatures of the non-Abelian time-evolution. 
\end{abstract}
\maketitle

Billiards are among the simplest physical systems that exhibit chaotic dynamics if the boundary is of sufficiently irregular shape \cite{Sinai1970}. When the motion of the particle in the billiard is coherent, the statistics of its energy eigenvalues follows a universal law \cite{BohigasPRL1984} that -- depending on the particle's spin and the presence of time-reversal symmetry -- derives from one of the three Wigner-Dyson random matrix ensembles \cite{Wigner1955, Dyson1962}. An ``electron billiard'' --- a quantum dot (QD) --- coupled to a superconductor confines quasiparticles by Andreev reflection \cite{Andreev1964}.
The chaotic quantum dynamics of such ``Andreev billiards'' \cite{AltlandPhysRevLett1996Apr, MelsenEurophysLett1996Jul, BelzigPhysRevB1996Oct, Beenakker2005} are described by novel random matrix ensembles, exhaustively contained in the tenfold-way classification \cite{AltlandPRB1997}. 

\begin{figure}[t]
\includegraphics[width=0.9\columnwidth]{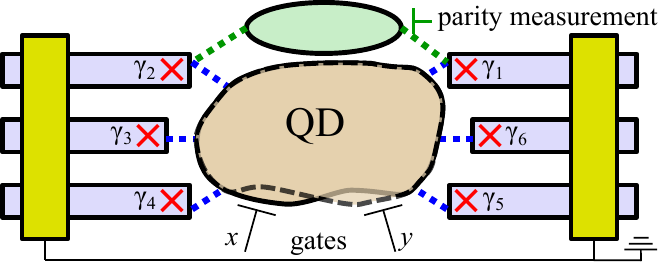}
\caption{
\label{fig:1} 
$N$ MZMs $\hat{\gamma}_j$ supported by a grounded topological superconductor are tunnel coupled to a chaotic QD. The dot shape can be changed by two gates with dimensionless voltages $x$ and $y$, whereas a third gate allows for a uniform shift of the dot potential.
A parity measurement involving two of the MZMs allows for the detection of non-Abelian evolution resulting from a closed loop in ``shape space''. 
}
\end{figure}

Here we consider a chaotic QD coupled to a {\em topological} superconductor, so that reflection at the boundary allows tunneling into Majorana zero modes (MZMs). We show that for weak coupling and close to a resonance, such a system has a degenerate ground state manifold with non-Abelian evolution under generic cyclic adiabatic changes of the dot shape if the number of coupled separated MZMs,  $N$, 
is five or larger.
The holonomy of closed loops in ``shape space'' inherits its statistical distribution from the universal statistics of QD wave functions. 

Demonstrating non-Abelian properties of MZMs remains an open milestone. Other approaches, such as braiding or fusion \cite{AasenPRX2016}, require strong direct manipulations of the MZMs. This may be challenging due to the sensitivity of MZMs to disorder \cite{PanPhysRevRes2020Mar, PradaNatRevPhys2020}. Our approach eases these requirements by coupling only perturbatively to the MZMs and driving nonlocally via the QD.

A schematic picture of the QD is shown in Fig.\ \ref{fig:1}. 
The MZMs $\hat{\gamma}_j,\ j = 1, ..., N$ are located at tunneling contacts to a topological superconductor. 
For weak coupling, if the Fermi energy is close to a QD resonance,
hybridization between MZMs and QD is 
dominated by a single 
dot 
level and may be described by the 
Hamiltonian \cite{RainisPhysRevB2013Jan, DengScience2016Dec}
\begin{equation}
  \hat H = \varepsilon \hat c^{\dagger} \hat c + \sum_{j=1}^{N} (v_j^* \hat c^{\dagger} - v_j \hat c) \hat \gamma_j, \label{eq:Hsimple}
\end{equation}
where $\hat c^{\dagger}$ and $\hat c$ are creation and annihilation operators of the resonant level, $\varepsilon$ is its energy 
(relative 
to the Fermi level), and the $v_j$ are complex coefficients describing tunneling between the dot level and the MZMs. 
Neglecting hybridization with non-resonant levels, only two linearly independent superpositions of MZMs 
hybridize with the dot mode and 
acquire a finite energy.
The remaining $N-2$ MZMs are ``dark'' modes, which remain at zero energy despite being coupled to the dot. The Hamiltonian (\ref{eq:Hsimple}) is easily diagonalized and one obtains the dark modes $\hat \Gamma_i$
as superpositions of the original MZMs
$\hat{\Gamma}_i = \sum_{j=1}^{N} o_{ji} \hat{\gamma}_j$,  $i=1,\ldots,N-2$,
with real coefficients $o_{ij}$ satisfying the (complex) orthogonality condition
\begin{equation}
  \sum_{j=1}^{N} o_{ji} v_{j} = 0. \label{eq:vortho}
\end{equation}

Changing the shape of the dot, while keeping $\varepsilon$ close to zero, changes its eigenmodes and, hence, the tunneling amplitudes $v_j$ between the resonant mode and the MZMs. 
(Keeping $\varepsilon$ close to zero requires adjusting a third gate voltage while performing the shape change.)
Via the orthogonality condition (\ref{eq:vortho}), adiabatically changing the dot shape therefore leads to an adiabatic change of the subspace spanned by the $N-2$ dark MZMs. A loop ${\cal C}$ in ``billiard shape space'', which returns the dot to its original shape, may nevertheless lead to a ``rotation'' in the space of the $N-2$ dark MZMs,
\begin{equation}
  \Gamma_j \to \sum_{k=1}^{N-2} W_{kj} \Gamma_k, \label{eq:GammaW}
\end{equation}
where the matrix $W \in \mbox{SO}(N-2)$ is known as the ``holonomy'' of the loop ${\cal C}$. 
For $N \geq 5$, the group $\text{SO}(N-2)$ is non-Abelian, so that holonomies of different loops in shape space generally do not commute \footnote{There is no non-Abelian evolution if $N=3$ or $N=4$: If $N=4$ the basis change (\ref{eq:GammaW}) amounts to a phase shift of the complex fermion $\Gamma_1 + i \Gamma_2$; If $N=3$, the dark MZM $\Gamma_1$ must be supplemented with an additional MZM not coupled to the dot and the same conclusion applies.}. 

Wave functions of a chaotic QD and their response to a change of the dot shape are random, with a universal statistical distribution described by random matrix theory \cite{altshuler1995,guhr1998,aleinerPR2002}. As a result, the holonomy $W$ of a loop in ``shape space'' is also a random quantity with universal statistics. We consider a QD with two parameters $x$ and $y$ that determine it shape. In random matrix theory, the "shape coordinates" $x$ and $y$ are dimensionless, normalized such that a change $\Delta x, y \sim 1$ corresponds to an effective "scrambling" 
of the spectrum \cite{EfetovPhysRevLett1995Mar,guhr1998,aleinerPR2002} \footnote{The dimensionless gate voltages $x, y$ can be identified experimentally by comparing conductance autocorrelation measurements as a function of gate voltage to predictions from random matrix theory.}. In the remainder of this letter, we determine the universal distribution of the holonomy $W$ for small and large loops in ``shape space''. We propose two observables 
of the 
non-Abelian holonomy
for $N \ge 5$ --- a ``fermion parity signature''
and a ``charge signature''
--- and show that these have universal statistics for a chaotic QD.

{\em Holonomy.---} We consider a QD with two parameters $x$ and $y$ determining its shape. Hence, ``shape space'' is the two-dimensional plane, parameterized by ``shape coordinates'' $x$ and $y$. A loop therein is parameterized by $x(\tau)$, $y(\tau)$, $0 \le \tau \le 1$, with $x(0) = x(1)=0$, $y(0) = y(1)=0$. The corresponding holonomy matrix $W \in \mbox{SO}(N-2)$ results from the Wilson loop operator\cite{wilcek1984}
$ {\cal W} = \mathcal{L}_\tau e^{-\int_0^1 d\tau {\cal P}(x,y) d{\cal P}(x,y)/d\tau} {\cal P},$
where ${\cal P}(x,y)$ is the projector on the $(N-2)$-dimensional subspace of dark MZMs, we abbreviated ${\cal P} = {\cal P}(0,0)$, and $\mathcal{L}_\tau$ denotes path ordering, such that factors with lower $\tau$ appear to the right of factors with higher $\tau$. 
To remain within the dark space, the loop in parameter space must be performed slowly on the time scale associated with coupling to the resonant level, but fast compared to the time scale related to the coupling to the non-resonant levels. 
For an infinitesimal shape-space loop with enclosed area $A$, the Wilson loop operator takes the form 
\begin{equation}
  {\cal W} \simeq e^{w A} {\cal P},\ \ 
  w = {\cal P}  [\partial {\cal P}/\partial y, \partial {\cal P}/\partial x] {\cal P} .
  \label{eq:WC1}
\end{equation}
where $w = {\cal P}( \partial_x {\cal A}_y - \partial_y {\cal A}_x +i [{\cal A}_x, {\cal A}_y] ) {\cal P}$ is the projected non-Abelian field strength associated to the gauge field ${\cal A}_\rho = {\cal P} \partial_{\rho} {\cal P},\, \rho=x,y$.
The dark subspace is defined by the orthogonality condition (\ref{eq:vortho}). The projection operator ${\cal P}$ can be expressed in terms of the $N$-component vector  $\vv = (v_1,\ldots,v_N)$,
$  {\cal P} = \openone_{N} - \frac{2 {\cal Q}}{\mbox{tr}\, {\cal Q}},$
where $\openone_N$ is the $N \times N$ unit matrix and
$  {\cal Q} = \mbox{Re}\, [ \vv (\vv^{\dagger} \vv) \vv^{\dagger} -
  \vv (\vv^{\dagger} \vv^*) \vv^{\rm T} ]$.
For the generator $w$ of the holonomy one then obtains (for details, see Supplemental Material \cite{Supplementary})
\begin{align}
  w =&\, \frac{2 {\cal P}}{\mbox{tr}\, {\cal Q}}
  \mbox{Re}\, [\vd_{x}^* (\vv^{\dagger} \vv) \vd_{y}^{\rm T}
    - \vd_{x} (\vv^{\dagger} \vv^*) \vd_{y}^{\rm T} - (x \leftrightarrow y) ] {\cal P},
  \label{eq:Wtriangle}
\end{align}
where we abbreviated $\vd_{x} = \partial \vv/\partial x$, $\vd_{y} = \partial \vv/\partial y$. 
The matrix elements $W_{kj}$ of Eq.\ \eqref{eq:GammaW} are found by the projection of ${\cal W}$ onto the dark MZM wave functions, $W_{kj} = \sum_{n,m = 1}^N o_{mk} {\cal W}_{mn} o_{nj}.$

{\em Statistics of the holonomy.---} The statistics of the coefficients $v_j$ and their dependence on the two shape parameters $x$ and $y$ can be obtained by modeling the Hamiltonian of the dot, without coupling to the superconductors, as an $M \times M$ random hermitian matrix \cite{guhr1998,altshuler1995},
\begin{equation}
  H(x,y) = H_0 + \frac{1}{\sqrt{M}} \left( x H_x + y H_y \right).
  \label{eq:Hxy}
\end{equation}
Assuming broken time-reversal and lifted spin degeneracy, {\em e.g.}, by the Zeeman coupling to an applied magnetic field \footnote{The GUE description also requires either the absence of spin-orbit coupling in the dot or the presence of sufficiently strong spin-orbit coupling. Without spin-orbit coupling the in-plane magnetic field must be large enough that tunneling between dot and superconductor is spin-polarized. If these conditions are not satisfied, the random matrix ensemble in the dot may deviate, see Refs.~\cite{HalperinPhysRevLett2001Mar, AleinerPhysRevLett2001Nov, CremersPhysRevB2003Sep} for a discussion of the random matrix theory for a dot with weak spin-orbit coupling}, the matrices $H_0$, $H_x$, and $H_y$ are statistically independent and taken from the Gaussian Unitary Ensemble (GUE). The normalization factors $\sqrt{M}$ are included such that the results become independent of $M$ as $M \to \infty$. The resonant dot mode is identified with the $m$th eigenstate of $H$, with eigenvalue $\varepsilon \equiv \varepsilon_m$ and $m \sim M/2$ near the center of its spectrum. Its eigenket $|\varepsilon_m\rangle$ determines the complex coefficients 
$  v_j = \eta_j \langle j | \varepsilon_m \rangle \sqrt{M},$
where $\eta_j$ is a proportionality factor depending on the QD level spacing $\delta$, superconducting gap $\Delta_j$ and normal-state transmission coefficient $T_j \ll 1$ of the $j$th tunnel contact to a MZM with wavefunction $|j\rangle$. The coefficients $v_j$ have independent Gaussian distributions with zero mean and with variance $\langle |v_j|^2 \rangle = \eta_j^2$. The derivatives $\vd_{x,y}$ follow from first-order perturbation theory in 
$H_x$ and $H_y$,
\begin{equation}
  \vd_{x,y} =  \frac{1}{\sqrt{M}}\sum_{k \neq m} \frac{
    \langle \varepsilon_k|H_{x,y}|\varepsilon_m\rangle
  }{\varepsilon_{k}-\varepsilon} \vv_{k},
\label{eq:v_xy}
\end{equation}
where $\varepsilon_k$, $k \neq m$, are the other eigenvalues of $H$ and $|\varepsilon_k\rangle$ and $\vv_k$ the corresponding eigenket and vector of coupling coefficients, respectively. The matrix elements $\langle \varepsilon_k|H_{x,y}|\varepsilon_m\rangle$
are independently distributed complex Gaussian random numbers with zero mean and
 with
variance
$M\delta^2/\pi^2$.
The 
vectors $\vv_k$ are statistically independent and have the same distribution as $\vv \equiv \vv_m$.

\begin{figure}
\includegraphics[width=\columnwidth]{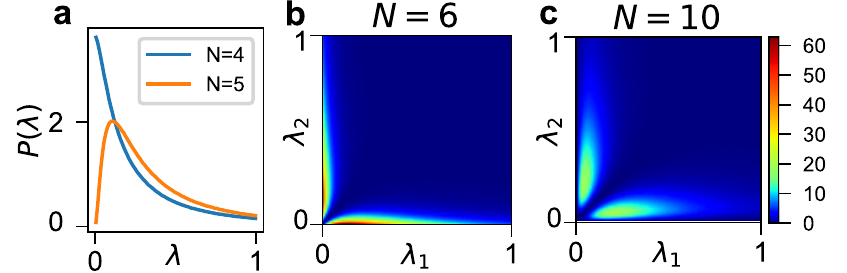}
\caption{
\label{fig:2}
PDFs $P(\lambda)$ (for $N=4$, $5$ MZMs coupled to the dot) and $P(\lambda_1,\lambda_2)$ (for $N=6$,$10$) of the principal values of the holonomy generator $w$. The statistical ensemble used to sample the distributions consists of $10^7$ realizations of the random matrix model (\ref{eq:Hxy}) with matrix size $M=20$. }
\end{figure}

To find the statistical distribution of the generator $w$ of the holonomy, we make the additional assumption that all $N$ tunnel contacts have the same proportionality constants $\eta \equiv \eta_j$.
Then, from a statistical point of view, the MZMs are interchangeable. With this simplifying assumption, the complex coefficients $v_j$ have Gaussian distributions with the same variance and the distributions of the $N$-component vector $\vv$ and its derivatives $\vd_{x,y}$ are separately invariant with respect to unitary transformations. It then follows that $w$
is an $(N-2)\times(N-2)$ antisymmetric real matrix of the form
\begin{equation}
  \label{eq:wform}
  w = \left\{ \begin{array}{ll} 0 & \mbox{if $N = 3$}, \\
    i \lambda \sigma_y & \mbox{if $N = 4$}, \\
    i \lambda O \mbox{diag}\,(\sigma_y,0) O^{\rm T} & \mbox{if $N = 5$}, \\
    i O \mbox{diag}\, (\lambda_1 \sigma_y,\lambda_2 \sigma_y) O^{\rm T}
    & \mbox{if $N = 6$},  \\
    i O \mbox{diag}\, (\lambda_1 \sigma_y,\lambda_2 \sigma_y,0,\ldots) O^{\rm T}
    & \mbox{if $N > 6$},\end{array} \right.
\end{equation}
where $\sigma_y$ is the Pauli matrix and $O \in \mbox{SO}\,(N-2)$ is uniformly distributed. In contrast, the principal values $\lambda_j$ of the holonomy generator $w$ have nontrivial and distinctive statistical distributions.
Figure \ref{fig:2} displays the numerically sampled probability distribution functions (PDFs) $P(\lambda)$ (for $N=4$, $5$) and $P(\lambda_1,\lambda_2)$ (for $N=6$, $10$).  
For $N=5$, $P(\lambda) \propto \lambda^2$ for small $\lambda$. For $N \geq 6$, $P(\lambda_1,\lambda_2)$ exhibits quadratic level repulsion $\propto |\lambda_2 - \lambda_1|^2$ for $\lambda_1 \approx \lambda_2$, in agreement with the result for real skew-symmetric matrices \cite{AltlandPRB1997}, and a power-law repulsion from the coordinate axes following $P(\lambda_1, \lambda_2) \propto \lambda_1^{N-6} \lambda_2^{N-6}$ for $|\lambda_1| \ll 1$ or $|\lambda_2| \ll 1$.
For large values, the PDFs of $\lambda$ (for $N=4$, $5$) and of $\lambda \equiv \max(\lambda_1,\lambda_2)$ have an algebraic tail $\propto \lambda^{-5/2}$ \footnote{ We expect that the algebraic decay to be cut off for $\lambda \gtrsim \delta^2/\eta^2$, because at this point the splitting of the MZMs from coupling to the nearest non-resonant level $\varepsilon_{m-1}$ or $\varepsilon_{m+1}$ is comparable to the splitting from the coupling to the resonant level $\varepsilon_m$.}, which can be traced back to rare events with a small spacing between $\varepsilon = \varepsilon_m$ and the neighboring dot levels $\varepsilon_{m-1}$ or $\varepsilon_{m+1}$ \cite{BerryJPA2018,BerryJSP2020,BerryJPA2020,PennerPRL2021}. 

For loops with small area in shape space, the distribution of the holonomy matrix $W$ is found by exponentiating the generator $w$, see Eq.\ (\ref{eq:WC1}), and its statistical distribution follows accordingly. To obtain a holonomy matrix that significantly differs from the identity, one may repeat a small-area loop $Z \gg 1$ times. In this case, $W = e^{Z A w}$ is of the form $O \mbox{diag}\, (e^{i \alpha_1 \sigma_y},e^{i \alpha_2 \sigma_y},1,\ldots) O^{\rm T}$ if $N \ge 4$, where $O$ is uniformly distributed in $\mbox{SO}(N-2)$ and the phase angles $\alpha_j = ZA \lambda_j \mod 2 \pi$, $j=1,2$, are uniformly and independently distributed in the interval $(-\pi,\pi]$ if $Z A \gg 1$. (The phase angle $\alpha_2 = 0$ for $N=4$, $5$.) A large holonomy may also be obtained simply by taking a shape-space loop with large area $A \gg 1$. In this case, the holonomy matrix $W$ itself becomes uniformly distributed in $\mbox{SO}(N-2)$ in the limit of large $A$.

The holonomy $W$ cannot be observed directly, because the dark MZMs involved in it are nonlocal modes without weight in the QD. Below we propose two measurement protocols that circumvent this problem. In the first protocol, which we refer to as "fermion parity signature", two dark MZMs are fully isolated from the chaotic dot and coupled to a small QD instead, which allows their parity to be measured. In the second protocol, referred to as "charge signature", a sudden change of the coupling parameters transfers weight from the dark MZMs to the chaotic QD, which leads to a change of the charge on the dot, which is then measured capacitively.

{\em Fermion parity signature.---} A pair of MZMs may be combined into a single (complex) fermion \cite{FlensbergPRL2011, PluggeNJP2017}. Figure \ref{fig:1} displays a setup that permits to observe a signature of the non-Abelian evolution through the measurement of the fermion occupation $\hat{f}_1^\dagger \hat{f}_1 = \frac{1}{2}(1 + i \hat{\gamma}_{1} \hat{\gamma}_{2})$ of two MZMs $\hat{\gamma}_1$ and $\hat{\gamma}_{2}$ by coupling both MZMs to a QD \cite{KarzigPRB2017, MunkPhysRevRes2020Aug, SteinerPhysRevRes2020Aug}. 
For initialization, we take $\hat{\gamma}_1$ and $\hat{\gamma_2}$ decoupled from the chaotic dot and set the occupation $\hat{f}_1^{\dagger}\hat{f}_1$ to zero by a projective measurement. The ``dark'' MZM operators $\hat \Gamma_i$, $i=1,\ldots,N-2$, then consist of $\hat \Gamma_1 = \hat \gamma_1$, $\hat \Gamma_2 = \hat \gamma_2$, as well as the operators $\hat \Gamma_3$, \ldots, $\hat \Gamma_{N-2}$, which are linear combinations of the MZMs $\hat \gamma_{3}$, \ldots, $\hat \gamma_{N}$. Their state can be initialized by letting the system relax to its unique ground state on time scales large enough that the coupling to non-resonant dot levels becomes relevant.
To bring about the non-Abelian evolution, we then (i) adiabatically increase the coupling strength of $\hat \gamma_{1,2}$ to the dot from $0$ to $\eta$, (ii) perform a loop in shape space, and (iii) again measure $\hat f^{\dagger}_1 \hat f_1$ after decoupling $\hat \gamma_{1,2}$ from the dot. 

Choosing the basis of the remaining dark-space MZMs such, that the initial state corresponds to the vacuum for the associated fermions $\hat f_{i} = (1/2)(\hat \Gamma_{2 i - 1} + i \hat \Gamma_{2 i})$, $i=2,\ldots,N/2$, 
the probability $\Delta p$ that the occupation $\hat f_{1}^{\dagger} \hat f_1$ has changed is \cite{Supplementary}
\begin{equation}
  \Delta  p = \frac{1}{2}-\frac{1}{2}\sum_{j=1}^{N/2-1}(W_{2j,2}W_{2j-1,1}-W_{2j-1,2}W_{2j,1}).
  \label{eq:Qf3_def}
\end{equation}
Since a rotation among the MZMs preserves the total fermion parity \cite{SauPRB2011, KarzigPRX2016}, the holonomy $W$ can have a nontrivial parity signature $\Delta p$ only if more than one fermion is associated with the MZMs in the dark subspace. This condition is met if $N \ge 5$.

\begin{figure}
\includegraphics[]{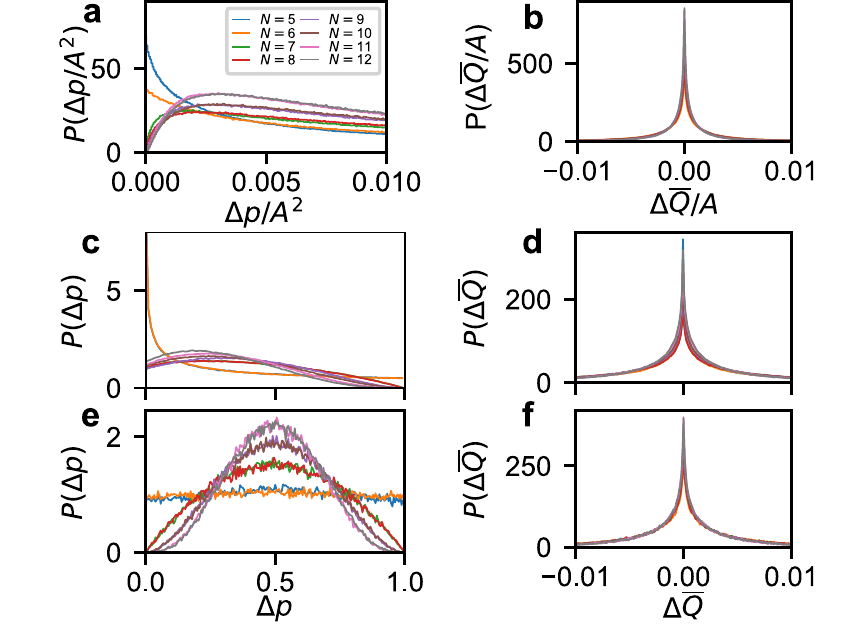}
\caption{
\label{fig:3}
Left: PDF of the parity signature $\Delta p$. Right: PDF of the charge signature $\Delta \overline{Q}$ for resonant energy  $\varepsilon/\eta = 2 \sqrt{20} \approx 9$, obtained by suddenly decoupling two MZMs from the dot. In (a), (b), the distributions of the normalized signatures $\Delta p / A$ and $\Delta \overline{Q} / A^2$ are shown for $N=5$, $6$, \ldots, $12$ and in the limit of small loop area $A \ll 1$. 
In (c), (d), distributions of $\Delta p$ and $\Delta \overline{Q}$ are shown for repeated infinitesimal loops with $Z A = 1000$. 
For (a), (b) and (c), (d), distributions are obtained from $10^7$ and $10^6$ realizations, respectively, of the random matrix model (\ref{eq:Hxy}) with matrix size $M = 20$.
The last row, (e), (f), shows distributions of $\Delta p$ and $\Delta \overline{Q}$ for large loop area $A \gg 1$, $M = 64$, sampled over $10^5$ realizations, and with $\epsilon/\eta = 16$. (See Ref.\ \onlinecite{Supplementary} for more details on the numerical simulations.)}
\end{figure}

For small areas $A$ of the enclosed loop in phase space, $\Delta p \propto A^2$ depends on the infinitesimal Wilson loop operator $w$ only. Figure \ref{fig:3}a displays the PDF of the normalized parity signature $\Delta p/A^2$ for $N = 5$, \ldots, $12$, using
the random-matrix model (\ref{eq:Hxy}).
The asymptotic distribution at small $\Delta p$ is proportional to $(\Delta p)^{\text{min}(N-4,8)/2 - 1}$ for $N$ even and $(\Delta p)^{\text{min}(N-3,8)/2 - 1}$ for $N$ odd, while it decays algebraically $\propto (\Delta p)^{-7/4}$ at large $\Delta p$.
For repeated small-area loops with cumulative area $ZA\gg 1$, the distribution converges towards the results shown in Fig.~\ref{fig:3}c. 
For $N = 5,6$, the distribution diverges as $(\Delta p)^{-1/2}$ for $\Delta p \to 0$ and is finite at $\Delta p = 1$. For larger $N$, the distribution is finite at $\Delta p = 0$ and zero at $\Delta p = 1$. 
In the limit of a large loop area $A$, $\Delta p$ is symmetrically distributed around $\Delta p = 0.5$, see Fig.~\ref{fig:3}e and becomes more peaked at $\Delta p = 1/2$ upon increasing $N$.

{\em Charge signature.---} An alternative method to observe the non-Abelian evolution involves a measurement of the charge on the QD. Hereto, (i) the system is initialized in a unique reference state, (ii) a shape-space loop is performed, and (iii) the time-averaged charge $\overline{Q}$ of the QD is measured after the coupling coefficients $v_j$ are diabatically changed, {\em e.g.}, by suddenly changing the transparency of some of the tunnel contacts to the MZMs \footnote{This charge signature is based on the non-equilibrium occupation of the QD, which is distinct from other approaches to measure the Majorana occupation based on an equilibrium energy shift of a QD level due to coherent tunneling events through a series of MZMs \cite{KarzigPRB2017, MunkPhysRevRes2020Aug, SteinerPhysRevRes2020Aug}.}. 
Measurement of the time-averaged charge on a QD is possible with established techniques \cite{SchoelkopfScience1998, AassimePRL2001, ElzermanPhysRevB2003, LiuPhysRevAppl2021Jul}.
The diabatic change of the $v_j$ is necessary, because it drives the system into a nonequilibrium state with a time-averaged dot charge that depends on the state of the dark MZMs before the quench.
The signature of the non-Abelian time evolution is the difference 
$  \Delta \overline{Q} = \overline{Q(W)} - \overline{Q(\id)} $
of the time-averaged dot charge with and without performing a closed loop in shape space at stage (ii).
An expression for $\Delta \overline{Q}$ in terms of $W$ is derived in \cite{Supplementary}. 

The charge signature depends on the resonant energy $\varepsilon$ 
and is maximal for $\varepsilon/\eta \sim 4$ \cite{Supplementary}. Unlike the parity signature $\Delta p$,
the charge signature $\Delta \overline{Q}$ is linear in $A$ for small $A$. The charge signature distribution $\Delta \overline{Q}$ 
depends on 
$N$ and on the type of diabatic quench used to obtain the nonequilibrium state. Figure~\ref{fig:3} (right) shows the PDF $\Delta \overline{Q}/A$ for selected values of $N$, for the case that diabatic change of the coupling coefficients $v_j$ is obtained by suddenly decoupling two MZMs from the dot. (The charge signature $\Delta \overline{Q} = 0$ if only one MZM is suddenly decoupled \cite{Supplementary}.) The distribution shows a logarithmic singularity at $\Delta \overline{Q} = 0$ and has a power-law dependence $\propto |\Delta \overline{Q}|^{-5/2}$ for large $|\Delta \overline{Q}|/A \gg 1$.
The logarithmic singularity persists for repeated infinitesimal loops with cumulative area $ZA \gg 1$ and for large loops with enclosed area $A \gg 1$, {\it c.f.} Fig.~\ref{fig:3}d and f. It dominates the distribution such that the dependence on $N$ is weak. Decoupling three or more but less than $N-2$ MZMs results in a non-divergent density at $\Delta \overline{Q} = 0$ \cite{Supplementary}.

{\em MZMs vs.\ Andreev bound states (ABSs).---} 
Superconductor contacts may host zero-energy ABS that mimic MZMs \cite{PanPhysRevRes2020Mar, PradaNatRevPhys2020}. 
Although ABSs need not be pinned to zero energy, for weak coupling to the dot, an ABS that is accidentally at zero energy $\ll \eta$ will remain there if the dot shape is changed. An ABS may be seen as consisting of two MZMs. 
The replacement of MZMs by ABSs has no effect 
if the two MZMs making up the ABS have very different coupling strength to the dot. 
If both MZMs 
couple to the dot with equal strength, their effect is to change the effective number $N$ of MZMs involved in the holonomy $W$.
A strong signature distinguishing MZMs from ABSs results from the charge signature: Pinching off one (two) contacts with MZMs has no charge signature (a logarithmic divergence around zero), while pinching off one (two) contacts with ABSs results in a logarithmic divergence (a finite density) around zero.
The parity signature strongly distinguishes five and six MZMs from more than six MZMs from its behavior around $\Delta p = 0,1$. The distinctions are based on the associated ground state degeneracy, similar to the topological Kondo effect \cite{BeriPhysRevLett2012Oct}.

{\em Concluding remarks.---}
We have shown that weakly coupling multiple Majorana zero modes to a chaotic QD leaves a dark subspace of zero-energy modes if the Fermi energy is close to a resonance of the QD. Tuning the shape of the chaotic QD leads to a nontrivial evolution inside the dark subspace, which has observable signatures with universal PDFs.
Since the single-particle QD states are non-degenerate,
our conclusions remain valid if Coulomb repulsion on the QD considered.
The same applies, qualitatively, for Andreev reflections at the QD boundaries, although the precise PDFs may differ in this case, because Andreev reflections change the symmetry class of the QD Hamiltonian to Cartan class D \cite{AltlandPRB1997}.

Estimating the relevant energy scales \cite{Supplementary}, we expect that our proposal is challenging but within reach of the InAs/Al platform \cite{QuantumPhysRevB2023Jun}. Another near-term experimental study could utilize ``poor man's'' Majorana fermions \cite{LeijnsePhysRevB2012Oct, FulgaNewJPhys2013Apr, DvirarXiv2022Jun}
or other types of degenerate qubits \cite{EarnestPhysRevLett2018Apr, LarsenPhysRevLett2020Jul, KalashnikovPRXQuantum2020Sep, Gyenis2021, GrimmNature2020Aug,Campagne-IbarcqNature2020Aug} interfaced by a single, chaotic element.

We acknowledge support by the European Research Council (ERC) under the European Union’s Horizon 2020 research and innovation program under grant agreement No.~856526, and from the Deutsche Forschungsgemeinschaft (DFG) project grant 277101999 within the CRC network TR 183 (subprojects A02, A03, C01, and C03), the Danish National Research Foundation, the Danish Council for Independent Research $\vert$ Natural Sciences, and a research
grant (Project 43951) from VILLUM FONDEN.

\nocite{QuantumPhysRevB2023Jun, SteinerPhysRevRes2020Aug, MunkPhysRevRes2020Aug, KarzigPhysRevLett2021Feb, BargerbosPRXQuantum2022Jul, SnizhkoPhysRevLett2019Aug, AltlandPRB1997, IthierPhysRevB2005Oct, AleinerPhysRevLett2001Nov}

\bibliographystyle{apsrev4-2}
\bibliography{refs_MajoranaBilliard}

\clearpage
\appendix
\onecolumngrid

\setcounter{figure}{0}
\renewcommand{\thefigure}{S.\arabic{figure}}

\section*{Supplementary material: Non-Abelian holonomy of Majorana zero modes coupled to a chaotic quantum dot}

\tableofcontents

\section{Derivation of the Wilson loop operator}

We start with an elementary derivation of Eq.\ (\refeqWC) and then discuss the derivation of Eq.\ (\refeqWtriangle) of the main text.
It is sufficient to derive Eq.\ (\refeqWC) for a square loop in shape space. For small loop area, we may approximate
\begin{equation}
  {\cal P}(x,y) = e^{x X + y Y} {\cal P} e^{-(x X + y Y)},
\end{equation}
where $X$ and $Y$ are real antisymmetric matrices and we abbreviated ${\cal P} = {\cal P}(0,0)$. The matrices $X$ and $Y$ are related to the derivatives $\partial {\cal P}/\partial x$ and $\partial {\cal P}/\partial y$ as
\begin{align}
  \frac{\partial {\cal P}}{\partial x} =&\, (1-{\cal P}) X {\cal P} - {\cal P} X (1-{\cal P}),\nonumber \\
  \frac{\partial {\cal P}}{\partial y} =&\, (1-{\cal P}) Y {\cal P} - {\cal P} Y (1-{\cal P}).
  \label{eq:Pderivatives}
\end{align}
For the Wilson loop operator for the loop $(0,0) \to (1,0) \to (1,1) \to (0,1) \to (0,0)$, we then obtain ${\cal W} {\cal P} = W_4 W_3 W_2 W_1$, with
\begin{align}
  {\cal W}_1 =&\, e^X e^{-{\cal P} X} {\cal P}, \nonumber \\
  {\cal W}_2 =&\, e^Y e^X {\cal P} e^{-e^{-X} Y e^{X} {\cal P}} e^{-X}, \nonumber \\
  {\cal W}_3 =&\, e^Y {\cal P} e^{e^{-Y} X e^Y {\cal P}} e^{-Y} e^{-X}, \nonumber \\
  {\cal W}_4 =&\, {\cal P} e^{Y {\cal P}} e^{-Y}.
\end{align}
Expanding ${\cal W}$ to leading order in $X$ and $Y$, we find
\begin{align}
  \label{eq:WP}
  {\cal W} {\cal P} =&\, {\cal P} + {\cal P} Y (1-{\cal P}) X {\cal P} - {\cal P} X (1-{\cal P}) Y {\cal P}  + \ldots \nonumber \\ =&\,
  {\cal P}
  + {\cal P} [\partial {\cal P}/\partial y,\partial {\cal P}/\partial x] {\cal P}
  + \ldots,
\end{align}
where the dots indicate terms of higher order in the derivatives $\partial {\cal P}/\partial x$ and $\partial {\cal P}/\partial y$ and we used Eq.\ (\ref{eq:Pderivatives}) to eliminate $X$ and $Y$ in favor of the partial derivatives $\partial {\cal P}/\partial x$ and $\partial {\cal P}/\partial y$.
Equation (\refeqWtriangle) of the main text then follows immediately.

The projector ${\cal P}(\vv)$ projects on the $(N-2)$-dimensional linear subspace of $\RR^N$ satisfying the orthogonality condition (\refeqvortho). As a function of the vector $\vv = (v_1,v_2,\ldots,v_N)$, the projection operator ${\cal P}(\vv)$ 
is defined above Eq.\ (\refeqWtriangle) in the main text.
For the partial derivative to the shape coordinate $x$ we then find
\begin{align}
  \frac{\partial {\cal P}}{\partial x} =&\,
  -\frac{2}{\mbox{tr}\, {\cal Q}} \mbox{Re}\,
  \left[ \vd_{x} (\vv^{\dagger} \vv) \vv^{\dagger} + \vv (\vv^{\dagger} \vv) \vd_x^{\dagger} - \vd_{x} (\vv^{\dagger} \vv^*) \vv^{\rm T} - \vv (\vv^{\dagger} \vv^*) \vd_x^{\rm T} \right] + \ldots
\end{align}
where $\vd_x = \partial \vv/\partial x$, ${\cal Q}$ 
is defined above Eq.\ (\refeqWtriangle) in the main text,
and the dots indicate terms that vanish upon left or right multiplication with ${\cal P}$. Similarly, for the partial derivative $\partial {\cal P}/\partial y$ we find
\begin{align}
  \frac{\partial {\cal P}}{\partial y} =&\,
  -\frac{2}{\mbox{tr}\, {\cal Q}} \mbox{Re}\,
  \left[ \vd_{y} (\vv^{\dagger} \vv) \vv^{\dagger} + \vv (\vv^{\dagger} \vv) \vd_y^{\dagger} - \vd_{y} (\vv^{\dagger} \vv^*) \vv^{\rm T} - \vv (\vv^{\dagger} \vv^*) \vd_y^{\rm T} \right] + \ldots,
\end{align}
again up to terms that vanish upon left or right multiplication with ${\cal P}$.
Upon calculation of the commutator in Eq.\ (\ref{eq:WP}), one obtains Eq.\ (\refeqWtriangle) of the main text with straightforward algebra.

\section{Principal values of the Wilson loop operator}

\subsection{Asymptotic distribution of the principal values}
\label{app:principal_values_asymptotic}

The infinitesimal Wilson loop operator $w$ is an antisymmetric $(N-2) \times (N-2)$ matrix acting in the basis of dark MZMs. It is obtained by projecting an antisymmetric real $N \times N$ matrix ${\cal D}$ down to the $(N-2)$-dimensional subspace of the dark Majoranas,
\begin{equation}
  w = {\cal P} {\cal D} {\cal P},
  \label{eq:Wtriangle2}
\end{equation}
with
\begin{equation}
  {\cal D} = \frac{2}{\mbox{tr}\, {\cal Q}}
  \mbox{Re}\, [\vd_{x}^* (\vv^{\dagger} \vv) \vd_{y}^{\rm T}
    - \vd_{x} (\vv^{\dagger} \vv^*) \vd_{y}^{\rm T} - (x \leftrightarrow y) ],
  \label{eq:Wtriangle3}
\end{equation}
compare with Eq.\ (\refeqWtriangle). Because the real part is the sum of dyadic products of four linearly independent real vectors, the matrix ${\cal D}$ has rank $\le 4$. Since the projection on the dark subspace does not increase the rank, the rank of the Wilson loop operator is also $\le 4$. This motivates the form (\refeqwform) of the Wilson loop operator $w$ with a single principal value $\lambda$ if $N=4$ or $N=5$ and two principal values $\lambda_{1,2}$ if $N \ge 6$.

The quadratic level repulsion of the principal value distribution for $N \geq 6$ is generic for real antisymmetric matrices \cite{AltlandPRB1997}. The power-law distributions around $\lambda=0$ (for $N=4$, $5$) and $\lambda_{1,2} = 0$ (for $N\ge 6$) require some discussion. If $N=4$ and $N=6$, $w$ is a generic real antisymmetric matrix of size $2 \times 2$ or $4 \times 4$, respectively. Random real antisymmetric matrices have no level repulsion at zero \cite{AltlandPRB1997}, so that $P(\lambda)$ and $P(\lambda_1,\lambda_2)$ are constant if $\lambda \to 0$ (for $N = 4$) or if $\lambda_1 \to 0$ while $\lambda_2 \neq 0$ or $\lambda_2 \to 0$ while $\lambda_1 \neq 0$ (for $N=6$). If $N=5$, there is only a single nonzero principal value $\lambda$. It is zero if and only if $w = 0$. Since a real antisymmetric $3 \times 3$ matrix has three degrees of freedom, it follows immediately that there is quadratic repulsion $P(\lambda) \propto \lambda^2$ for small $\lambda$.
Finally, if $N \ge 6$, a vanishing principal value implies that the infinitesimal Wilson loop operator $w$ has a rank $\le 2$. This can happen if the real antisymmetric $N \times N$ matrix ${\cal D}$ has rank $\le 2$ or if ${\cal D}$ has rank $4$ and the projection ${\cal P}$ onto the dark subspace lowers the rank. The matrix ${\cal D}$ has rank $\le 2$ if the real and imaginary parts of the vectors $\vd_x$ and $\vd_y$ are linearly dependent, which requires to fine tune $N - 3$ parameters. A lowering of the rank upon projection to the dark subspace, on the other hand, only requires fine tuning of $N - 5$ parameters. The latter scenario dominates and we conclude that $P(\lambda_1,\lambda_2) \propto \lambda_1^{N-6}$ if $\lambda_1 \to 0$ and $\lambda_2 \neq 0$. Similarly, $P(\lambda_1,\lambda_2) \propto \lambda_2^{N-6}$ if $\lambda_2 \to 0$ and $\lambda_1 \neq 0$. (Note that for $N=6$ one may also use the alternative argument given above.)

The asymptotic distribution for large principal values follows from the observation that asymptotically $w \propto s^{-2}$, where $s$ is the spacing to the nearest energy level, see Eq.\ (\refeqvxy). Since the probability density to find small level spacings $s$ is proportional to $s^2$ for small $s$, it follows that the largest principal value $\lambda_{\rm max} = \max(\lambda_1,\lambda_2)$ of $w$ has a probability distribution $\propto \lambda_{\rm max}^{-5/2}$ for large $\lambda_{\rm max}$.

\subsection{Numerical results}

\begin{figure*}
\includegraphics[width=1\columnwidth]{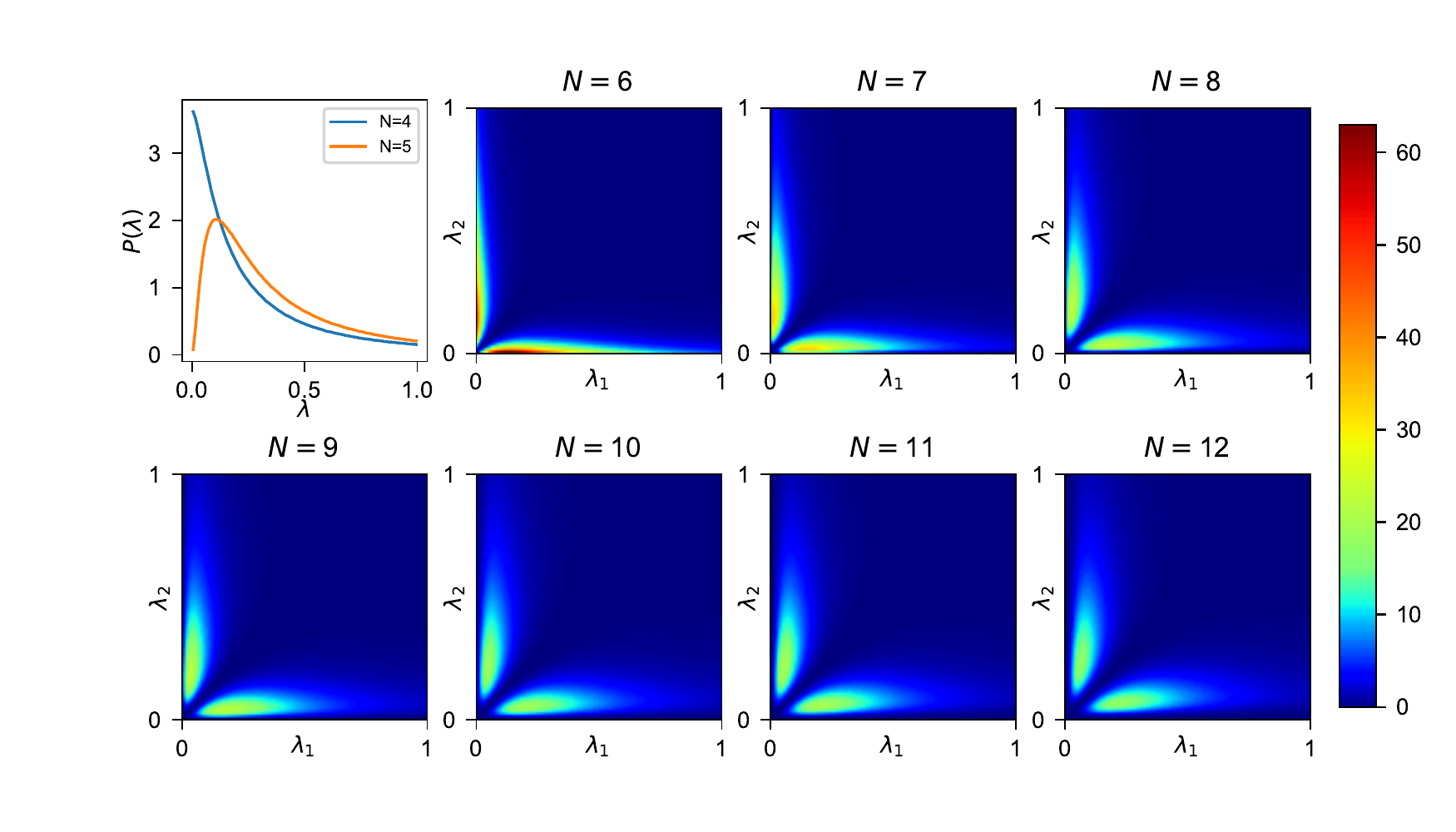}
\caption{Probability distribution for the principal value $\lambda$ of the Wilson loop operator (\refeqWtriangle) for $N=4$, $5$ (upper left) and joint probability distribution for the principal values $\lambda_{1,2}$ for $N=6$,\ldots,$12$ (remaining panels). The distributions are obtained by numerically sampling $10^7$ realizations of random matrices as in Eq.~(\refeqHxy) with $M = 20$. The coupling coefficients $\eta_j = \eta$ have been set equal for all $N$ contacts. 
\label{fig:S1}}
\end{figure*}

\begin{figure*}
\begin{tabular}{c}
      \includegraphics[width=0.9\columnwidth]{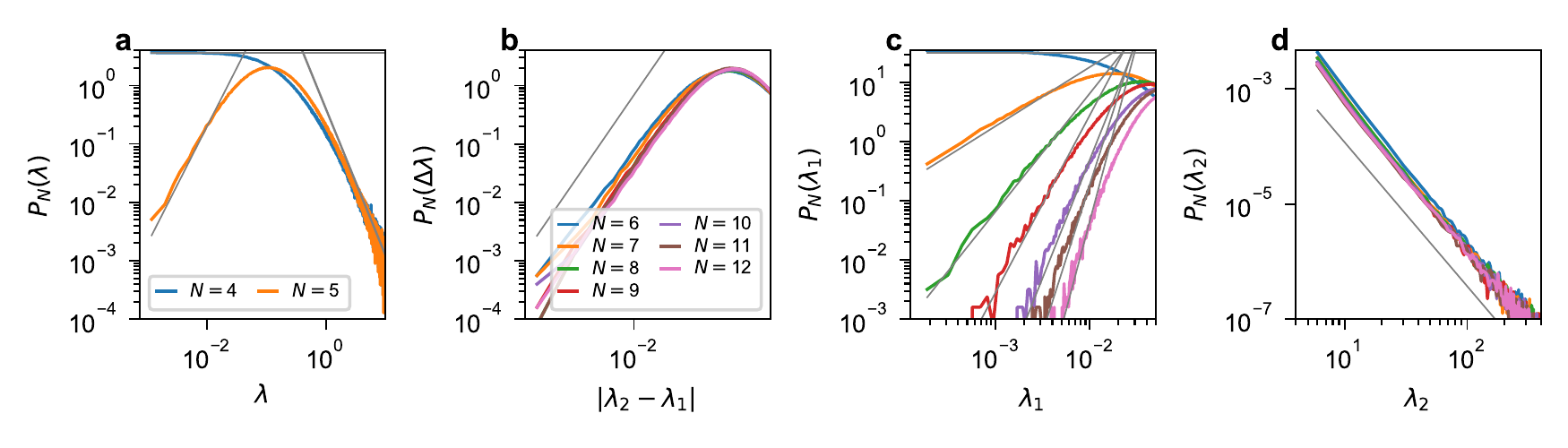} \\
\end{tabular}
\caption{Asymptotics of the distribution of the principal values of the Wilson loop operator Eq.~(\refeqWtriangle) displayed in Fig.~\ref{fig:S1}. The probability distribution $P_N(\lambda)$ for $N=4$, $5$ is shown in panel (a), together with power law fits $P_5(\lambda) \propto \lambda^2$ for $\lambda \ll 1$ and $P_{N}(\lambda) \propto \lambda^{-5/2}$ for $\lambda \gg 1$. Panels (b)--(d) show marginal distributions obtained from the joint probability distribution $P_N(\lambda_1,\lambda_2)$ for $N=6$, \ldots, $12$: the probability distribution $P_N(\Delta \lambda) = \int_0^\infty d\lambda_1 d\lambda_2 P_N(\lambda_1, \lambda_2) \delta(\Delta \lambda - \lambda_1 + \lambda_2) \propto (\Delta \lambda)^2$ near a degeneracy, the probability distribution $P_N(\lambda_1) = \int_{\lambda_1}^\infty d\lambda_2 P_N(\lambda_1, \lambda_2) \propto \lambda_1^{N-6}$ near $\lambda_1 = 0$ and generic $\lambda_2 > \lambda_1$ (c), and the probability distribution $P_N(\lambda_2) = \int_0^{\lambda_2} d\lambda_1 P_N(\lambda_1, \lambda_2) \propto \lambda_2^{-5/2}$, for $\lambda_2 \gg 1$ and generic $\lambda_1 < \lambda_2$ (d). The grey lines indicate the asymptotic power laws. Numerical data were obtained by sampling $10^7$ independent realizations of the random matrix model (\refeqHxy) with $M=20$.
\label{fig:S2}}
\end{figure*}

To obtain statistical distributions numerically, we generate $M \times M$ random matrices $H_0$, $H_x$, and $H_y$ with $M = 20$ and diagonalize $H_0$ to obtain the vectors $\vv$, $\vd_x$, and $\vd_y$ as described in the main text.
The numerically sampled joint distribution function of the principal values of the Wilson loop operator for $N = 4,5,...,12$ of MZMs coupled to the quantum dot with equal coupling strengths $\eta = \eta_j$ is shown in Fig.~\ref{fig:S1}.
The asymptotic distribution around a degeneracy $\lambda_1 = \lambda_2$ for $N \geq 6$ and for small and large principal values $\lambda_i$ is shown in Fig.~\ref{fig:S2}(a)-(c).
The numerical results agree with the analytical estimates in the previous subsection.
We verified that the power laws for small and large principal values $\lambda_{1,2}$ and the level repulsion around a degeneracy $\lambda_1 = \lambda_2$ do not depend on the assumption of equal coupling strengths for all $N$ tunnel contacts (data now shown). We also verified that the numerical results do not change when the matrix dimension $M$ is increased.

\subsection{Wilson loop operator for large loops $A \gg 1$} 
\label{app:Wilson_infiniteA}

To avoid a systematic change of the mean level spacing while performing a loop in shape space, for large loops the random matrix model (\refeqHxy) is replaced by the model
\begin{equation}
    H(x,y) = H_0 \cos \theta + H_x \sin \theta \cos \varphi + H_y \sin \theta \sin \phi, 
    \label{eq:H_infiniteA}
\end{equation}
with $\sin \theta \cos \varphi = x/\sqrt{M}$ and $\sin \theta \sin \varphi = y/\sqrt{M}$, which manifestly preserves the level density upon varying the parameters $\theta$ and $\varphi$.
The Wilson loop operator (\refeqWC) can be expressed as the path-ordered product of projectors along a loop ${\cal C}$ in shape space,
\begin{equation}
    W = {\cal L} \prod_{j} {\cal P}(\vv_{(\theta_j,\varphi_j)}),
\end{equation}
where the ``shape-space points'' $(\theta_j, \varphi_j)$ form are a dense set of points along the loop ${\cal C}$ and $\vec{v}_\vec{(\theta_j,\varphi_j)}$ is the coupling vector obtained by exactly diagonalizing the random matrix model Eq.~\eqref{eq:H_infiniteA}. Formally, the product should be taken over an infinite number of projectors with infinitesimal spacing to the next projection along the loop. For numerical evaluation, we choose a finite discretization of the loop. Due to the finite overlap between adjacent projectors ${\cal P}$, the magnitude of the singular values of the multiplied projectors shrinks. As a consequence thereof, the nonzero principal values of the resulting Wilson loop become smaller than one and the product of projection operators loses orthogonality. We correct this numerical error by performing a singular value decomposition of the multiplied projectors after each step and reset the magnitude of the nonzero singular values back to 1. With this procedure, we find convergence after $3 \times 2^8$ steps along the loop.

\begin{figure*}
\includegraphics[width=0.9\columnwidth]{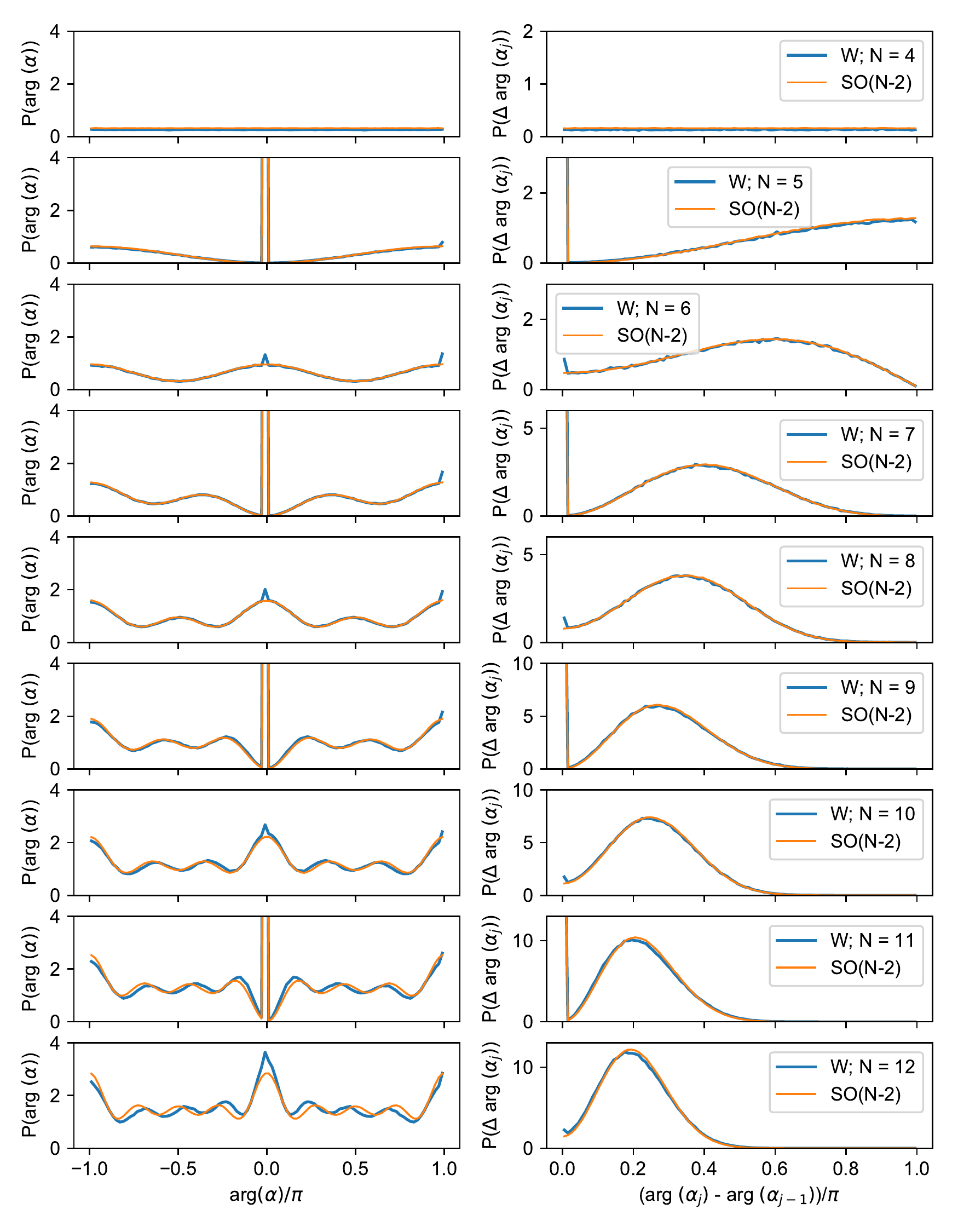}
\caption{Probability density for angle $\alpha_j$ (left) and spacing distribution (right) of the angles $\alpha_j$ of the eigenvalues $e^{\pm i \alpha_j}$ of the Wilson loop operator $W$ for a shape-space loop with area $A \gg 1$. Blue curves are obtained from the model Eq.~\eqref{eq:H_infiniteA} using the curve ${\cal C} = (\theta(t), \varphi(t))^{\rm T} = \frac{\pi}{2}\Theta(1-t) (t, 0)^{\rm T}  + \frac{\pi}{2}\Theta(t-1) \Theta(2-t) (1, t-1)^{\rm T} + \frac{\pi}{2}\Theta(t-2) (3-t, 1)^{\rm T},\ t \in [0,3]$ where $\Theta(t)$ is the Heaviside step function. Orange curves give distributions for uniformly distributed $W \in \mbox{SO}(N-2)$. Distributions are obtained by sampling $10^5$ realizations of the random matrix model \eqref{eq:H_infiniteA} with matrix size $M = 64$. The distributions for the $\text{SO}(N-2)$ random matrix model are obtained by sampling $10^6$ realizations. 
\label{fig:S_Wilson}}
\end{figure*}

In Fig.~\ref{fig:S_Wilson} we compute the density of the angle $\alpha$ of the eigenvalues $e^{\pm i \alpha}$ of the Wilson loop operator as well as their spacing and compare the result to a prediction from $\text{SO}(N-2)$, where $N$ is the total number of MZM coupled to the dot during the shape space loop. 
For $N$ odd, there is always an eigenvalue $1$, which is seen in the density and spacing distribution in Fig.~\ref{fig:S_Wilson}. We find that the angle density and spacing distribution agree well with that of a uniformly distributed special orthogonal matrix $\in \text{SO}(N-2)$, up to an increased density near $\alpha = 0$, $\pi$, which we attribute to a residual numerical error due to the discretization of the loop and a bias of the numerical diagonalization algorithm. 
The matrix dimension $M$ at which convergence is reached depends on the dimension of the Wilson loop operator $N-2$. In the results shown in Fig.~\ref{fig:S_Wilson}, we use $M = 64$ where we find very good agreement with $\text{SO}(N-2)$ for $N \leq 9$ and small deviations for $N \geq 10$. We verified that these deviations becomes smaller as $M$ is increased to $M = 96$, where only a slight deviation for $N = 12$ remains (results not shown).

\clearpage
\section{Parity signature}
\label{app:parity}

\subsection{Derivation of Eq.\ (\refeqQf)}
\label{app:parity_1}

In the main text, the parity signature is discussed for the setup of Fig.\ \reffigone, in which two MZMs $\hat \gamma_1$ and $\hat \gamma_2$ are adiabatically coupled and decoupled from the quantum dot. The MZMs  $\hat \gamma_1$ and $\hat \gamma_2$ are initialized, such that the associated complex fermion $\hat f_1 = (\hat \gamma_1 + i \hat \gamma_2)/2$ is unoccupied. The parity signature $\Delta p$ corresponds to the change of the occupation $\hat f_1^{\dagger} \hat f_1$ after (i) adiabatically coupling $\hat \gamma_1$ and $\hat \gamma_2$ to the quantum dot, (ii) executing a loop in shape space with holonomy $W$, and (iii) again adiabatically decoupling $\hat \gamma_1$ and $\hat \gamma_2$.

Since the total number of MZMs must be even, the system must harbor additional MZMs not coupled to the quantum dot if $N$ is odd. We denote the number of additional MZMs with $N'$ and require that $N+N'$ is even. By construction, the additional MZMs not coupled to the quantum dot are always in the dark subspace. The $N'$ additional MZMs will appear explicitly in the equations in this Section. They are left implicit in the corresponding equations in the main text.

The two MZMs $\hat \gamma_1$ and $\hat \gamma_2$ are initialized such that the fermion $\hat f_1 = (1/2) (\hat \gamma_1 + i \hat \gamma_2)$ is unoccupied,
\begin{equation}
  p_{\rm i} \equiv \langle \hat f_1^{\dagger} \hat f_1 \rangle_{\rm i} = 0
\end{equation}
or, equivalently, 
\begin{equation}
  \langle \hat{\gamma}_{2} \hat{\gamma}_{1} \rangle_{\rm i} = -i.
\end{equation}
Here, the subscript ``i'' refers to the expectation value after initialization. The dark MZMs $\hat{\Gamma}_{n},n=3,4,...N + N'-2$ are initialized in the vacuum state of the fermions $\hat{f}_{p}=\frac{1}{2}(\hat{\Gamma}_{2p-1}+i\hat{\Gamma}_{2p})$, {\em i.e.}, so that $\langle\hat{f}_{p}^{\dagger}\hat{f}_{p}\rangle_{\rm i}=0$ or, equivalently,
\begin{equation}
  \langle\hat{\Gamma}_{2p}\hat{\Gamma}_{2p-1}\rangle_{{\rm i}}=-i.
\end{equation}
The loop in shape space evolves the expectation values according to Eq.\ (\refeqGammaW),
\begin{equation}
\langle\hat{\Gamma}_{k}\hat{\Gamma}_{l}\rangle_{\rm f}=-i\sum_{j=1}^{(N+N^{\prime}-2)/2}(W_{2j,k}W_{2j-1,l}-W_{2j-1,k}W_{2j,l}), \label{eq:app_exp_GammaW}
\end{equation}
where $\langle...\rangle_{\rm f}$ denotes the expectation value after the shape space loop and, to keep the notation simple, we wrote $\hat{\Gamma}_{1}$ and $\hat{\Gamma}_2$ instead of $\hat{\gamma}_{1}$ and $\hat{\gamma}_{2}$, respectively. Thus, the fermion parity $p_{\rm f} = \langle \hat f_1^{\dagger} \hat f_1 \rangle_{\rm f}$ reads
\begin{align}
  p_{\rm f} 
  =&\, \frac{1}{2}-\frac{i}{2}\langle\hat{\gamma}_{2}\hat{\gamma}_{1}\rangle_{\rm f}\nonumber \\
  =&\, \frac{1}{2}-\frac{1}{2}\sum_{j=1}^{(N+N^{\prime}-2)/2}(W_{2j,2}W_{2j-1,1}-W_{2j-1,2}W_{2j,1}).\label{eq:app_p_bothMZM}
\end{align}
The {\em parity signature} $\Delta p = p_{\rm f} - p_{\rm i}$ is the difference of the fermion parity $\langle \hat f_1^{\dagger} \hat f_1 \rangle$ with and without performing the loop in phase space. Since the MZMs $\hat \gamma_1$ and $\hat \gamma_2$ were initialized such that $p_{\rm i} = 0$, $\Delta p$ equals $p_{\rm f}$. This gives Eq.\ (\refeqQf) of the main text.
For small loop area $A\ll1$, the parity signature $\Delta p$ may be expressed in terms of the infinitesimal Wilson loop operator $w$. To lowest order in $A$, one finds
\begin{equation}
  \Delta p=\frac{1}{4}A^{2}\sum_{p=2}^{(N+N^{\prime}-2)/2}
  \left[ (w_{2p-1,1} - w_{2p,2})^{2} +
  (w_{2p-1,2} + w_{2p,1})^{2} \right]. \label{eq:app_p_bothMZM_smallA}
\end{equation}

\subsection{Alternative protocol for parity signature}
\label{app:parity_2}

\begin{figure}[t]
\includegraphics[width=0.4\columnwidth]{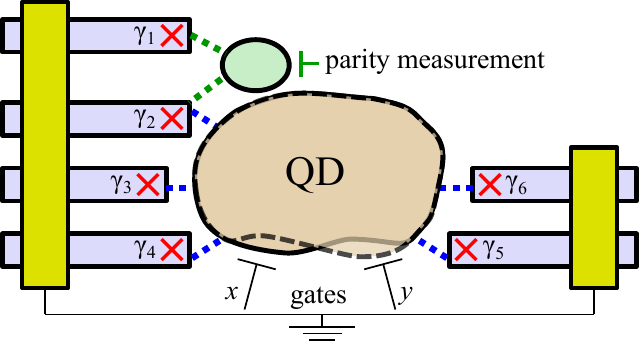}
\caption{
  \label{fig:1b}
Chaotic quantum dot coupled to $N-1$ Majorana zero modes (MZMs) $\hat{\gamma}_j$. The alternative protocol for a parity signature involves a measurement of the change of the occupation of the fermion $\hat f_1 = (\hat \gamma_1 + i \hat \gamma_2)/2$ upon switching the coupling $\hat \gamma_2$ to the quantum dot is adiabatically on and off before and after a loop in shape space is executed, whereas $\hat \gamma_1$ remains decoupled from the quantum dot at all times. }
\end{figure}

A variation of the protocol the the parity signature makes use of the setup shown in Fig.\ \ref{fig:1b}. Of the two MZMs $\hat \gamma_1$ and $\hat \gamma_2$, only the coupling of $\hat \gamma_2$ to the quantum dot is switched on and off, whereas $\hat \gamma_1$ remains decoupled from the quantum dot at all times. In this case the total number of MZMs coupled to the quantum dot is $N-1$, not $N$.

Since $\hat \gamma_1$ is not involved in the phase-space loop, for the holonomy $W$ one then has $W_{j,1}=W_{1,j}=\delta_{j,1}$. From Eq.\ (\ref{eq:app_p_bothMZM}) we then immediately obtain that
\begin{equation}
  \Delta p=\frac{1}{2}-\frac{1}{2}W_{2,2}.\label{eq:app_p_singleMZM}
\end{equation}
For small loop area $A\ll1$ this simplifies to
\begin{equation}
  \Delta p=\frac{1}{4}A^{2} \sum_{p=3}^{N+N^{\prime}-2}w_{p,2}^{2}.\label{eq:app_p_singleMZM_smallA}
\end{equation}

\subsection{Distribution of the parity signature for large $\Delta p$}
\label{app:parity_large_p}

We now show that for small areas $A \ll 1$ of the phase space loop the probability distribution has an algebraic tail $\propto \Delta p^{-7/4}$ for large $\Delta p$. As the perturbative expressions 
\eqref{eq:app_p_bothMZM_smallA} and \eqref{eq:app_p_singleMZM_smallA} 
are proportional to the absolute square of matrix elements of $w$, the parity signature contains terms proportional to the fourth power of the inverse level spacing. At rare events where a neighboring level approaches the resonant level, these terms dominate the parity signature and give rise to the algebraic tail of the distribution at large $\Delta p$. Since the probability density of small level spacings $s$ (normalized to the average level spacing) is proportional to $s^2$ if time-reversal symmetry is broken, the power law of the algebraic tail is 
\begin{align}
    P(\Delta p) & \propto\, \
  \int ds s^{2}\delta(\Delta p-\frac{1}{s^{4}}) \nonumber \\
  & \propto\,
  \Delta p^{-7/4}.
\end{align}

\subsection{Distribution of the parity signature for small $\Delta p$}
\label{app:parity_small_p}

In this appendix, we derive the power laws of the probability distribution $P(\Delta p)$ for $\Delta p \ll 1$. As before, we consider small areas $A \ll 1$ of the shape space loop. We consider the cases that $\hat \gamma_1$ remains decoupled during the phase space loop (alternative protocol), for which $\Delta p$ is given by Eq.\ (\ref{eq:app_p_singleMZM_smallA}), and that both $\hat{\gamma}_1$ and $\hat \gamma_2$ are coupled during the phase space loop (protocol of the main text), for which $\Delta p$ is given by Eq.\ (\ref{eq:app_p_bothMZM_smallA}), separately. In the latter case we also have to distinguish even $N$ and odd $N$.

{\em $\hat{\gamma}_{1}$ decoupled during phase-space loop.---}
Because $\hat \gamma_1$ remains decoupled from the quantum dot, the total number of MZMs coupled to the quantum dot is $N-1$ and the total number of ``dark'' MZMs coupled to the quantum dot is $N-3$. The total number of MZMs not coupled to the quantum dot is $N' + 1$. 
Since the MZMs $\hat \gamma_1$ and $\hat \Gamma_{p}$ with $p = N-1, \ldots, N+N'-2$ are not involved in the phase space loop, the corresponding elements $w_{1,2}$ and $w_{p,2}$, $p = N-1, \ldots, N+N'-2$, of the infinitesimal Wilson loop operator are automatically zero and they do not enter into the expression (\ref{eq:app_p_singleMZM_smallA}) for the parity signature $\Delta p$. Hence, the parity signature $\Delta p = 0$ if and only if
\begin{equation}
  w_{p,2} = 0,\ \ p=3,\ldots,N-2.
  \label{eq:wpeq}
\end{equation}
Since $\Delta p$ is quadratic in the matrix elements $w_{p,2}$, see Eq.\ (\ref{eq:app_p_singleMZM_smallA}), for a generic antisymmetric matrix $w$ one therefore expects that the $N-4$ constraints (\ref{eq:wpeq}) lead to $P(\Delta p) \propto (\Delta p)^{(N-4)/2-1}$. However, the infinitesimal Wilson loop operator $w$ is not a generic antisymmetric matrix, as can be seen from the principal-value decomposition (\refeqwform). This leads to a modification of the small-$\Delta p$ power law for $N \ge 8$. 

With $\hat \gamma_1$ decoupled from the quantum dot, the elements of the $(N-3) \times (N-3)$ orthogonal matrix $O$ in the principal-value decomposition (\refeqwform) are labeled $O_{p,j}$, with $p=2,3,4,\ldots,N-2$ and $j=1,2,\ldots,N-3$. From of Eq.\ (\refeqwform) we then have
\begin{equation}
  w_{p,2} = O_{p,1} \lambda_1 O_{2,2} - O_{p,2} \lambda_1 O_{2,1}
  + O_{p,3} \lambda_2 O_{2,4} - O_{p,4} \lambda_2 O_{2,3}.
\end{equation}
For generic values of $\lambda_1$ and $\lambda_2$ and of the coefficients $O_{p,1}$, $O_{p,2}$, $O_{p,3}$, and $O_{p,4}$, $p = 3,\ldots,N-2$, the equations (\ref{eq:wpeq}) can be seen as a set of $N-4$ coupled linear equations for the four coefficients $O_{2,1}$, $O_{2,2}$, $O_{2,3}$, and $O_{2,4}$. Since the number of constraints imposed on four coefficients cannot exceed four, the number of constraints imposed by these equations is $\min(N-4,4)$. As $\Delta p$ depends quadratically on the $w_{p,1}$, one thus finds
\begin{equation}
  P(\Delta p) \propto (\Delta p)^{\min(N-4,4)/2-1},\ \ \Delta p \ll 1.
  \label{eq:Psmallp}
\end{equation}
One verifies that imposing constraints on the coefficients $O_{p,1}$, $O_{p,2}$, $O_{p,3}$, and $O_{p,4}$, $p = 3,\ldots,N-2$, or setting $\lambda_1 = 0$ and/or $\lambda_2 = 0$ may lift some of the constraints on the coefficients $O_{2,1}$, $O_{2,2}$, $O_{2,3}$, and $O_{2,4}$, but only at the cost of additional constraints on these other coefficients, so that the effective number of constraints and, hence, the exponent of $\Delta p$ in Eq.\ (\ref{eq:Psmallp}) is not lowered.

{\em Both $\hat{\gamma}_{1}$ and $\hat{\gamma}_{2}$ coupled during the phase-space loop; $N$ even.---}
If both $\hat{\gamma}_{1}$ and $\hat{\gamma}_{2}$ are coupled during the phase-space loop and $N$ is even, $\Delta p = 0$ if and only if 
\begin{align}
  &w_{2p-1,1} - w_{2p,2} =\, 0,\nonumber \\
  &w_{2p-1,2} + w_{2p,1} =\, 0,\ \ 
  p=2,3,\ldots,(N-2)/2.
  \label{eq:wcond}
\end{align}
(As before, we need not consider coefficients $w_{p,1}$ and $w_{p,2}$ with $p > N-2$ because these correspond to MZMs that are decoupled from the quantum dot at all times.) Substituting the principal-value decomposition (\refeqwform), we obtain the $N-4$ equations
\begin{align}
  &\, O_{2p-1,1} \lambda_1 O_{1,2} - O_{2p-1,2} \lambda_1 O_{1,1}
  + O_{2p-1,3} \lambda_2 O_{1,4} - O_{2p-1,4} \lambda_2 O_{1,3}
  \nonumber \\ &\, \ \ \ \ \ \ \ \ \ \ \mbox{} 
  - O_{2p,1} \lambda_1 O_{2,2} + O_{2p,2} \lambda_1 O_{2,1}
  - O_{2p,3} \lambda_2 O_{2,4} + O_{2p,4} \lambda_2 O_{2,3} = 0, 
  \nonumber \\
  &\, O_{2p-1,1} \lambda_1 O_{2,2} - O_{2p-1,2} \lambda_1 O_{2,1}
  + O_{2p-1,3} \lambda_2 O_{2,4} - O_{2p-1,4} \lambda_2 O_{2,3}
   \nonumber \\ &\, \ \ \ \ \ \ \ \ \ \ \mbox{} 
 + O_{2p,1} \lambda_1 O_{1,2} - O_{2p,2} \lambda_1 O_{1,1}
  + O_{2p,3} \lambda_2 O_{1,4} - O_{2p,4} \lambda_2 O_{1,3} = 0, 
  \label{eq:Oeq}
\end{align}
with $p=2,3,\ldots,(N-2)/2$. These equations may be seen as $(N-4)$ coupled linear equations for the eight coefficients $O_{1,2}$, $O_{1,2}$, $O_{1,3}$, $O_{1,4}$, $O_{2,1}$, $O_{2,2}$, $O_{2,3}$, and $O_{2,4}$. For generic values of the other coefficients, the equations (\ref{eq:Oeq}) imply $\min(N-4,8)$ constraints on the coefficients $O_{1,2}$, $O_{1,2}$, $O_{1,3}$, $O_{1,4}$, $O_{2,1}$, $O_{2,2}$, $O_{2,3}$, and $O_{2,4}$, which gives 
\begin{equation}
  P(\Delta p) \propto (\Delta p)^{\min(N-4,8)/2-1}\ \ \mbox{if $\Delta p \ll 1$}, \label{eq:Pdp1}
\end{equation}
since $\Delta p$ depends quadratically on the elements of the infinitesimal Wilson-loop operator $w$. Again, one verifies that imposing constraints on the coefficients $O_{2p-1,j}$ and $O_{2p,j}$ with $p \ge 2$ and $j=1,2,3,4$ or requiring that $\lambda_1 = 0$ or $\lambda_2 = 0$ does not reduce the total number of constraints imposed and, hence, does not lower the exponent of $\Delta p$ in Eq.\ (\ref{eq:Pdp1}).

{\em Both $\hat{\gamma}_{1}$ and $\hat{\gamma}_{2}$ coupled during the phase-space loop; $N$ odd.---} In this case, the matrix elements $w_{N-1,1}$ and $w_{N-1,2}$ are both zero, so that the conditions (\ref{eq:wcond}) are replaced by
\begin{align}
  & w_{2p-1,1} - w_{2p,2} =\, 0,\nonumber \\
  & w_{2p-1,2} + w_{2p,1} =\, 0,\ \ 
  p=2,3,\ldots,(N-3)/2, \nonumber \\
  & w_{N-2,1} = 0,\nonumber \\ & w_{N-2,2} = 0.
  \label{eq:wcond_Nodd}
\end{align}
Proceeding in the same manner as before, one finds that the number of constraints imposed on the coefficients $O_{1,2}$, $O_{1,2}$, $O_{1,3}$, $O_{1,4}$, $O_{2,1}$, $O_{2,2}$, $O_{2,3}$, and $O_{2,4}$ is $\min(N-3,8)$, so that
\begin{equation}
  P(\Delta p) \propto (\Delta p)^{\min(N-3,8)/2-1}\ \ \mbox{if $\Delta p \ll 1$}. \label{eq:Pdp2}
\end{equation}

\subsection{Numerical results}
\label{app:num_p}

\begin{figure}
\includegraphics[width=0.8\columnwidth]{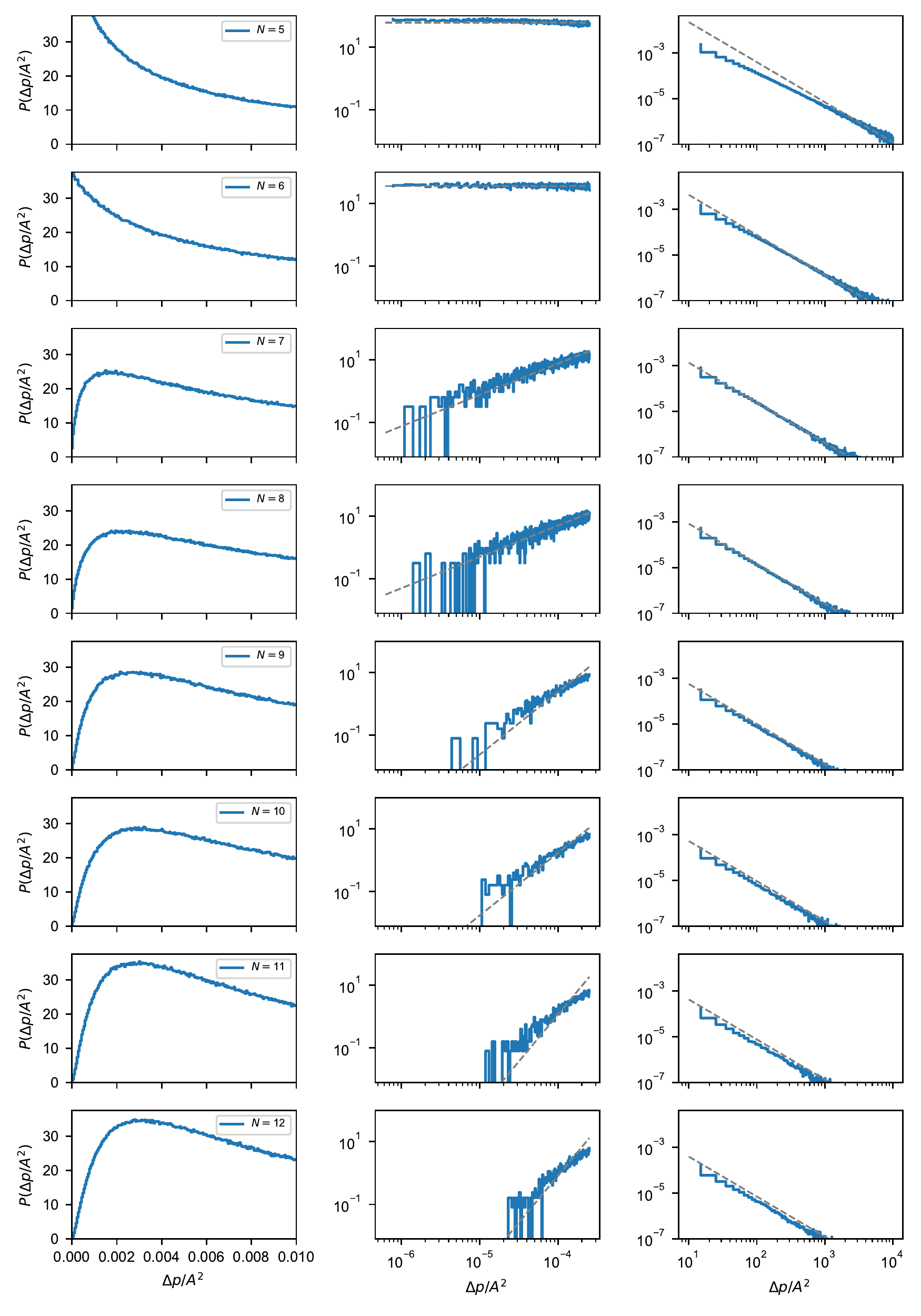}
\caption{Numerically sampled probability distribution of the parity signature $\Delta p$ in the limit of small area $A$ of the phase space loop. Both MZMs $\hat \gamma_1$ and $\hat \gamma_2$ are coupled to the quantum dot during the phase space loop (protocol from the main text and Sec.\ \ref{app:parity_1}, see Fig.\ \reffigone). The total number $N$ of MZMs coupled to the quantum dot during the phase space loop runs from $N=5$ (top) to $N=12$ (bottom). The left panels are identical to the curves shown in Fig.~\reffigthree (left) of the main text. The middle and right panels show the asymptotic distributions for $\Delta p/A^2 \ll 1$ and $\Delta p/A^2 \gg 1$, respectively, with asymptotic power laws indicated by grey lines. The distributions are obtained by sampling over $10^7$ independent realizations of the random matrix model (\refeqHxy).
\label{fig:SM_Qf3_NMzero2}}
\end{figure}

\begin{figure}
\includegraphics[width=0.8\columnwidth]{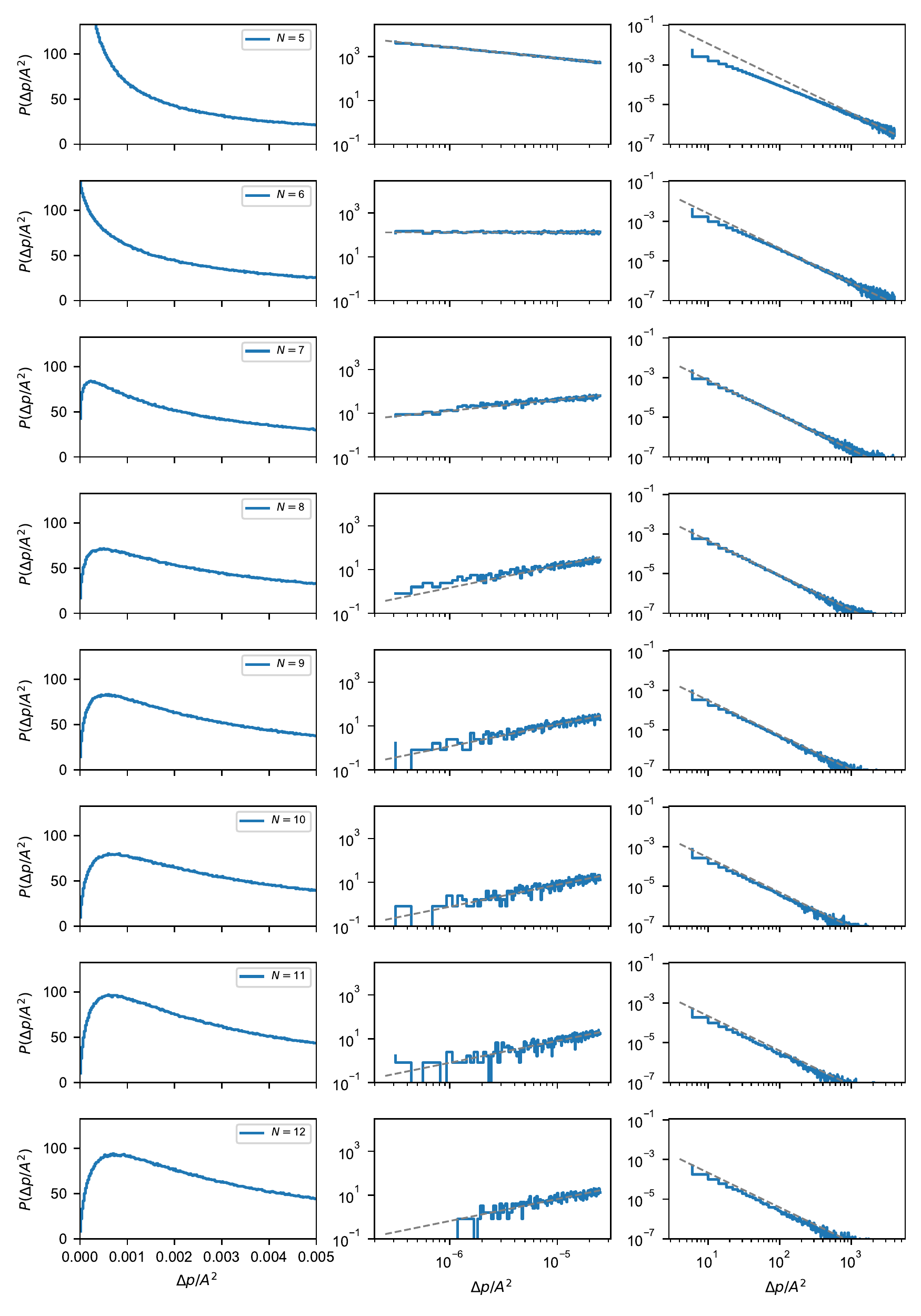}
\caption{
Numerically sampled probability distribution of the parity signature $\Delta p$ in the limit of small area $A$ of the phase space loop (alternative protocol of Sec.\ \ref{app:parity_2}). Of the two MZMs $\hat \gamma_1$ and $\hat \gamma_2$ involved in the parity measurement, only $\hat \gamma_2$ is coupled to the quantum dot during the phase space loop, while $\hat \gamma_1$ remains decoupled, see Fig.\ \ref{fig:1b}. The total number $N-1$ of MZMs coupled to the quantum dot during the phase space loop runs from $N=5$ (top) to $N=12$ (bottom).
The middle and right panels show the asymptotic distributions for $\Delta p/A^2 \ll 1$ and $\Delta p/A^2 \gg 1$, respectively, with asymptotic power laws indicated by grey lines. The distributions are obtained by sampling over $10^7$ independent realizations of the random matrix model (\refeqHxy).
\label{fig:SM_Qf3_NMzero1}}
\end{figure}

Figures \ref{fig:SM_Qf3_NMzero2} and \ref{fig:SM_Qf3_NMzero1} show numerically sampled probability distributions of the parity signature $\Delta p$ for $N = 5,6,...,12$ MZMs involved in the shape space loop in the limit of small loop area $A$, where two MZMs (Fig.~\ref{fig:SM_Qf3_NMzero2}, protocol from main text) and one MZM (Fig.~\ref{fig:SM_Qf3_NMzero1}, alternative protocol) are coupled to the quantum dot after initialization.
The size $M$ of the random matrices is $M=40$. Grey lines indicate power-law fits to the asymptotic distribution for small and large $\Delta p/A^2$ as discussed in the previous subsections.

\begin{figure}
\includegraphics[]{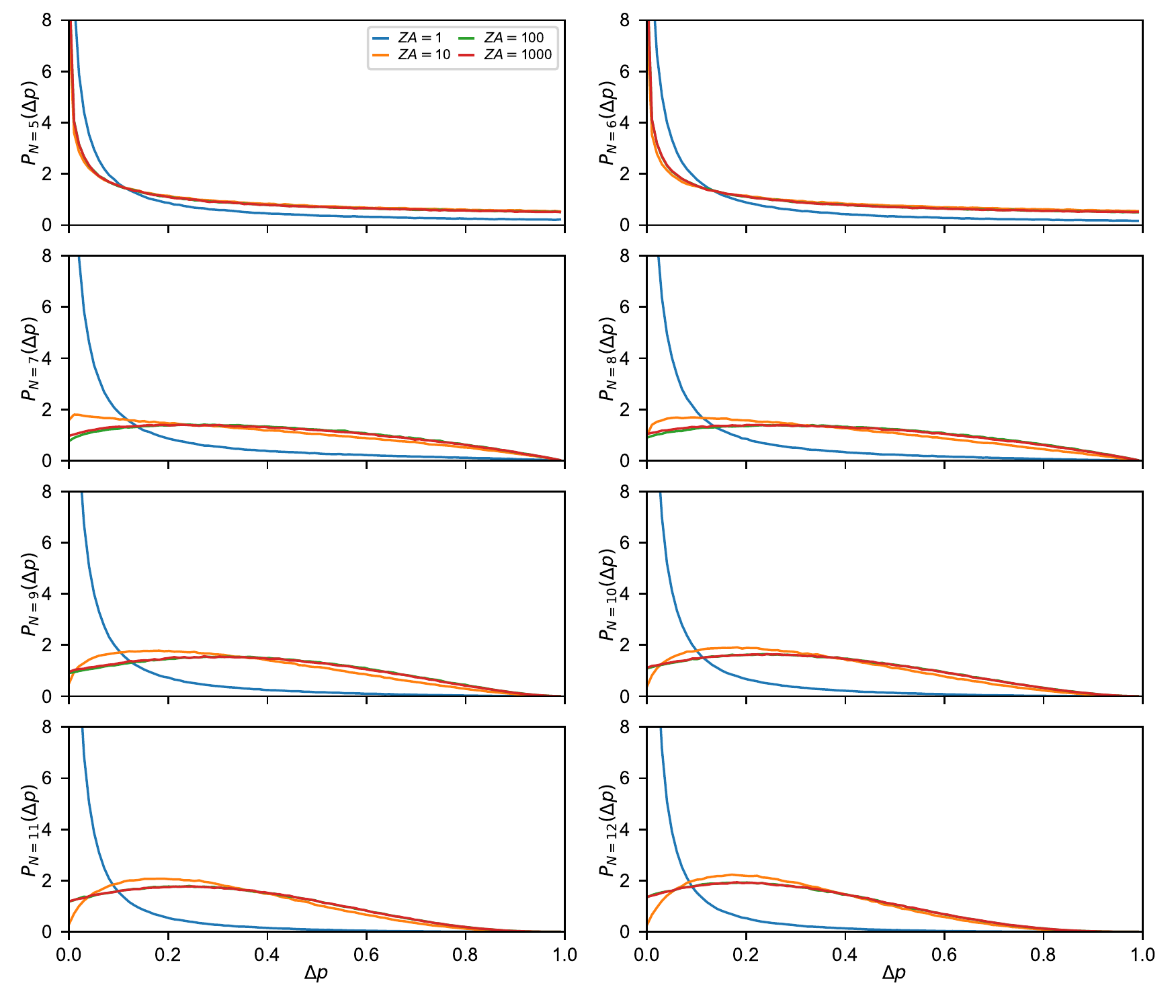}
\caption{Numerically sampled probability distributions of the parity signature $\Delta p$ for different effective area $Z A$ of the phase space loop for the protocol of the main text and Sec.\ \ref{app:parity_1} obtained by $Z$ times repeating a loop with infinitesimal area $A$.
Distributions are obtained by sampling over $10^6$ realizations of the random matrix model (\refeqHxy).}
\label{fig:S_p_ZA_MainProtocol}
\end{figure}

\begin{figure}
\begin{tabular}{cc}
\includegraphics[]{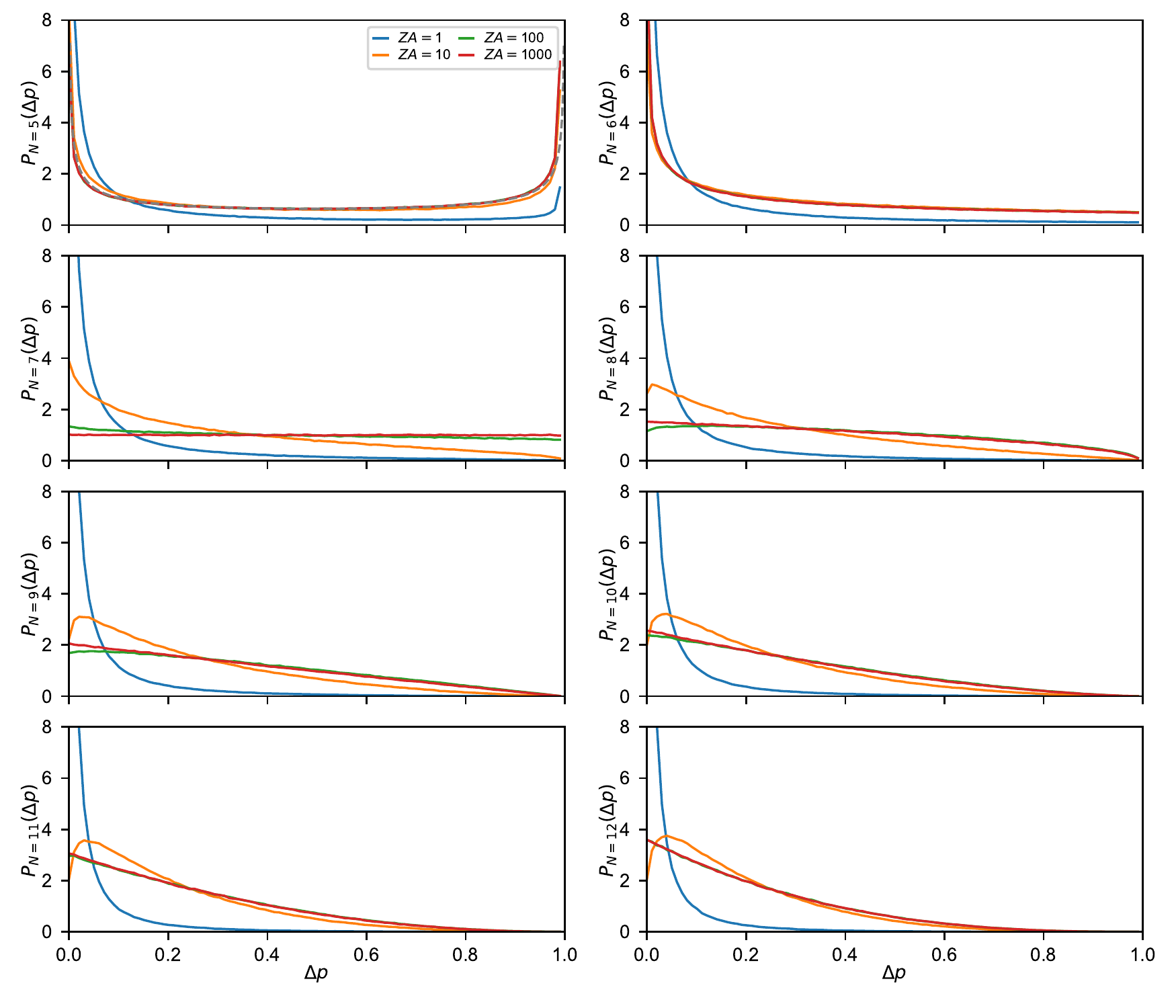}
\end{tabular}
\caption{Numerically sampled probability distributions of the parity signature $\Delta p$ for different effective areas $A$ of the phase space loop for the alternative protocol of Sec.\ \ref{app:parity_2} obtained by $Z$ times repeating a loop with infinitesimal area $A$. For $N = 5$, we include the analytical result for $ZA \gg 1$ as the gray dashed line.
Distributions are obtained by sampling over $10^6$ realizations of the random matrix model (\refeqHxy).}
\label{fig:S_p_ZA_AlternativeProtocol}
\end{figure}

Figures \ref{fig:S_p_ZA_MainProtocol} and \ref{fig:S_p_ZA_AlternativeProtocol} show the dependence of the parity signature on the cumulative area $Z A$ obtained by $Z$ times repeating a loop with small area $A \ll 1$. For cumulative areas $ZA > 1$, the distribution acquires significant weight for all values $0 \le \Delta p < 1$. The algebraic tail at large $\Delta p$ that was found for small areas $A$ smoothly connects to $P(\Delta p = 1) = 0$. For intermediate $Z A \lesssim 10$, a repulsion from $\Delta p = 0$ remains. The converged distribution in the limit of large $ZA$ has finite weight at $\Delta p = 0$.

For the alternative protocol, the asymptotic distribution $P(\Delta p)$ for $N=5$ has a simple analytical form. As $\hat{\gamma}_1$ remains decoupled from the quantum dot and for large $Z A \gg 1$, the Wilson loop operator $W$ describing the rotation is of the form $W = e^{i \alpha \sigma_2}$ with $\alpha$ evenly distributed on the interval $(-\pi, \pi]$. Then the element $W_{22} = \cos \alpha$ is distributed as
$$
    P_{N=5}(W_{22}) = 
  \frac{1}{\pi} \frac{1}{\sqrt{1 - W_{22}^2}} .
$$
With $\Delta p = (1 - W_{22})/2$ we obtain the asymptotic distribution of the parity signature
\begin{equation}
    P_{N=5}(\Delta p) = \frac{2}{\pi} \frac{1}{\sqrt{1 - (2 \Delta p - 1)^2}} 
  \ \ \mbox{for $Z A \gg 1$}.
    \label{eq:app_p_ZA_N5}
\end{equation}

\begin{figure}
\includegraphics[width=0.8\columnwidth]{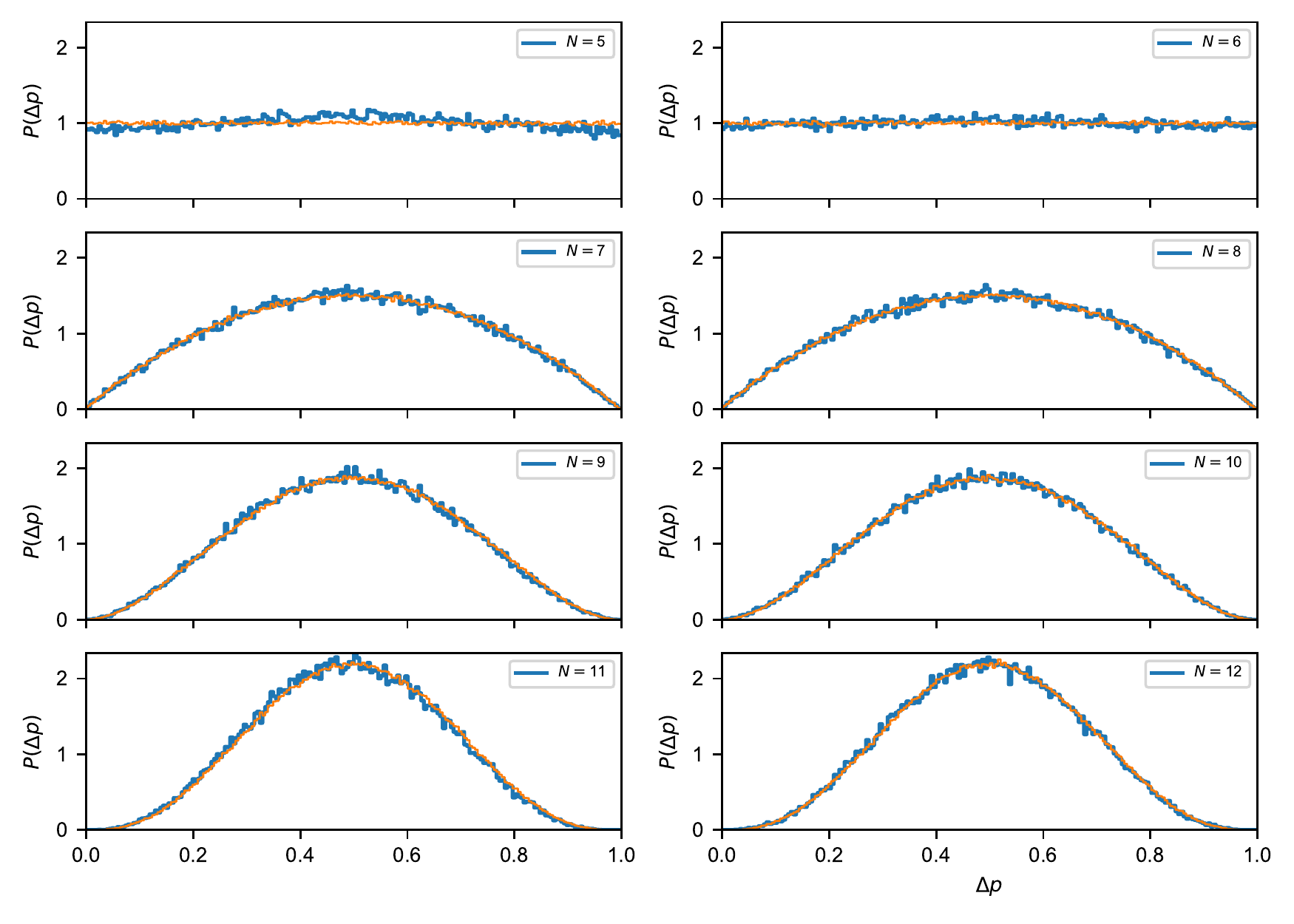}
\caption{
Numerically sampled probability distribution (blue lines) of the parity signature $\Delta p$ for large enclosed area $A \gg 1$ of the phase space loop using the random matrix model \eqref{eq:H_infiniteA} for the protocol of the main text, where both MZMs $\hat \gamma_1$ and $\hat \gamma_2$ are coupled to the quantum dot during the phase space loop, see Fig.\ \reffigone). The orange lines are results obtained by sampling the Wilson loop operator from $\text{SO}(N-2)$.  Here we sampled $10^5$ realizations of the loop $W$.
\label{fig:SM_V311}}
\end{figure}

\begin{figure}
\includegraphics[width=0.8\columnwidth]{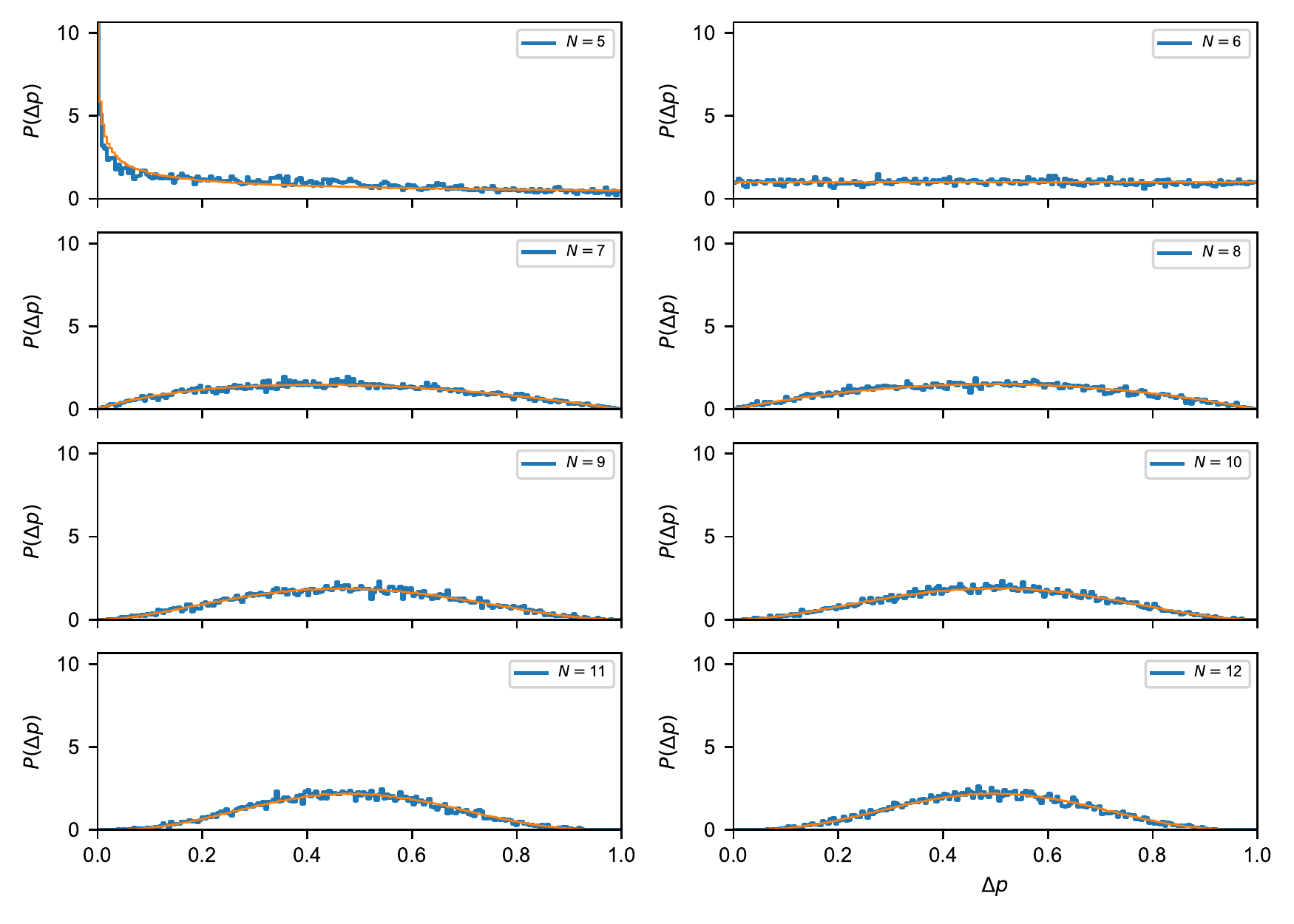}
\caption{
Numerically sampled probability distribution (blue lines) of the parity signature $\Delta p$ for large enclosed area $A \gg 1$ of the phase space loop using the random matrix model \eqref{eq:H_infiniteA} for the alternative protocol, where of the two MZMs $\hat \gamma_1$ and $\hat \gamma_2$ involved in the parity measurement, only $\hat \gamma_2$ is coupled to the quantum dot during the phase space loop, while $\hat \gamma_1$ remains decoupled, see Fig.\ \ref{fig:1b}. The orange lines are results obtained by sampling the Wilson loop operator as $W = O (\tilde{W} \oplus 1) O^T$ with $\tilde{W} \in \text{SO}(N-3)$ and $O \in \text{SO}(N-2)$ uniformly distributed, see the discussion in Sec.\ \ref{app:num_p}. Here we sampled $10^4$ realizations of the loop $W$.
\label{fig:SM_V310}}
\end{figure}

Figures~\ref{fig:SM_V311} and \ref{fig:SM_V310} show the parity signature for large areas $A\to\infty$ using the random matrix model \eqref{eq:H_infiniteA} obtained from the protocol of the main text and the alternative protocol of Sec.\ \ref{app:parity_2}, respectively. The figures also show a comparison of the parity signature obtained from a uniformly distributed Wilson loop operator $W$. Since the MZM $\hat{\gamma}_1$ remains decoupled from the billiard during the shape space loop for the alternative protocol, the number of MZM coupled to the billiard during the shape space loop is $N-1$, not $N$. In this case, in the limit of large loop area, the random matrix model for the Wilson loop operator is given by $W = O (\tilde{W} \oplus 1) O^T$, where "$\oplus$" denotes the direct matrix sum, $O \in \text{SO}(N-2)$ is an orthogonal matrix that describes the random basis after projection onto the Majorana dark space, and $\tilde{W} \in \text{SO}(N-3)$ describes the random rotation of the MZM coupled to the billiard. Both $O$ and $W$ are uniformly distributed. This model accurately describes the asymmetric distribution of the parity signature of the alternative protocol for odd $N$, see Fig.~\ref{fig:SM_V310}.

\clearpage 

\section{Charge signature}

\label{app:charge}

For a quantitative description, we find it advantageous to recast
the model Hamiltonian (\refeqHsimple) in terms of Majorana operators
$\hat{\beta}_{1}=\hat{c}+\hat{c}^{\dagger}$ and $\hat{\beta}_{2}=-i\hat{c}+i\hat{c}^{\dagger}$,
\begin{align}
\hat{H}=\frac{i}{2}\varepsilon\hat{\beta}_{2}\hat{\beta}_{1}-i\sum_{j=1}^{N}\left(\mbox{Im}\,v_{j}\hat{\beta}_{1}\hat{\gamma}_{j}+\mbox{Re}\,v_{j}\hat{\beta}_{2}\hat{\gamma}_{j}\right),
  \label{eq:E1}
\end{align}
 where we omitted a constant term that does not depend on the state
of the system. To keep the notation simple, we take $N$ to be the
total number of MZMs, including the ``additional'' MZMs that are not
coupled to the dot. (The coupling coefficients $v_{j}$ are set
to zero for these MZMs.) Hence, in this appendix we take $N$ to be
even.

\subsection{Sudden change of coupling coefficients}

\label{app:charge_1}

To observe the charge signature, the system is prepared with
coupling coefficients $v_{j}$, $j=1$, $\ldots$, $N$. The coupling
coefficients are then abruptly changed to values $v_{j}'$, $j=1$,$\ldots$,
$N$. This is when the charge read-out takes place. The charge signature
$q$ involves a comparison of dot charges with and without performing
a closed loop in shape space during after initialization, but before
the quench. We first describe the read-out stage and then turn to
the preparation stage. 

\emph{Read-out phase.} After changing the coupling coefficients to
the values $v_{j}'$, the Hamiltonian reads 
\begin{align}
  \hat{H}' =\frac{i}{2}\varepsilon\hat{\beta}_{2}\hat{\beta}_{1}-i\sum_{j=1}^{N}\left(\mbox{Im}\,v_{j}^{\prime}\hat{\beta}_{1}\hat{\gamma}_{j}+\mbox{Re}\,v_{j}^{\prime}\hat{\beta}_{2}\hat{\gamma}_{j}\right).
  \label{eq:E2}
\end{align}
By a change of basis, it may be brought to the canonical form 
\begin{align}
  \hat{H}' =\frac{i}{2}E'_{1}\hat{B}'_{2}\hat{B}'_{1}+\frac{i}{2}E'_{2}\hat{B}'_{4}\hat{B}'_{3},
  \label{eq:E3}
\end{align}
where $E_{i}' > 0$, $i=1,2$ and the Majorana operators $\hat{B}'_{1}$, $\hat{B}'_{2}$, $\hat{B}'_{3}$,
and $\hat{B}'_{4}$ are orthonormal linear combinations of the operators
$\hat{\beta}_{1}$, $\hat{\beta}_{2}$, and the $\hat{\gamma}_{i}$,
$i=1$,$\ldots$, $N$. 
To describe the basis transformation between the Majorana operators $\hat \beta_1$, $\hat \beta_2$, $\gamma_1$, \ldots, $\gamma_N$ appearing in Eq.\ (\ref{eq:E2}) and the operators $\hat B_1'$, $\hat B_2'$, $\hat B_3'$, $\hat B_4'$, appearing in Eq.\ (\ref{eq:E3}) and the dark Majorana operators $\Gamma_1'$, \ldots, $\Gamma_{N-2}'$, we introduce the two-component vector $\hat \beta = (\hat \beta_1,\hat \beta_2)^{\rm T}$, the four-component vector $\hat B' = (\hat B_1',\hat B_2',\hat B_3',\hat B_4')^{\rm T}$, the $(N-2)$-component vector $\hat \Gamma' = (\hat \Gamma_1',\ldots,\hat \Gamma_{N-2}')^{\rm T}$, and the $N$-component vector $\hat \gamma = (\hat \gamma_1,\ldots,\hat \gamma_N)^{\rm T}$. The required basis transformation is then of the form
\begin{equation}
  \begin{pmatrix}
  \hat \beta \\
  \hat \gamma
  \end{pmatrix}
  =
  \begin{pmatrix}
  \openone_2 & 0 \\
  0 & R' \end{pmatrix}
  \begin{pmatrix}
  O' & 0 \\ 0 & \openone_{N-2}
  \end{pmatrix}
  \begin{pmatrix}
  \hat B' \\
  \hat \Gamma' 
  \end{pmatrix},
  \label{eq:E4}
\end{equation}
where $R'$ and $O'$ are orthogonal matrices of size $N-2$ and $4$, respectively. In particular, the operators $\hat{\beta}_{1}$ and $\hat{\beta}_{2}$
may be written as linear combinations of the operators $\hat{B}'_{1}$,
$\hat{B}'_{2}$, $\hat{B}'_{3}$, and $\hat{B}'_{4}$,
\begin{align}
\hat{\beta}_{i}=\sum_{j=1}^{4}O_{ij}^{\prime}\hat{B}'_{j},\ \ i=1,2,\label{eq:beta}
\end{align}
The dark Majorana operators $\Gamma'_{i}$, $i=1$, $\ldots$, $N-2$,
do not appear in this expression and in the Hamiltonian (\ref{eq:E3}), because they have no weight in the dot. 

The operator for the dot charge $Q$ is, written in the Majorana
basis and up to a constant that does not depend
on the state of the system,
\begin{align}
\hat{Q}-\frac{1}{2}= & -\frac{i}{2}\hat{\beta}_{2}\hat{\beta}_{1}\nonumber \\
= & -\frac{i}{2}\sum_{j=1}^{4}\sum_{k=1}^{4}O_{2j}^{\prime}O_{1k}^{\prime}\hat{B}'_{j}\hat{B}'_{k},
\end{align}
 After time averaging, only the products $\hat{B}'_{2}\hat{B}'_{1}$
and $\hat{B}'_{4}\hat{B}'_{3}$ give a contribution that depends on
the state of the system. Hence, we obtain 
\begin{align}
\overline{Q}-\frac{1}{2}= & \,-\frac{i}{2}(O_{22}^{\prime}O_{11}^{\prime}-O_{21}^{\prime}O_{12}^{\prime})\langle\hat{B}'_{2}\hat{B}'_{1}\rangle-\frac{i}{2}(O_{24}^{\prime}O_{13}^{\prime}-O_{23}^{\prime}O_{14}^{\prime})\langle\hat{B}'_{4}\hat{B}'_{3}\rangle.\label{eq:Qavg1}
\end{align}

\emph{Preparation stage.} In the preparation stage, the canonical
form of the Hamiltonian reads 
\begin{align}
\hat{H}= & \,\frac{i}{2}E_{1}\hat{B}_{2}\hat{B}_{1}+\frac{i}{2}E_{2}\hat{B}_{4}\hat{B}_{3},
\end{align}
where $E_{i} > 0$, $i=1,2$, and the Majorana operators $\hat{B}_{1}$, $\hat{B}_{2}$, $\hat{B}_{3}$, and $\hat{B}_{4}$ are orthonormal linear combinations of the operators $\hat{\beta}_{1}$, $\hat{\beta}_{2}$, and the $\hat{\gamma}_{i}$,
$i=1$, $\ldots$,$N$. With the notation $\hat B = (\hat B_1, \hat B_2, \hat B_3, \hat B_4)^{\rm T}$, $\hat \Gamma = (\hat \Gamma_1, \ldots, \hat \Gamma_{N-2})^{\rm T}$, we may write this basis transformation as (compare with Eq.\ (\ref{eq:E4}))
\begin{equation}
  \begin{pmatrix}
  \hat \beta \\
  \hat \gamma
  \end{pmatrix}
  =
  \begin{pmatrix}
  \openone_2 & 0 \\
  0 & R \end{pmatrix}
  \begin{pmatrix}
  O & 0 \\ 0 & \openone_{N-2}
  \end{pmatrix}
  \begin{pmatrix}
  \hat B \\
  \hat \Gamma 
  \end{pmatrix},
  \label{eq:E9}
\end{equation}
where $R$ and $O$ are orthogonal matrices of size $N-2$ and $4$, respectively. As before, the ``dark'' Majorana operators $\Gamma_{i}$, $i=1, \ldots, N-2$ do not appear in the Hamiltonian. 

The system is initialized in its ground state, which is unique if
the coupling to the non-resonant dot-levels is taken into account.
We choose the basis of the dark states such, that the ground state
corresponds to the vacuum for the fermions $\hat{f}_{i}=\frac{1}{2}(\hat{\Gamma}_{2i-1}+i\hat{\Gamma}_{2i})$.
With that convention, the only products of different Majorana operators
with nonzero expectation value are
\begin{align}
\langle\hat{B}_{2}\hat{B}_{1}\rangle=\langle\hat{B}_{4}\hat{B}_{3}\rangle=-i\label{eq:app_exp_B}
\end{align}
 and 
\begin{align}
\langle\hat{\Gamma}_{2j}\hat{\Gamma}_{2j-1}\rangle=-i,\ \ j=1,\ldots,N/2-1.\label{eq:app_exp_Gamma}
\end{align}
If, after initializing the system, a loop in shape space is performed,
the expectation values of products of the dark Majorana operators
change according to Eq.\ (\refeqGammaW), 
\begin{align}
\langle\hat{\Gamma}_{k}\hat{\Gamma}_{l}\rangle=-i\sum_{j=1}^{(N-2)/2}(W_{2j,k}W_{2j-1,l}-W_{2j-1,k}W_{2j,l}).\label{eq:S10}
\end{align}

\emph{Charge signature.} The charge signature $\Delta \overline{Q} = \overline{Q(W)} - \overline{Q(\id)}$ is defined as the difference of the time-averaged dot charges with and
without performing a loop in shape space in the preparation stage,
see also the discussion in the main text.
To calculate the expectation values $\langle\hat{B}_{2}'\hat{B}_{1}'\rangle$
and $\langle\hat{B}_{4}'\hat{B}_{3}'\rangle$ appearing in the expression
(\ref{eq:Qavg1}) for the time-averaged charge $\overline{Q}$, one
has to express the Majorana operators $\hat{B}_{1}'$, $\hat{B}_{2}'$,
$\hat{B}_{3}'$, and $\hat{B}_{4}'$ in terms of the operators $\hat{B}_{1}$,
$\hat{B}_{2}$, $\hat{B}_{3}$, $\hat{B}_{4}$, and the dark Majorana
operators $\hat{\Gamma}_{j}$, $j=1$,$\ldots$,$N-2$, with the help of
Eqs.\ (\ref{eq:E4}) and (\ref{eq:E9}). The expectation
values can then be calculated with the help of Eqs.\
(\ref{eq:app_exp_B}) and (\ref{eq:S10}) if a loop in
phase space is performed during the preparation stage and with the
help of (\ref{eq:app_exp_B}) and (\ref{eq:app_exp_Gamma}) if no
loop in shape space is performed after initialization. Explicitly,
we have for the difference of the expectation values with and without performing the phase space loop
\begin{align}
  \Delta \langle \hat B_{2l}' \hat B_{2l-1}' \rangle =&\,
  -i
  \sum_{j=1}^{N-2} \sum_{k=1}^{N-2} 
  {\cal M}_{2l,j+4} {\cal M}_{2l-1,k+4}\,
  \Delta \langle \hat \Gamma_{j} \hat \Gamma_{k} \rangle,
  \label{eq:E13}
\end{align}
where
\begin{align}
  \Delta \langle \hat \Gamma_{j} \hat \Gamma_{k} \rangle =&\,
  \sum_{p=1}^{(N-2)/2}
  \left( W_{2p,j} W_{2p-1,k} - W_{2p,k} W_{2p-1,j}
  - \delta_{2p,j} \delta_{2p-1,k} + \delta_{2p,k} \delta_{2p-1,j}
  \right),
  \label{eq:E13b}
\end{align}
and
\begin{equation}
  \label{eq:E14}
  {\cal M} = 
  \begin{pmatrix}
  O'^{\rm T} & 0 \\ 0 & \openone_{N-2}
  \end{pmatrix}
  \begin{pmatrix}
  \openone_2 & 0 \\
  0 & R'^{\rm T} R \end{pmatrix}
  \begin{pmatrix}
  O & 0 \\ 0 & \openone_{N-2}
  \end{pmatrix}
\end{equation}
Unlike the parity signature, the charge signature $\Delta \overline{Q}$ is linear in the area $A$ enclosed by the phase-space loop for small $A$.

We close with a discussion of the orthogonal transformations (\ref{eq:E4}) and (\ref{eq:E9}) and the derivation of an approximate expression for $\Delta \overline Q$ in the limit of weak coupling between the MZMs and the quantum dot. For definiteness, we focus on Eq.\ (\ref{eq:E9}), and note that the transformation (\ref{eq:E4}) can be obtained upon replacing the coupling parameters $\vv$ by $\vv'$. Writing $\va = \mbox{Re}\, \vv$, $\vb = \mbox{Im}\, \vv$, we write the Hamiltonian (\ref{eq:E1}) in matrix form,
\begin{equation}
  H = -i \begin{pmatrix} 0 & \varepsilon/2 & \va^{\rm T} \\
  - \varepsilon/2 & 0 & \vb^{\rm T} \\
  -\va & -\vb & 0_N \end{pmatrix}.
  \label{eq:E16}
\end{equation}
With a suitably chosen basis change of the Majoranas, in a first step $H$ may be brought to the form
\begin{equation}
  H = \begin{pmatrix} \openone_2 & 0 \\ 0 & R \end{pmatrix}
  \begin{pmatrix} h & 0 \\ 0 & 0_{N-2} \end{pmatrix}
  \begin{pmatrix} \openone_2 & 0 \\ 0 & R^{\rm T} \end{pmatrix},
  \label{eq:E17}
\end{equation}
where $R$ is an $N \times N$ orthogonal matrix and
\begin{equation}
  h = -i \begin{pmatrix} 0 & \varepsilon/2 & a & 0 \\
  -\varepsilon/2 & 0 & b \cos \varphi & b \sin \varphi \\
  - a & -b \cos \varphi & 0 & 0 \\
  0 & -b \sin \varphi & 0 & 0 & 
  \label{eq:E18}
 \end{pmatrix}
\end{equation}
with $a^2 = \va^{\rm T} \va$, $b^2 = \vb^{\rm T} \vb$, $a b \cos \varphi = \va^{\rm T} \vb$, and $0 \le \varphi \le \pi$.
Only the first two rows of the matrix $R$ are determined by the coupling parameters $\vv$,
\begin{align}
  R_{j1} =&\, a_j/a, \nonumber \\
  R_{j2} =&\, b_j/b \sin \varphi - (a_j/a) \cot \varphi, \ \ j=1,\ldots,N.
  \label{eq:E19}
\end{align}
For the remaining rows, we note that right-multiplication of $R$ with an orthogonal matrix of the form $\mbox{diag}\,(\openone_2,S)$, where $S$ is an orthogonal matrix of size $N-2$, leaves Eq.\ (\ref{eq:E17}) invariant. This degree of freedom is reduced to matrices $S$ that represent a pairwise permutation of rows and columns by the additional constraint that the exact ground state --- which involves coupling between the MZMs and quantum dot levels at higher energy --- is described by Eq.\ (\ref{eq:app_exp_Gamma}).
In the same way, a matrix $R'$ can be defined that brings the Hamiltonian $H'$ in the read-out phase to a form analogous to Eq.\ (\ref{eq:E17}). This matrix is again determined up to right-multiplication with an orthogonal matrix of the form $\mbox{diag}\,(\openone_2,S')$, where $S'$ is an orthogonal matrix of size $N-2$. However, in this case this additional degree of freedom does not affect the charge signature $\Delta \overline{Q}$, as it does not enter into the relevant elements of the matrix ${\cal M}$, see Eqs.\ (\ref{eq:E13}) and (\ref{eq:E14}).

In a second step, we bring the $4 \times 4$ antisymmetric matrix $h$ to canonical form,
\begin{equation}
  h = -i \, O \begin{pmatrix} 0 & E_1 \\ -E_1 & 0 \\ && 0 & E_2 \\ && -E_2 & 0 \end{pmatrix} O^{\rm T},
\end{equation}
where $O$ is an orthogonal $4 \times 4$ matrix, which is unique up to right multiplication with matrices of the form $\mbox{diag}\, (o_1, o_2)$, with $o_1$ and $o_2$ special orthogonal $2 \times 2$ matrices. In the limit of weak coupling between the MZMs and the quantum dot, one may obtain approximate expressions for $O$ and the principal values $E_1$ and $E_2$,
\begin{equation}
  E_1 \approx \varepsilon/2,\ \
  E_2 \approx \frac{2 a b}{\varepsilon} \sin \varphi,
\end{equation}
and
\begin{equation}
  \label{eq:E22}
  O \approx \begin{pmatrix} 1 & 0 & (2 b/\varepsilon) \cos \varphi & (2 b/\varepsilon) \sin \varphi \\ 
  0 & 1 & -2 a/\varepsilon & 0 \\
  - (2 b/\varepsilon) \cos \varphi & 2 a/\varepsilon & 1 & 0 \\
  - (2 b/\varepsilon) \sin \varphi & 0 & 0 & 1 \end{pmatrix}.
\end{equation}
The expressions for $E_1$ and $E_2$ are valid to leading order in $v/\varepsilon$, where $v^2 = \vv^{\dagger} \cdot \vv = a^2 + b^2$. The expression for $O$ is valid to to sub-leading order in $v/\varepsilon$. In the same way, for the read-out phase the corresponding $4 \times 4$ matrix $h'$ is brought to canonical form with principal values $E_1'$ and $E_2'$ by an orthogonal matrix $O'$, which can analogously be expressed in terms of $\varepsilon$ and the coupling constants $a'$, $b'$, and $\varphi'$. 

To obtain an expression for the charge signature $\Delta \overline Q$ in the limit of weak coupling, we need the expectation value $\Delta \langle \hat B_2' \hat B_1' \rangle$ to order $(v/\varepsilon)^2$ and the expectation value $\Delta \langle \hat B_4' \hat B_3' \rangle$ to order $(v/\varepsilon)^0$. From Eqs.\ (\ref{eq:E13}), (\ref{eq:E14}), (\ref{eq:E19}), and (\ref{eq:E22}) we obtain
\begin{align}
  \Delta \langle \hat B_2' \hat B_1' \rangle =&\, 
   4 i \sum_{p, q = 1}^{N} \sum_{j,k=1}^{N-2}
  \frac{b_p' a_q'}{\varepsilon^2}
  R_{p,j+2} R_{q,k+2}\,
  \Delta \langle \hat \Gamma_{j} \hat \Gamma_{k} \rangle,
  \label{eq:E23}
  \\
  \Delta \langle \hat B_4' \hat B_3' \rangle =&\,
    i \sum_{p, q = 1}^{N} \sum_{j,k =1}^{N-2}
    \frac{a_p' a_q' (b' / a') \cos \varphi' - a_p' b_q'}{a' b' \sin \varphi'}
  R_{p,j+2} R_{q,k+2}\,
  \Delta \langle \hat \Gamma_{j} \hat \Gamma_{k} \rangle.
  \label{eq:E24}
\end{align}
For the charge signature we then find from Eq.\ (\ref{eq:Qavg1})
\begin{align}
  \Delta \overline{Q} =&\,
  2 \sum_{p, q = 1}^{N} \sum_{j,k =1}^{N-2}
  \frac{b_p' a_q' - a_p' b_q'}{\varepsilon^2}
  R_{p,j+2} R_{q,k+2}\,
  \Delta \langle \hat \Gamma_{j} \hat \Gamma_{k} \rangle,
  \label{eq:E25}
\end{align}
with $\Delta \langle \hat \Gamma_{j} \hat \Gamma_{k} \rangle$ given in Eq.\ (\ref{eq:E13b}). As discussed above, the matrix $R$ is determined by the details of the ground state in the preparation stage, which not only depend on the coupling $\vv$ between the MZMs and the resonant quantum dot level, but also on couplings to non-resonant levels.

The charge signature $\Delta \overline{Q} = 0$ when pinching off only a single contact or all but one of the contacts. To see this, we rewrite Eq.\ (\ref{eq:E25}) as
\begin{equation}
  \Delta \bar{Q} = \frac{2}{\varepsilon^2} \sum_{j,k=1}^{N-2} \vec{R}_{j+2}^{\rm T} (\vec{b}' \wedge \vec{a}') \vec{R}_{k+2} \Delta \langle \hat{\Gamma}_j \hat{\Gamma}_k \rangle
\label{eq:E36}
\end{equation}
where $(\vec{R}_j)_n = R_{nj}$ is the $j$th column of the matrix $R$ and the wedge product $\vb' \wedge \va'$ is short-hand notation for $\vec{b}' \wedge \vec{a}' = b_p' a_q' - b_q' a_p' = \vec{b}^{\prime {\rm T}} \vec{a}' - \vec{a}^{\prime {\rm T}} \vec{b}'$. If only the $p$th contact is pinched off, the coupling vectors change as
\begin{align}
    \vec{a}' & = \vec{a} - a_p \hat{\ve}_p, \nonumber \\
    \vec{b}' & = \vec{b} - b_p \hat{\ve}_p.
\end{align}
where $\hat{\ve}_p$ is the unit vector in the $p$th direction. For the wedge product $\vb' \times \va'$ this implies
\begin{align}
    \vec{b}' \wedge \vec{a}' & = \vec{b} \wedge \vec{a} - (b_p \vec{a} - a_p \vec{b}) \wedge \hat{\ve}_p.
    \label{eq:E37} 
\end{align}
Since the first two columns $\vR_1$ and $\vR_2$ of the  matrix $R$ are spanned by linear combinations of $\va$ and $\vb$, see Eq.\ (\ref{eq:E19}), the remaining columns of $R$ are orthogonal to $\va$ and $\vb$ because $R$ is an orthogonal matrix. Hence
\begin{align}
    \vec{R}_{j+2}^{\rm T} (\vec{b} \wedge \vec{a}) \vec{R}_{k+2}  = 
    \vec{R}_{j+2}^{\rm T} ((b_p \vec{a} - a_p \vec{b}) \wedge \hat{\ve}_p) \vec{R}_{k+2} & = 0,  
    \label{eq:E38} 
\end{align}
so that $\Delta \overline{Q} = 0$ in this case. The same conclusion is reached if all contacts are pinched off except for the $p$th contact, so that $\va' = a_p \hat \ve_p$, $\vb' = b_p \hat \ve_p$.

\subsection{Alternative protocol for charge signature: Sudden change of the resonant level}

\label{app:charge_2}

The sudden change of the coupling coefficients required to observe
the charge signature can be brought about by a sudden change of the
backgate voltage, such that the resonance condition shifts from dot
level $m$ to the neighboring level $m\pm1$. As a result, there are
a few changes to the theoretical description of the protocol. For
definiteness, we take the neighboring level to be level $m+1$, which
has energy $\varepsilon_{m+1}>\varepsilon_{m}$.

Neglecting the coupling to non-resonant levels, we may describe this
process using a generalization of the Hamiltonian (\refeqHsimple),
\begin{align}
\hat{H}=(\varepsilon-\varepsilon_{{\rm g}})\hat{c}_{0}^{\dagger}\hat{c}_{0}+(\varepsilon'-\varepsilon_{{\rm g}})\hat{c}_{1}^{\dagger}\hat{c}_{1}+\sum_{j=1}^{N}(v_{j}^{*}\hat{c}_{0}^{\dagger}+v_{j}'^{*}\hat{c}_{1}^{\dagger}-v_{j}\hat{c}_{0}-v_{j}'\hat{c}_{1})\hat{\gamma}_{j},\label{eq:Hgate}
\end{align}
 where $\varepsilon=\varepsilon_{m}$, $\varepsilon'=\varepsilon_{m+1}$,
$\varepsilon_{{\rm g}}$ is the potential shift from the backgate,
$\hat{c}_{0}$ and $\hat{c}_{1}$ are annihilation operators
of the $m$th and $(m+1)$th dot mode, and the complex coefficients $v_{n,j}$
describe the coupling between the $(m+n)$th dot mode and the MZMs.
As before, we take $N$ to be the total number of MZMs, including
the ``additional'' MZMs that are not coupled to the dot, so that
$N$ is even. Writing $\hat{c}_{0}=\frac{1}{2}(\hat{\beta}_{1}+i\hat{\beta}_{2})$,
$\hat{c}_{1}=\frac{1}{2}(\hat{\beta}_{3}+i\hat{\beta}_{4})$, where
the operators $\hat{\beta}_{i}$, $i=1,2,3,4$, are Majorana operators,
the Hamiltonian (\ref{eq:Hgate}) reads 
\begin{align}
\hat{H}=\frac{i}{2}(\varepsilon-\varepsilon_{{\rm g}})\hat{\beta}_{2}\hat{\beta}_{1}+\frac{i}{2}((\varepsilon'-\varepsilon_{{\rm g}})\hat{\beta}_{4}\hat{\beta}_{3}-i\sum_{j=1}^{N}\left(\mbox{Im}\,v_{j}\hat{\beta}_{1}\hat{\gamma}_{j}+\mbox{Im}\,v_{j}'\hat{\beta}_{3}\hat{\gamma}_{j}+\mbox{Re}\,v_{j}\hat{\beta}_{2}\hat{\gamma}_{j}+\mbox{Re}\,v_{j}'\hat{\beta}_{4}\hat{\gamma}_{j}\right),
\end{align}
 up to a constant term that does not depend on the state of the system.

To observe the charge signature, the system is prepared while
$|\varepsilon-\varepsilon_{{\rm g}}|$ is much smaller than the
level spacing $\delta$ in the dot, so that only the dot level $\varepsilon$
hybridizes significantly with the MZMs. This is the situation described
in the main text. After the quench, one has $|\varepsilon'-\varepsilon_{{\rm g}}|\ll\delta$,
so that now only the dot level $\varepsilon_1$ hybridizes with the
MZMs. This is when the charge read-out takes place. As before, we
describe the read-out stage first.

\emph{Read-out phase.--} Neglecting the hybridization of the dot
level $\varepsilon_0$ with the MZMs after the quench, the Hamiltonian
$\hat{H}$ may be brought to the canonical form 
\begin{align}
\hat{H}\approx\frac{i}{2}(\varepsilon-\varepsilon_{{\rm g}})\hat{\beta}_{2}\hat{\beta}_{1}+\frac{i}{2}E'_{1}\hat{B}'_{2}\hat{B}'_{1}+\frac{i}{2}E'_{2}\hat{B}'_{4}\hat{B}'_{3},
\label{eq:E28}
\end{align}
where the Majorana operators $\hat{B}'_{1}$, $\hat{B}'_{2}$, $\hat{B}'_{3}$,
and $\hat{B}'_{4}$ are orthonormal linear combinations of the operators
$\hat{\beta}_{3}$, $\hat{\beta}_{4}$, and the $\hat{\gamma}_{i}$,
$i=1$, $\ldots$, $N$. 
With the notation $\hat \beta_j = (\hat \beta_{2j-1}, \hat \beta_{2j})^{\rm T}$, $\hat B' = (\hat B_1', \hat B_2', \hat B_3', \hat B_4')^{\rm T}$, $\hat \Gamma' = (\hat \Gamma_1', \ldots, \hat \Gamma_{N-2}')^{\rm T}$, with $\Gamma_j'$, $j=1,\ldots,N-2$ the ``dark'' Majorana operators after the quench, we may write this basis transformation as (compare with Eq.\ (\ref{eq:E4}))
\begin{equation}
  \begin{pmatrix}
  \hat \beta_1 \\
  \hat \beta_2 \\
  \hat \gamma
  \end{pmatrix}
  =
  \begin{pmatrix}
  \openone_2 & 0 & 0 \\
  0 & \openone_2 & 0 \\
  0 & 0 & R' 
  \end{pmatrix}
  \begin{pmatrix}
  \openone_2 & 0 & 0\\
  0 & O' & 0 \\
  0 & 0 & \openone_{N-2} 
  \end{pmatrix}
  \begin{pmatrix}
  \hat \beta_1 \\
  \hat B' \\
  \hat \Gamma' 
  \end{pmatrix},
  \label{eq:E29}
\end{equation}
where $R'$ and $O'$ are orthogonal matrices of size $N-2$ and $4$, respectively.
As before, the matrix $R'$ is defined uniquely only up to right multiplication with an orthogonal matrix of the form $\text{diag}(\openone_2, S')$. This ambiguity does not affect the charge signature $\Delta \overline{Q}$, however.
Inverting the
transformation (\ref{eq:E29}), the operators $\hat{\beta}_{3}$ and $\hat{\beta}_{4}$
may be written as linear combinations of the operators $\hat{B}'_{1}$,
$\hat{B}'_{2}$, $\hat{B}'_{3}$, and $\hat{B}'_{4}$ as in Eq.~(\ref{eq:beta}).
The six nonzero energies of the system consisting
of quantum dot and Majorana modes are $\pm(\varepsilon-\varepsilon_{{\rm g}})$,
$\pm E'_{1}$, and $\pm E'_{2}$. 

The operator for the dot charge is, written in the Majorana basis,
\begin{align}
\hat{Q} - 1= & \,-\frac{i}{2}(\hat{\beta}_{2}\hat{\beta}_{1}+\hat{\beta}_{4}\hat{\beta}_{3})\nonumber \\
= & \,-\frac{i}{2}\hat{\beta}_{2}\hat{\beta}_{1}+\frac{i}{2}\sum_{j=1}^{3}\sum_{k=1}^{3}O_{2j}^{\prime}O_{1k}^{\prime}\hat{B}'_{j}\hat{B}'_{k}.
\label{eq:E31}
\end{align}
After time averaging, only
the products $\hat{\beta}_{2}\hat{\beta}_{1}$, $\hat{B}'_{2}\hat{B}'_{1}$,
and $\hat{B}'_{4}\hat{B}'_{3}$ remain. Hence, we obtain 
\begin{align}
\overline{Q} - 1 = & \,-\frac{i}{2}\langle\hat{\beta}_{2}\hat{\beta}_{1}\rangle-\frac{i}{2}(O_{22}^{\prime}O_{11}^{\prime}-O_{21}^{\prime}O_{12}^{\prime})\langle\hat{B}'_{2}\hat{B}'_{1}\rangle-\frac{i}{2}(O_{24}^{\prime}O_{13}^{\prime}-O_{23}^{\prime}O_{14}^{\prime})\langle\hat{B}'_{4}\hat{B}'_{3}\rangle.\label{eq:Qavg}
\end{align}
 In this expression, the expectation value $\langle\hat{\beta}_{2}\hat{\beta}_{1}\rangle$
does not depend on the state of the dark Majorana modes before the
quench.

\emph{Preparation stage.--} In the preparation stage, only the dot
level $\varepsilon$ hybridizes with the MZMs. Hence, the diagonal
form of the Hamiltonian reads 
\begin{align}
\hat{H}\approx\frac{i}{2}(\varepsilon'-\varepsilon_{{\rm g}})\hat{\beta}_{4}\hat{\beta}_{3}+\frac{i}{2}E_{1}\hat{B}_{2}\hat{B}_{1}+\frac{i}{2}E_{2}\hat{B}_{4}\hat{B}_{3},
\label{eq:E32}
\end{align}
 where the Majorana operators $\hat{B}_{1}$, $\hat{B}_{2}$, $\hat{B}_{3}$,
and $\hat{B}_{4}$ are (orthonormal) linear combinations of the operators
$\hat{\beta}_{1}$, $\hat{\beta}_{2}$, and the $\hat{\gamma}_{i}$,
$i=1$, $\ldots$, $N$. 
With the notation $\hat \beta_j = (\hat \beta_{2j-1}, \hat \beta_{2j})^{\rm T}$, $\hat B = (\hat B_1, \hat B_2, \hat B_3, \hat B_4)^{\rm T}$, $\hat \Gamma = (\hat \Gamma_1, \ldots, \hat \Gamma_{N-2})^{\rm T}$, where the operators $\hat \Gamma_j$, $J=1,2,\ldots,N-2$ represent the dark MZMs before the quench, we have (compare with Eq.\ (\ref{eq:E9}) and note the ording of the vectors $\hat \beta_2$ and $\hat \beta_1$ in the column vectors on the left and right side of the equation)
\begin{equation}
  \begin{pmatrix}
  \hat \beta_2 \\
  \hat \beta_1 \\
  \hat \gamma
  \end{pmatrix}
  =
  \begin{pmatrix}
  \openone_2 & 0 & 0 \\
  0 & \openone_2 & 0 \\
  0 & 0 & R 
  \end{pmatrix}
  \begin{pmatrix}
  \openone_2 & 0 & 0\\
  0 & O & 0 \\
  0 & 0 & \openone_{N-2} 
  \end{pmatrix}
  \begin{pmatrix}
  \hat \beta_2 \\
  \hat B \\
  \hat \Gamma 
  \end{pmatrix},
  \label{eq:E29b}
\end{equation}
where $R$ and $O$ are orthogonal matrices of size $N-2$ and $4$, respectively. The energies of the system consisting of
quantum dot and Majorana modes are $\pm(\varepsilon_1-\varepsilon_{{\rm g}})$,
$\pm E_{1}$, and $\pm E_{2}$, where we use the convention that $E_{1}$,
$E_{2}>0$. 

As before, the system is initialized in its unique ground state and
we choose the basis of the dark states such, that the ground state
corresponds to the vacuum for the fermions $\hat{f}_{i}=\frac{1}{2}(\hat{\Gamma}_{2i-1}+i\hat{\Gamma}_{2i})$.
With that convention, the only products of different Majorana operators
with nonzero expectation value are given by Eqs.~(\ref{eq:app_exp_B})
and (\ref{eq:app_exp_Gamma}) above. If, after initializing the system,
a loop in shape space is performed, the expectation values of products
of the dark Majorana operators are given by 
Eq.~(\ref{eq:S10}). 

\emph{Charge signature.}-- To calculate expectation values $\langle\hat{\beta}_{2}\hat{\beta}_{1}\rangle$,
$\langle\hat{B}_{2}'\hat{B}_{1}'\rangle$, and $\langle\hat{B}_{4}'\hat{B}_{3}'\rangle$
appearing in the expression (\ref{eq:Qavg1}) for the time-averaged
charge $\overline{Q}$, one has to express the Majorana operators
$\hat{\beta}_{1}$, $\hat{\beta}_{2}$, $\hat{B}_{1}'$, $\hat{B}_{2}'$,
$\hat{B}_{3}'$, and $\hat{B}_{4}'$ in terms of the operators $\hat{\beta}_{4}$,
$\hat{\beta}_{3}$, $\hat{B}_{1}$, $\hat{B}_{2}$, $\hat{B}_{3}$,
$\hat{B}_{4}$, and the dark Majorana operators $\hat{\Gamma}_{j}$,
$j=1$, $\ldots$, $N-2$. The expectation values can then be calculated
with the help of Eqs.~(\ref{eq:app_exp_B}) and (\ref{eq:S10})
if a loop in phase space is performed during the preparation stage
and with the help of (\ref{eq:app_exp_B}) and (\ref{eq:app_exp_Gamma})
if no loop in shape space is performed after initialization. Since
the expectation value $\langle\hat{\beta}_{2}\hat{\beta}_{1}\rangle$
in Eq.~(\ref{eq:Qavg1}) is the same with or without
execution of a closed loop in shape space in the preparation stage,
for the calculation of the charge signature $\Delta \overline{Q}$, it is sufficient
to calculate the expectation values $\langle\hat{B}_{2}'\hat{B}_{1}'\rangle$
and $\langle\hat{B}_{4}'\hat{B}_{3}'\rangle$. 
Explicitly, we have for the difference of the expectation values with and without performing the loop,
\begin{align}
  \Delta \langle \hat B_{2l}' \hat B_{2l-1}' \rangle =&\,
  -i
  \sum_{j=1}^{N-2} \sum_{k=1}^{N-2} 
  {\cal M}_{2l+2,j+6} {\cal M}_{2l+1,k+6}\,
  \Delta \langle \hat \Gamma_{j} \hat \Gamma_{k} \rangle,
  \label{eq:E34}
\end{align}
where $\Delta \langle \hat \Gamma_{j} \hat \Gamma_{k} \rangle$ is given by Eq.~(\ref{eq:E13b})
and
\begin{equation}
{\cal M} = 
  \begin{pmatrix}
  \openone_2 & 0 & 0\\
  0 & O'^{\rm T} & 0 \\
  0 & 0 & \openone_{N-2} 
  \end{pmatrix}
  \begin{pmatrix}
  0 & \openone_2 & 0 \\
  \openone_2 & 0 & 0 \\
  0 & 0 & R'^{\rm T}R 
  \end{pmatrix}
  \begin{pmatrix}
  \openone_2 & 0 & 0\\
  0 & O & 0 \\
  0 & 0 & \openone_{N-2} 
  \end{pmatrix}
  \label{eq:E35}
\end{equation}
As in App.~\ref{app:charge_1}, the matrices $O$ and $O'$ are expressed by Eq.~(\ref{eq:E22}) in the limit of weak coupling, whereas the first two rows of $R$ and $R'$ are expressed by Eq.~(\ref{eq:E19}). From Eqs.~(\ref{eq:E19}), (\ref{eq:E22}), (\ref{eq:E31}), (\ref{eq:E34}), (\ref{eq:E35}), we again obtain Eqs.~(\ref{eq:E23}) and (\ref{eq:E24}) for $\Delta \langle \hat{B}_2' \hat{B}_1' \rangle$ and $\Delta \langle \hat{B}_4' \hat{B}_3' \rangle$ and Eq.~(\ref{eq:E25}) for the charge signature $\Delta \overline{Q}$. That the same result for the charge signature is obtained as in App.~\ref{app:charge_1} can be understood by noting that the exchange of dot levels, as implemented by the structure of the matrix in the middle of the r.h.s.\ of Eq.\ (\ref{eq:E35}),
acts outside of the space of dark MZM $\hat{\Gamma}_j$, whereas the charge signature is sensitive to rotations in the space of dark MZMs only.  

\subsection{Limiting behavior of the distribution of the charge signature}
\label{app:charge_limiting_behavior}

In this appendix, we discuss the limiting behavior of the charge signature $\Delta \overline{Q}$ at small and large $|\Delta \overline{Q}|$. The asymptotic at small $|\Delta \overline{Q}|$ is different for the case where two or more contacts (but not more than all contacts but two) are pinched off and for the case of a sudden shift of the resonant level. (In Sec.\ \ref{app:charge_1} it was shown that $\Delta \overline{Q} = 0$ if pinching off one contact or if pinching off all contacts but one.) The limiting behavior for large $|\Delta \overline{Q}|$ is the same in both cases. Starting point of our considerations is Eq.\ (\ref{eq:E36}). 

{\em Pinching off of two contacts, small $|\Delta \overline{Q}|$.---}
When pinching off two contacts to MZM $p$ and $q$, the coupling vectors change as
\begin{align}
    \vec{a}'	&=\vec{a}-a_{p}\hat{\ve}_{p}-a_{q}\hat{\ve}_{q}, \nonumber \\
    \vec{b}'	&=\vec{b}-b_{p}\hat{\ve}_{p}-b_{q}\hat{\ve}_{q},
\end{align}
so that
\begin{align}
    \vec{a}'\wedge\vec{b}'	&=\vec{a}\wedge\vec{b}-\sum_{r=p,q}(a_{r}\vec{b}-b_{r}\vec{a})\wedge\hat{\ve}_{r}-(a_{q}b_{p}-b_{q}a_{p})\hat{\ve}_{q}\wedge\hat{\ve}_{p},
\label{eq:E39}
\end{align}
where we use the notation introduced at the end of Sec.\ \ref{app:charge_1}.
Using that $\vec{R}_{k+2}$ is perpendicular to $\vec{a}$ and $\vec{b}$ for $k \ge 1$, we find
\begin{equation}
    \Delta \overline{Q}=-2 \frac{a_{q}b_{p}-b_{q}a_{p}}{\varepsilon^{2}}\sum_{j,k=1}^{N-2}(\vec{R}_{j+2})^{T}(\hat{\ve}_{q}\wedge\hat{\ve}_{p})(\vec{R}_{k+2})\Delta\langle\hat{\Gamma}_{j}\hat{\Gamma}_{k}\rangle.
    \label{eq:E40}
\end{equation}
This expression for the charge signature is a product of two random variables that can be zero with finite probability. Such distributions exhibit a logarithmic singularity for small $|\Delta \overline{Q}|$.

{\em Pinching off of all but two contacts, small $|\Delta \overline{Q}|$.---}
When closing all but two contacts, the coupling vectors can be written as
\begin{align}
    \vec{a}'	&=a_{p}\hat{\ve}_{p}+a_{q}\hat{\ve}_{q}, \nonumber \\
    \vec{b}'	&=b_{p}\hat{\ve}_{p}+b_{q}\hat{\ve}_{q}.
\end{align}
The resulting charge signature has the same form as Eq.~\eqref{eq:E40} up to a minus sign. Hence, the distribution for this case also exhibits a logarithmic singularity for small $|\Delta \overline{Q}|$.

{\em Pinching off other numbers of contacts or sudden change of the resonant level, small $|\Delta \overline{Q}|$.---}
Upon closing any other number of contacts, 
or upon a sudden change of the resonant level,
one finds that the charge signature $\Delta \overline{Q}$ is a sum of random variables distributed symmetrically around zero. The value $\Delta \overline{Q} = 0$ requires cancellation of terms in the sum, which results in a constant weight of the distribution around zero.

{\em Large $|\Delta \overline{Q}|$.}
The limiting behavior at large $\Delta \overline{Q}$ arises from large values of the factor $\Delta \langle \hat{\Gamma}_j \hat{\Gamma}_k \rangle$, which, in turn, follow from large values of the principal values $\lambda_{1,2}$ of the Wilson loop operator $w$. Since 
$\Delta \langle \hat{\Gamma}_j \hat{\Gamma}_k \rangle$ is a linear function of $\lambda_{1,2}$,
the distribution of the largest principal value determines the limiting behavior of the charge signature at large $\Delta \overline{Q}$. With the limiting behavior $P(\lambda_{1},\lambda_{2})\propto\max(\lambda_{1},\lambda_{2})^{-5/2}$ for large $\lambda_{1},\lambda_{2}$ (see App.~\ref{app:principal_values_asymptotic}), we thus have $P(\Delta \overline{Q}) \propto |\Delta \overline{Q}|^{-5/2}$ at large $|\Delta \overline{Q}|$.

\subsection{Numerical results}
\label{app:num_q}

\begin{figure}
\includegraphics[width=\columnwidth]{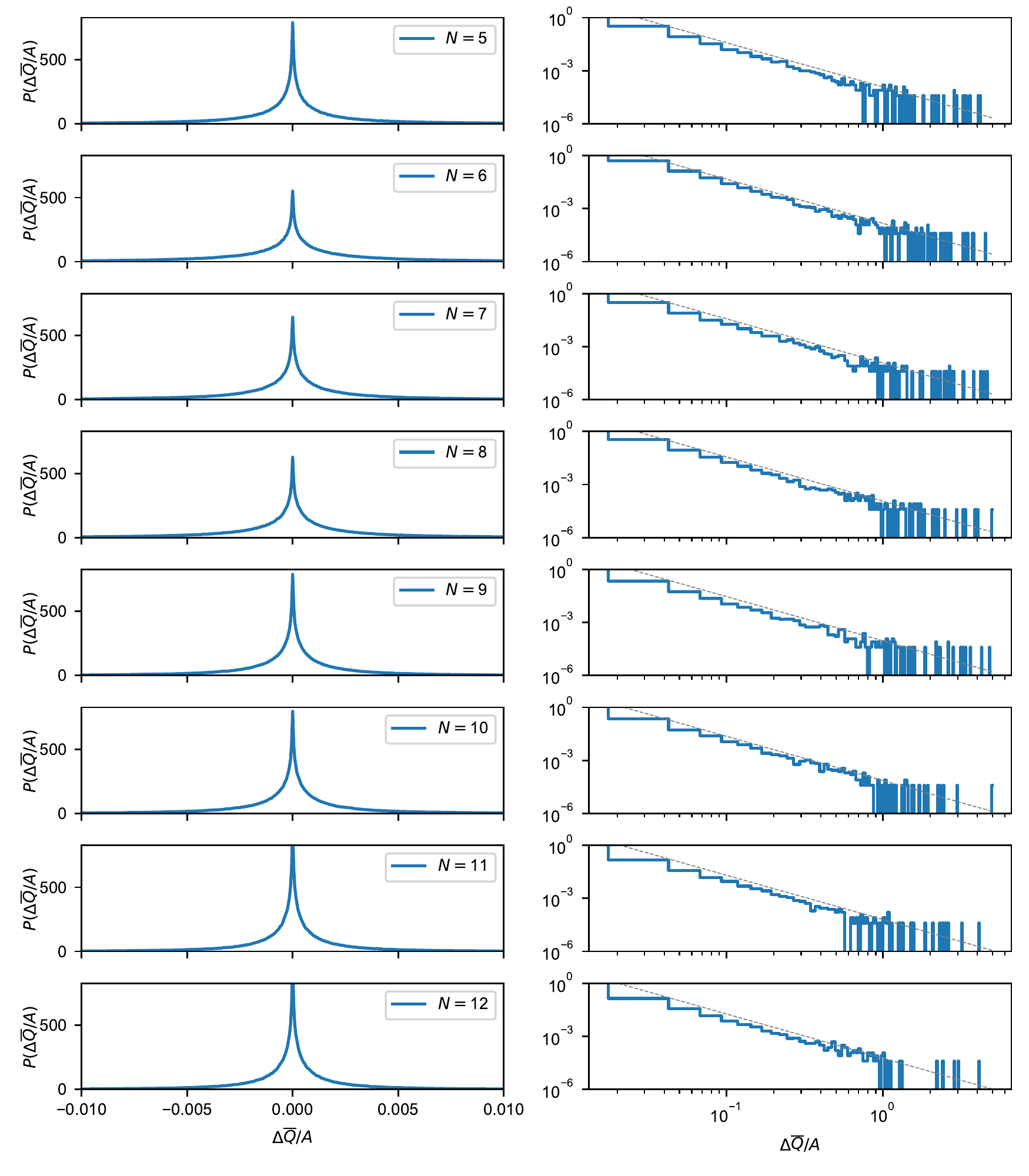}
\caption{Numerically sampled probability distributions of the normalized charge signature $\Delta \overline{Q} / A$, obtained by a sudden pinching off of the contacts to two of the MZMs for infinitesimal loops. The energy $\varepsilon$ of the resonant level is $\varepsilon/\eta = 2 \sqrt{20} \approx 9$. The left and right panels show asymptotic behavior for small and large $|\Delta \overline{Q}|$, respectively, with fit to a power law $P(|\Delta \overline{Q}|) \propto |\Delta \overline{Q}|^{-5/2}$ in the right panels. The numerical distributions were obtained by sampling over $10^7$ independent realizations. 
\label{fig:SM_QQD_quench_MZM}}
\end{figure}

\begin{figure}
\includegraphics[width=\columnwidth]{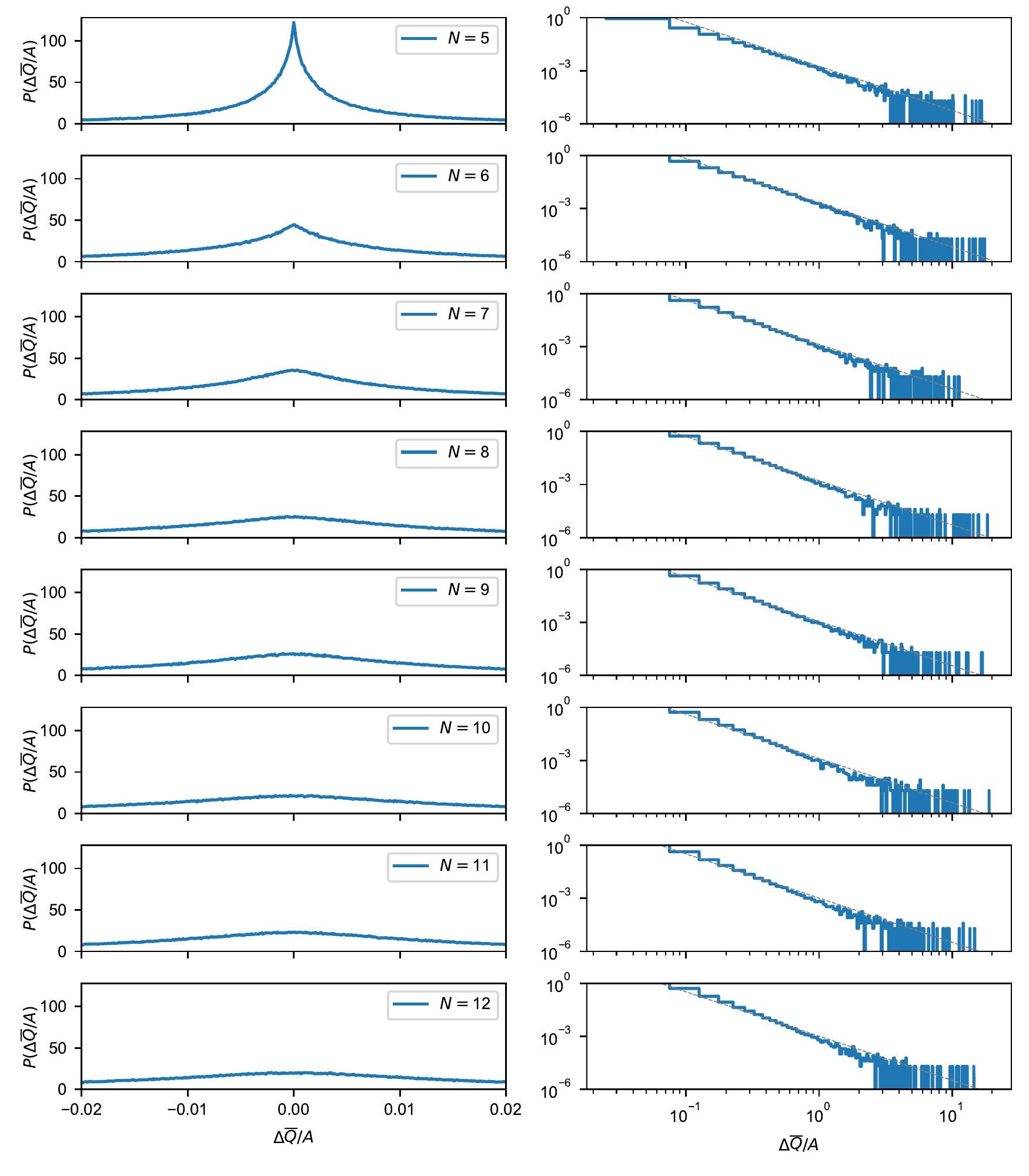}
\caption{Numerically sampled probability distributions of the normalized charge signature $\Delta \overline{Q} / A$, obtained by a sudden shift of the resonant level for infinitesimal loops. The energies of the resonant levels are such that $(\varepsilon-\varepsilon_{\rm g})/\eta = 2 \sqrt{20} \approx 9$ before the quench and $(\varepsilon'-\varepsilon_{\rm g})/\eta = 2 \sqrt{20} \approx 9$ after the quench. The right panels show the asymptotic behavior of the probability distribution for $\Delta \overline{Q}/A \gg 1$ with fit to a power law $P(|\Delta \overline{Q}|) \propto |\Delta \overline{Q}|^{-5/2}$. The numerical distributions were obtained by sampling over $10^6$ independent realizations. 
\label{fig:SM_QQD_shift_level}}
\end{figure}

Figure \ref{fig:SM_QQD_quench_MZM} shows the numerically sampled distribution function of the normalized charge signature $\Delta \overline{Q} / A$ for a quantum dot coupled to $N=5,6,...,12$ MZMs, obtained by suddenly switching off the coupling to two of the MZMs. Figure \ref{fig:SM_QQD_shift_level} shows the numerically sampled probability distributions of $\Delta \overline{Q} / A$, obtained by a sudden shift of the resonant level. In the numerical simulations we set $M = 20$. The energy of the resonant level before and after the quench was set to $\varepsilon / \eta = 2 \sqrt{20} \approx 9$.

\begin{figure}
\includegraphics[width=\columnwidth]{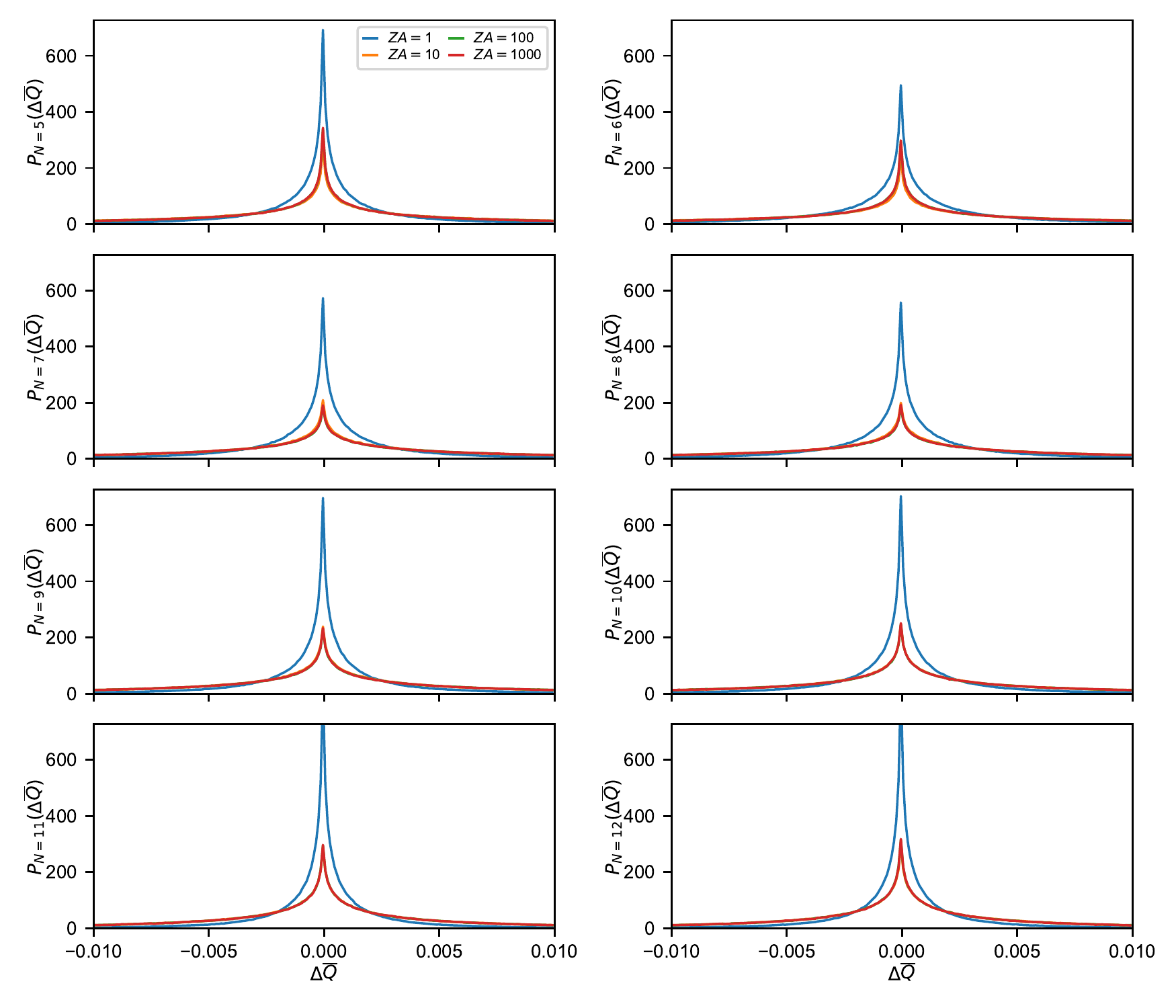}
\caption{Numerically sampled probability distributions of the charge signature $\Delta \overline{Q}$, obtained by a sudden pinching off of the contacts to two of the MZMs for infinitesimal loops repeated $Z$ times with cumulative area $Z A$. The energy $\varepsilon$ of the resonant level is $\varepsilon/\eta = 2 \sqrt{20} \approx 9$. Curves for $ZA=10$, $100$, and $1000$ overlap strongly. The numerical distributions were obtained by sampling over $10^7$ independent realizations. 
\label{fig:SM_QQD_quench_MZM_ZA}}
\end{figure}

\begin{figure}
\includegraphics[width=\columnwidth]{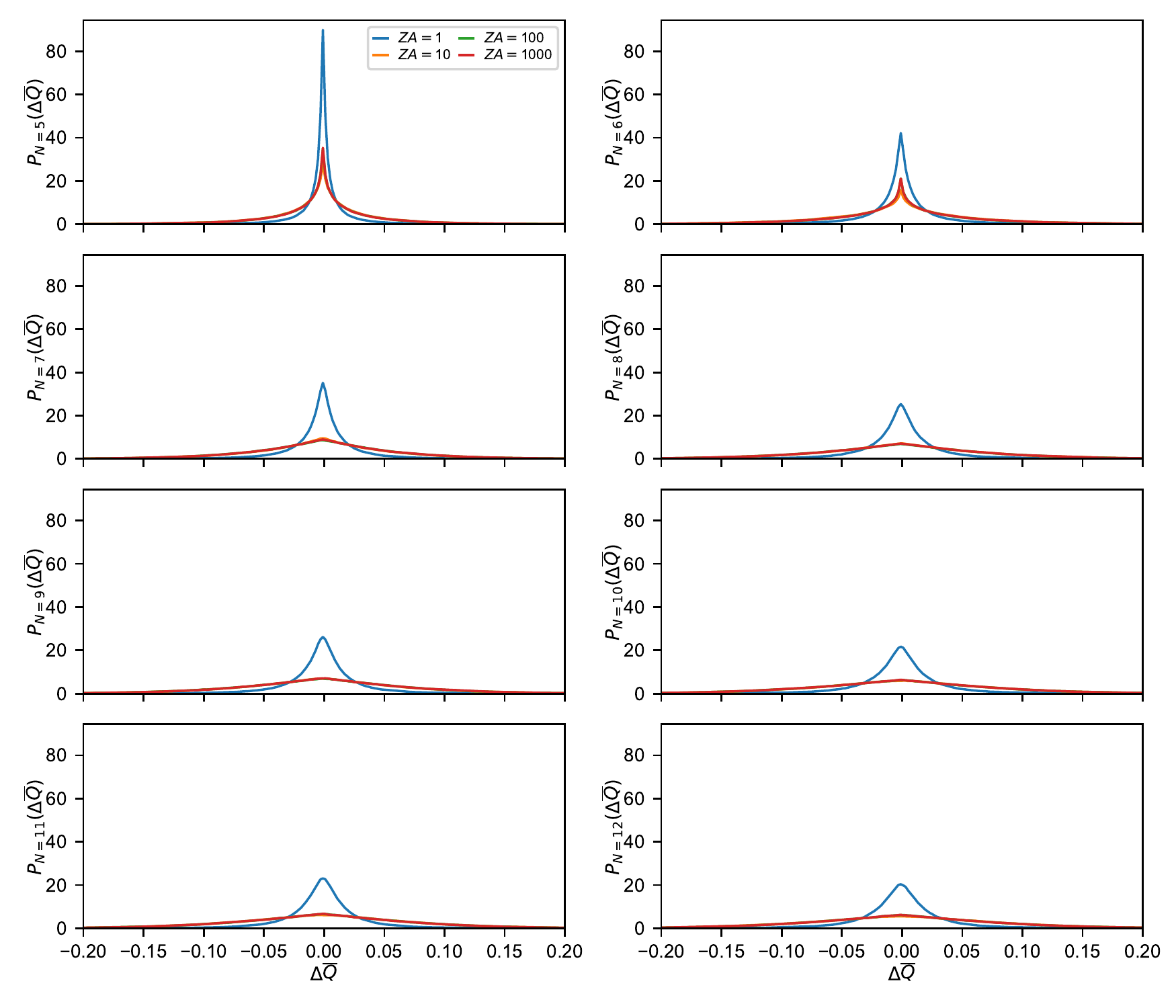}
\caption{Numerically sampled probability distributions of the charge signature $\Delta \overline{Q}$, obtained by a sudden shift of the resonant level for infinitesimal loops repeated $Z$ times with cumulative area $Z A$. The energies of the resonant levels are such that $(\varepsilon-\varepsilon_{\rm g})/\eta = 2 \sqrt{20} \approx 9$ before the quench and $(\varepsilon'-\varepsilon_{\rm g})/\eta = 2 \sqrt{20} \approx 9$ after the quench. Curves for $ZA=10$, $100$, and $1000$ overlap strongly. The numerical distributions were obtained by sampling over $10^6$ independent realizations. 
\label{fig:SM_QQD_shift_level_ZA}}
\end{figure}

\begin{figure}
\includegraphics[width=\columnwidth]{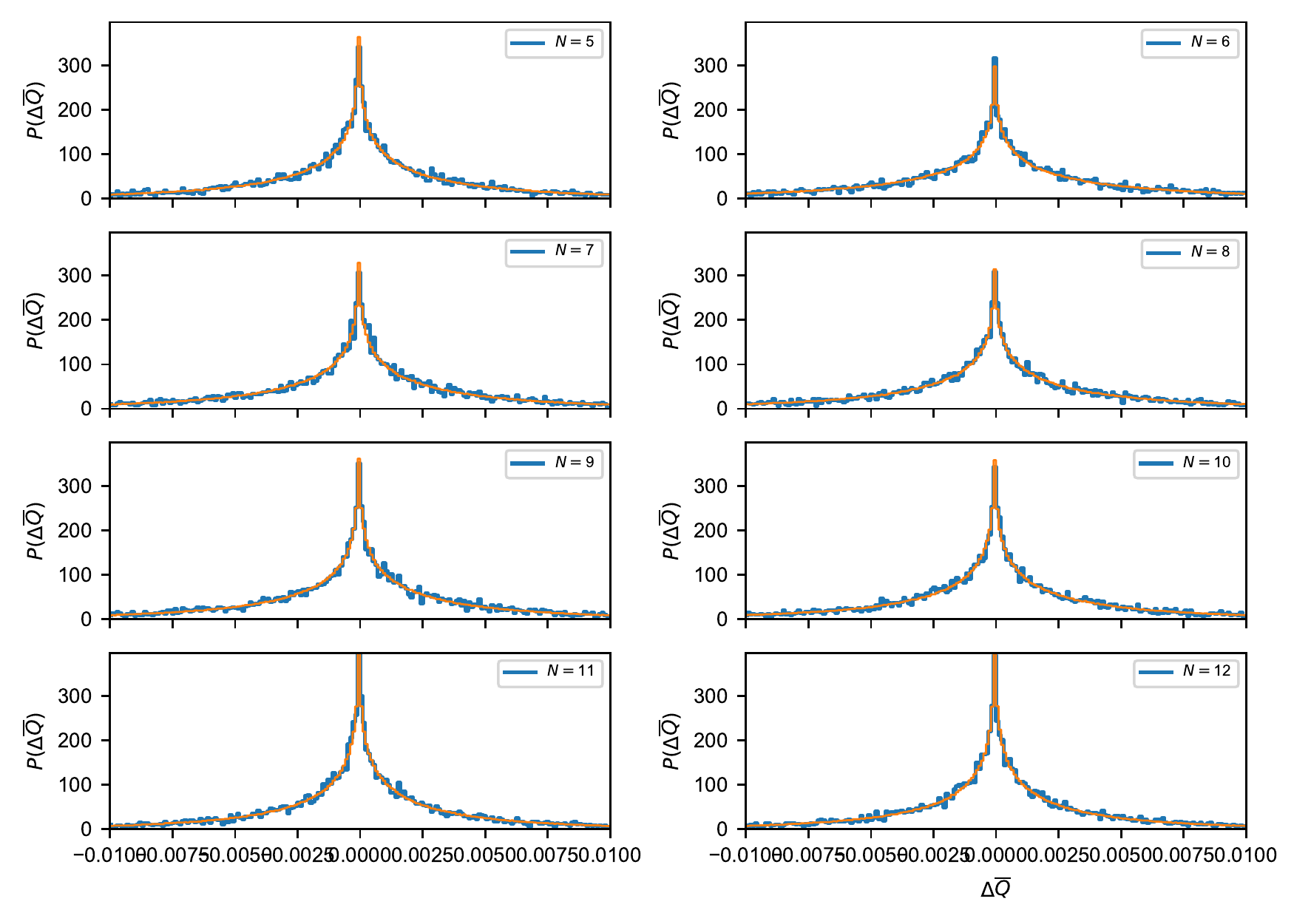}
\caption{Numerically sampled probability distributions of the charge signature $\Delta \overline{Q}$, obtained by a sudden pinching off of the contacts to two of the MZMs for loops with large enclosed area $A \gg 1$ using the random matrix model Eq.~\eqref{eq:H_infiniteA}. The energy $\varepsilon$ of the resonant level is $\varepsilon/\eta = 16$.  The numerical distributions were obtained by sampling over $10^5$ independent realizations. The orange lines were obtained calculating the charge signature using a Wilson loop operator sampled from $\text{SO}(N-2)$.
\label{fig:SM_QQD_quench_MZM_A}}
\end{figure}

\begin{figure}
\includegraphics[width=\columnwidth]{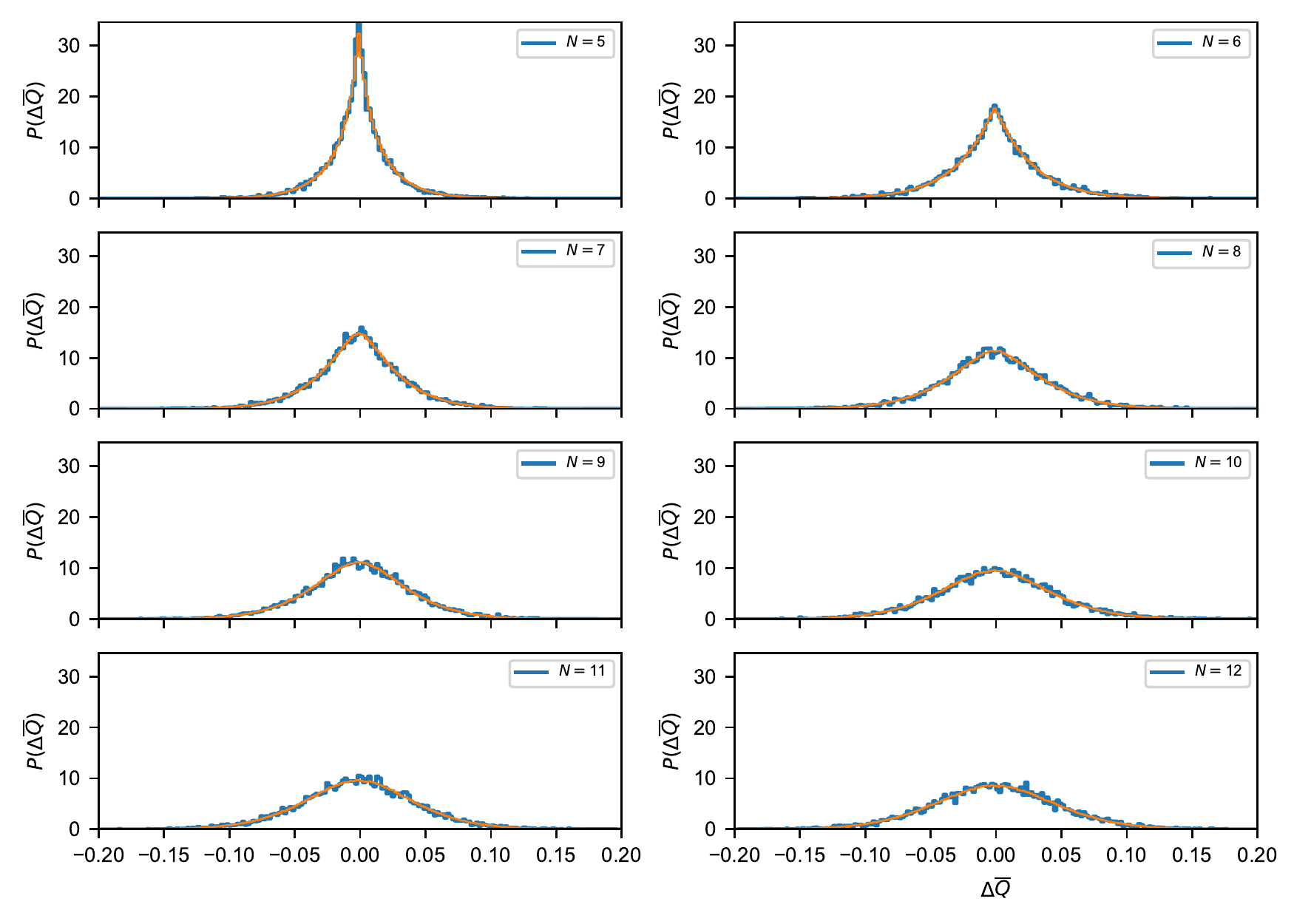}
\caption{Numerically sampled probability distributions of the charge signature $\Delta \overline{Q}$, obtained by a sudden shift of the resonant level for loops with large enclosed area $A \gg 1$ using the random matrix model Eq.~\eqref{eq:H_infiniteA}. The energies of the resonant levels are such that $(\varepsilon-\varepsilon_{\rm g})/\eta = 16$ before the quench and $(\varepsilon'-\varepsilon_{\rm g})/\eta = 16$ after the quench. The numerical distributions were obtained by sampling over $10^4$ independent realizations. The orange lines were obtained calculating the charge signature using a Wilson loop operator sampled from $\text{SO}(N-2)$.
\label{fig:SM_QQD_shift_level_A}}
\end{figure}

Figures \ref{fig:SM_QQD_quench_MZM_ZA} and \ref{fig:SM_QQD_shift_level_ZA} show the numerically sampled distribution functions for the two protocols for small-area loops repeated $Z$ times. The distributions converge for large cumulative area $ZA \gtrsim 10$. Here, we also use $M = 20$ and a resonant level energy $\varepsilon / \eta = 2 \sqrt{20} \approx 9$ before and after the quench. For the protocol of pinching off the coupling to two MZMs, the logarithmic singularity around $\Delta \overline{Q} = 0$ remains. Results for shape-space loops with large dimensionless area $A$ are shown in Fig.\ \ref{fig:SM_QQD_quench_MZM_A} and \ref{fig:SM_QQD_shift_level_A}. 

\begin{figure}
\begin{tabular}{cc}
     \includegraphics[width=0.5\columnwidth]{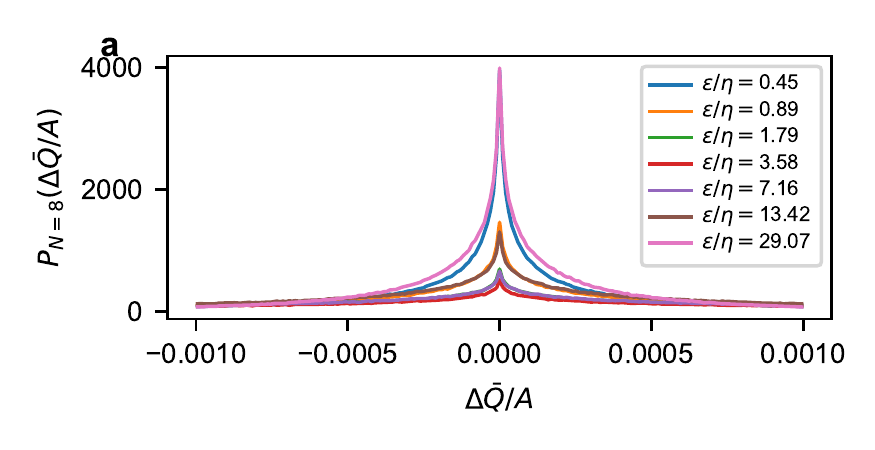}  &
     \includegraphics[width=0.5\columnwidth]{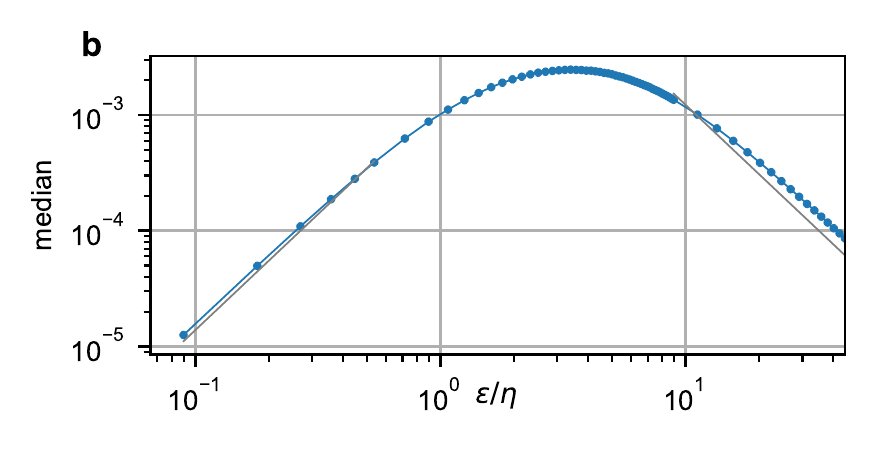} 
\end{tabular}
\caption{Probability distributions of the normalized charge signature $\Delta \overline{Q}/A$ for small $A$, for various values of the resonant energy $\varepsilon$ (left) and median of $|\Delta Q|/A$ vs.\ $\varepsilon$ (right). For both panels the number of MZMs coupled to the quantum dot is $N=8$; the charge signature is obtained by suddenly pinching off two contacts to MZMs. The grey lines in the right panel indicate power laws $\propto (\varepsilon/\eta)^2$ and $(\eta/\varepsilon)^2$. A numerical sampling over $10^6$ realizations was used to obtain the distributions. }
\label{fig:SM_QQD_FWHM}
\end{figure}

In Fig.\ \ref{fig:SM_QQD_FWHM} we compare distributions of the charge signature $\Delta \overline{Q}$ for different values of the energy $\varepsilon$ of the resonant level. We have taken $N=8$ and obtain the charge signature by pinching off two MZMs. We observe that the width of the distribution of $\Delta \overline{Q}$ is maximal at $\varepsilon/\eta \approx 4$ and goes to zero proportional to $(\eta/\varepsilon)^2$ for $\varepsilon \gg \eta$ and proportional $(\varepsilon/\eta)^2$ to for $\varepsilon/\eta \to 0$. The same qualitative conclusions are found for different values of $N$ and for the case of a sudden change of the resonant level (data not shown).

\clearpage

\section{Experimental feasibility}

We discuss the relevant energy and time scales for the operation of
the system, the inclusion of additional quantum dot levels beyond the low-energy Hamiltonian in Eq. (1), sources of noise, and their effect on the observables.
To address experimental feasibility, we focus on the InAs/Al platform as a concrete example.

\subsection{Typical parameters for InAs/Al platform}

\begin{table}
\begin{tabular}{cc|c|ccc}
\hline \hline
\multicolumn{2}{c|}{superconductors} & contacts & \multicolumn{3}{c}{quantum dot}\tabularnewline
$k_{B}T_{e}$ & $\Delta_{{\rm t}}$ & $\eta$ & $m^{*}$ & $A_{{\rm QD}}$ & $\delta$\tabularnewline
$k_{B}50{\rm mK}=4\mu{\rm eV}$ & $60\mu{\rm eV}$ & $\approx\Delta_{{\rm t}}/4=15\mu{\rm eV}$ & $0.023m_{e}$ & $0.1\mu{\rm m}^{2}$ & $100\mu{\rm eV}$\tabularnewline
\hline \hline
\multicolumn{2}{c|}{hybridized system} & \multicolumn{4}{c}{protocol}\tabularnewline
$E_{1}$ & splitting & \multicolumn{2}{c}{$T_{{\rm loop}}$} & $T_{{\rm read}}$ & $T_{{\rm init}}$\tabularnewline
$\approx\eta=15\mu{\rm eV}$ & $\frac{\eta^{2}}{\delta}=2\mu{\rm eV}$ & \multicolumn{2}{c}{$1{\rm ns}$} & $1\mu{\rm s}$ & $1\mu{\rm s}$\tabularnewline
\hline \hline
\end{tabular}\caption{\label{tab:parameters_InAl/Al} \textbf{Typical parameters for InAs/Al
platform.} From top left to bottom right: $k_{B}T_{e}$ the typical
electron temperature of an InAs/Al system in a dilution refridgerator,
$\Delta_{{\rm t}}$ the spectral gap in the topological superconductors,
$\eta$ the coupling strength between MZMs and quantum dot, $m^{*}$
the effective mass of the InAs 2DEG forming the quantum dot, $A_{{\rm QD}}$
the area of the quantum dot, $\delta$
the mean level spacing of the quantum dot, $E_{1}$ the energy
of lowest-energy eigenstate of the hybridized system described by
Hamiltonian Eq. (1) in the main text for a resonant level with energy
$\varepsilon\ll\eta$, $\frac{\eta^{2}}{\delta}$ the typical splitting
of the Majorana dark space due to coupling to the non-resonant quantum
dot levels, $T_{{\rm loop}}$ the runtime for a loop of area $A\approx1$, $T_{{\rm read}}$
the read-out time of the charge and parity signature, and $T_{{\rm init}}$
the initialization time for the Majorana dark space.}
\end{table}

Table \ref{tab:parameters_InAl/Al} summarizes typical parameters
for the InAs/Al material platform as a candidate to realize our proposal.
We comment on the individual parameters below.

\emph{Electron temperature $k_{B}T_{e}$, spectral gap $\Delta_{{\rm t}}$
in the superconductors and effective mass $m^{*}$ in the quantum
dot.---} These parameters are typical for the InAs/Al hybrid platform\cite{QuantumPhysRevB2023Jun}. 

\emph{Coupling strength} $\eta$.--- The coupling strength $\eta$
between MZMs and quantum dot is tunable with electrostatic gates setting
the transparency of the contact. An upper bound for $\eta$ is given
by the spectral gap $\Delta_{t}$ in the topological superconductors
in order to exclude these states from the low-energy dynamics described
by Hamiltonian Eq. (1) in the main text. We estimate $\eta\approx\Delta_{{\rm t}}/4$
as a realistic value.

\emph{Quantum dot area} $A_{{\rm QD}}$ \emph{and level spacing }$\delta$.---
The area $A_{{\rm QD}}$ of the quantum dot sets the level spacing
$\delta=\frac{1}{2}\frac{h^{2}}{2\pi m^{*}A_{{\rm QD}}}$ of the quantum
dot \cite{AleinerPhysRevLett2001Nov}, where a factor $1/2$ accounts for spin-split
levels. We require $\delta\gg\eta$ to achieve a separation of scales
between the energy $E_{1}\approx\eta$ of the state formed by hybridizing
the MZMs with the resonant level and the splitting $\frac{\eta^{2}}{\delta}$
of the Majorana dark space due to hybridization with the non-resonant
levels. For $\eta\approx15\mu{\rm eV}$, we estimate $\delta\gtrapprox100\mu{\rm eV}$
which corresponds to an area $A_{{\rm QD}}\lessapprox0.1\mu{\rm m}^{2}$. 

\emph{Time $T_{{\rm loop}}$ to perform a loop}.--- The time to perform
a loop is constrained from below and above by adiabaticity requirements.
The velocity of the dimensionless gate voltage $\dot{x}$ (as defined in the main text) should be
adiabatic on the scale of $E_{1}\approx1$ but fast on the scale of
the residual splitting $\frac{\eta^{2}}{\delta}$ due to the non-resonant
levels. For a loop of area $A\approx1$, this constrains the runtime
as $2{\rm ns}\approx\frac{h\delta}{\eta^{2}}\gg T_{{\rm loop}}\gg\frac{h}{\eta}\approx0.3{\rm ns}$. 

\emph{Readout time $T_{{\rm read}}$.}--- The charge of a quantum
dot can be read out with RF reflectometry within $1\mu s$ \cite{LiuPhysRevAppl2021Jul}. Since readout proposals of Majorana qubits are also based on a charge
readout of a single quantum dot \cite{SteinerPhysRevRes2020Aug,MunkPhysRevRes2020Aug},
we estimate similar readout times there too.

\emph{Initialization time $T_{{\rm init}}$.}--- The Majorana subspace
can be initialized by measurement of the fermion parity in pairs of
Majorana modes. For this procedure, we estimate the initialization
time to be similar to the read-out time. Alternatively, the Majorana
dark space can be initialized by coupling the system to a bath with
temperature $k_{B}T\ll\frac{h\delta}{\eta^{2}}$ much smaller than
the splitting due to coupling to the non-resonant levels. The time
scale for this process depends on the system-bath coupling. 

\subsection{Temperature}

\label{subsec:Temperature}

The temperature should be much smaller than the energy $E_{1}\approx\eta$
of the hybridized mode composed out of resonant level and external
MZMs (the lowest-energy eigenstate of Hamiltonian Eq. (1) for $\varepsilon\ll\eta$).
This avoids occupation and evolution of high-energy modes outside
of the Majorana dark space. 

A further energy scale is the splitting $\frac{\eta^{2}}{\delta}$
of the Majorana dark space due to coupling to non-resonant levels
with energy $\approx\delta$ of the order of the level spacing of
the quantum dot. This splitting imposes a requirement on temperature
$k_{B}T\ll\frac{\eta^{2}}{\delta}$ only if the system is supposed
to be initialized by thermal relaxation, such as we suggested for
the charge signature. If initialization by thermal relaxation 
is attempted with $k_{B}T\gtrsim\frac{\eta^{2}}{\delta}$,
subsequent measurements of the charge signature become uncorrelated
and the distribution becomes gaussian centered around $\Delta\bar{Q}=0.5$.
Instead, the system can be initialized using measurement of the external
MZMs in pairs. Furthermore, the parity signature does not require
such small temperatures: There, a decoupled pair of MZMs is initialized,
then coupled to the quantum dot, the loop is performed, and the pair
is then decoupled and measured. This does not require a initialization
of the remaining MZMs coupled to the QD.

\emph{InAs/Al platform.}--- In InAs/Al, the typical electron temperature
$k_{B}T_{e}\approx k_{B}50{\rm mK}=4\mu e{\rm V}$ is much smaller
than our estimate $\eta\approx\Delta_{{\rm t}}/4\approx15\mu{\rm eV}$
but larger than the residual splitting $\frac{\eta^{2}}{\delta}\approx2\mu e{\rm V}$.
We thus expect that the dynamics can be constrained to the Majorana
dark space. However, initialization of the Majorana dark space by
thermal relaxation would require novel techniques.

\subsection{Quasi-particle poisoning}

To reduce quasiparticle poisoning of the Majorana bound states, the
temperature $k_{B}T\ll\Delta_{{\rm t}}$ needs to be much smaller
than the spectral gap in the external topological superconductors
hosting the Majorana bound states. This is a weaker constraint than
$k_{B}T\ll\eta\lesssim\Delta_{{\rm t}}$ discussed in Sec. \ref{subsec:Temperature}
above. 

\emph{Effect on observables.---} If quasiparticle poisoning of the
Majorana bound state parity is faster than the time to initialize,
run, and measure, then the parity signature becomes uniformly distributed
between 0 and 1, independent on whether a loop was performed. Similarly,
the charge signature becomes uniformly distributed between zero and
one because the state of the Majorana dark state for the two quenches
becomes uncorrelated.

\emph{InAs/Al platform.}--- A detailed study \cite{KarzigPhysRevLett2021Feb} finds for typical electron temperatures of $T=50$mK
in dilution refrigerators suggests that at this temperature the poisoning
rate should happen at the rate of quasiparticle generation, which
the authors estimate to be of the order of seconds or even larger.
Other poisoning events, where an excited quasiparticle enters the
quantum dot, can be suppressed by charging energy on the quantum dot
\cite{BargerbosPRXQuantum2022Jul}.

\subsection{Adiabaticity}

The velocity of the dimensionless gate voltages $\partial_{t}x,\ \partial_{t}y$
should be adiabatic with respect to $E_{1}\approx\eta$, but fast
compared to the residual splitting $\frac{\eta^{2}}{\delta}$. By
driving fast compared to $\frac{\eta^{2}}{\delta}$, the residual
splitting can be neglected. 

\emph{Effect on observables.--- }If the drive is slow compared to
$\frac{\eta^{2}}{\delta}$, the evolution becomes adiabatic with respect
to the residual splitting. This effectively reduces the dimension
of the Majorana dark space by $2n_{{\rm adiabatic}}$ where $n_{{\rm adiabatic}}$
is the number of levels the drive is adiabatic to. When the dimension
of the Majorana dark space is two or less, the time-evolution becomes
abelian and the Majorana parities remain unchanged.

If the drive is fast compared to $E_{1}\approx\eta$, high-energy
states outside of the Majorana dark state are populated. For the charge
signature, this broadens the distribution because the occupation of
$E_{1}$ has shared weight on the quantum dot and Majorana subspace
which increases the difference of the quantum dot charge with and
without loop following the quench. For the parity signature, we expect
a similar broadening of the distribution.

\emph{InAs/Al platform.}--- For a loop of area $A\approx1$, we estimate
a run-time of the loop of $2{\rm ns}\approx\frac{h\delta}{\eta^{2}}\gg T_{{\rm loop}}\gg\frac{h}{\eta}\approx0.3{\rm ns}$.
This lies within the achievable range of current arbitrary waveform
generators.

\subsection{Resonant level energy and loop size}

The energy of the resonant level $|\varepsilon|$ is required to be much smaller than the level spacing $\delta$ of the quantum dot. This ensures a clear separation of scales between the energy $E_{1}$ of MZMs hybridized with the resonant level and the splitting $\frac{\eta^2}{\delta}$ of the Majorana dark space due to hybridization with the non-resonant levels. This separation of scales is largest when $|\varepsilon|<\eta$. However, a sufficient separation of scales is also achieved as long as the resonant dot level energy $|\varepsilon|\ll\delta$ is much smaller than the level spacing of the quantum dot.

A selected quantum dot level can be kept in resonance during the loop by using a calibrated third gate that offsets its energy. This allow for an arbitrary loop size. Without a compensating third gate, the changes of the dimensionless gate voltages are restricted to $x,y\ll1$ where first-order perturbation theory is valid and the changes of the level energies are small compared to the level spacing.

 We expect that the contributions from nonresonant levels cut off the algebraic decay of the distribution functions obtained for small loops on scales $\lambda \gtrsim \delta^2/\eta^2$ because at these scales the splitting of the MZMs from coupling to the nearest non-resonant level $\varepsilon_{m-1}$ or $\varepsilon_{m+1}$ is comparable to the splitting from the coupling to the resonant level $\varepsilon_m$.

\subsection{Fluctuations of the electrostatic environment}

Fluctuations of the electrostatic environment effectively lead to
a change of the quantum-dot shape with time. We need to require that
these fluctuations are small on time scales to measure a charge or
parity expectation value, because many experiments with the same loop
need to be performed to estimate an expectation value. If fluctuations
are large on these time scales, then subsequent measurements are statistically
independent and the distribution converges to a normal distribution
centered around the mean of the respective distributions by the central
limit theorem. This effect is similar to the dephasing mechanism discussed
in \cite{SnizhkoPhysRevLett2019Aug}, with the addition that
fluctuations during initialization and readout need to be considered
as well. Fluctuations on longer time scales are actually helpful for
our proposal as they contribute to obtain statistical independent
realization by randomizing the quantum dot shape.

\emph{InAs/Al platform.}--- For our example parameters in the InAs/Al
platform, we estimated that the time to perform a loop $\sim{\cal O}(1){\rm ns}$
is much shorter than the read-out and initialization time $\sim{\cal O}(1)\mu{\rm s}$.
We thus need to require that fluctuations of the electostatic environment
are weak ($\langle|\eta_{j}(t')-\eta_{j}(t)|^{2}\rangle\ll\eta^{2}$ for
$|t'-t|\leq rT$) on time scales $rT_{{\rm P}}\sim{\cal O}(1){\rm ms}$
with $r\sim\mathcal{O}(10^{3})$ the number of repetitions to estimate
the expectation value and $T_{{\rm P}}\sim{\cal O}(1)\mu{\rm s}$
the runtime of the protocol to acquire a single measurement of the
parity or charge. 

Electric field fluctuations in InAs/Al hybrid platforms mostly originate
from fluctuations of the electrostatic gates. Their noise power spectral
density has a characteristic form $S(f)\sim1/f^{\beta}$ with $\beta\sim1$
within a range from the order of several ${\rm kHz}$ up around $1{\rm GHz}$
where the power spectral density crosses over to $S(f)\sim f$ at
large frequencies $f\approx1{\rm GHz}$ and approaches a constant
for small frequencies $f\lessapprox1{\rm kHz}$ \cite{IthierPhysRevB2005Oct}. The effect of electric field fluctuations from
electrostatic gates can be reduced by designing the system such that
the resulting fluctuations of the dimensionless gate voltage $x,y$
is much smaller than $1$. This can be achieved by reducing the size
of the quantum dots or increasing the distance between gates and quantum
dot.

\subsection{Number of realizations to distinguish MZMs and ABSs}

We estimate the number of statistically independent measurements of
charge or parity expectations values need to be performed to distinguish
ABSs from MZMs. The distinction is based on identifying the reference
distribution $P(x)$ that best describes an empirical distribution
$\tilde{P}_{n}(x)$ obtained from $n$ independent measurements. A
mathematical procedure to perform the identification is the Kolmogorov-Smirnov
test. The test is based on the distance measure
\begin{equation}
D_{n}={\rm sup}_{x}\left|\tilde{P}_{n}(x)-P(x)\right|\label{eq:KS_distance}
\end{equation}
which is the largest absolute difference of the two distribution functions.
In case the empirical data follows the reference distribution $P(x)$,
the distance $D_{n}$ converges almost surely to zero as $n^{-1/2}$
in the limit $n\to\infty$. 

\begin{figure}
\begin{tabular}{ll}
(a) & (b)\tabularnewline
\includegraphics[width=8cm]{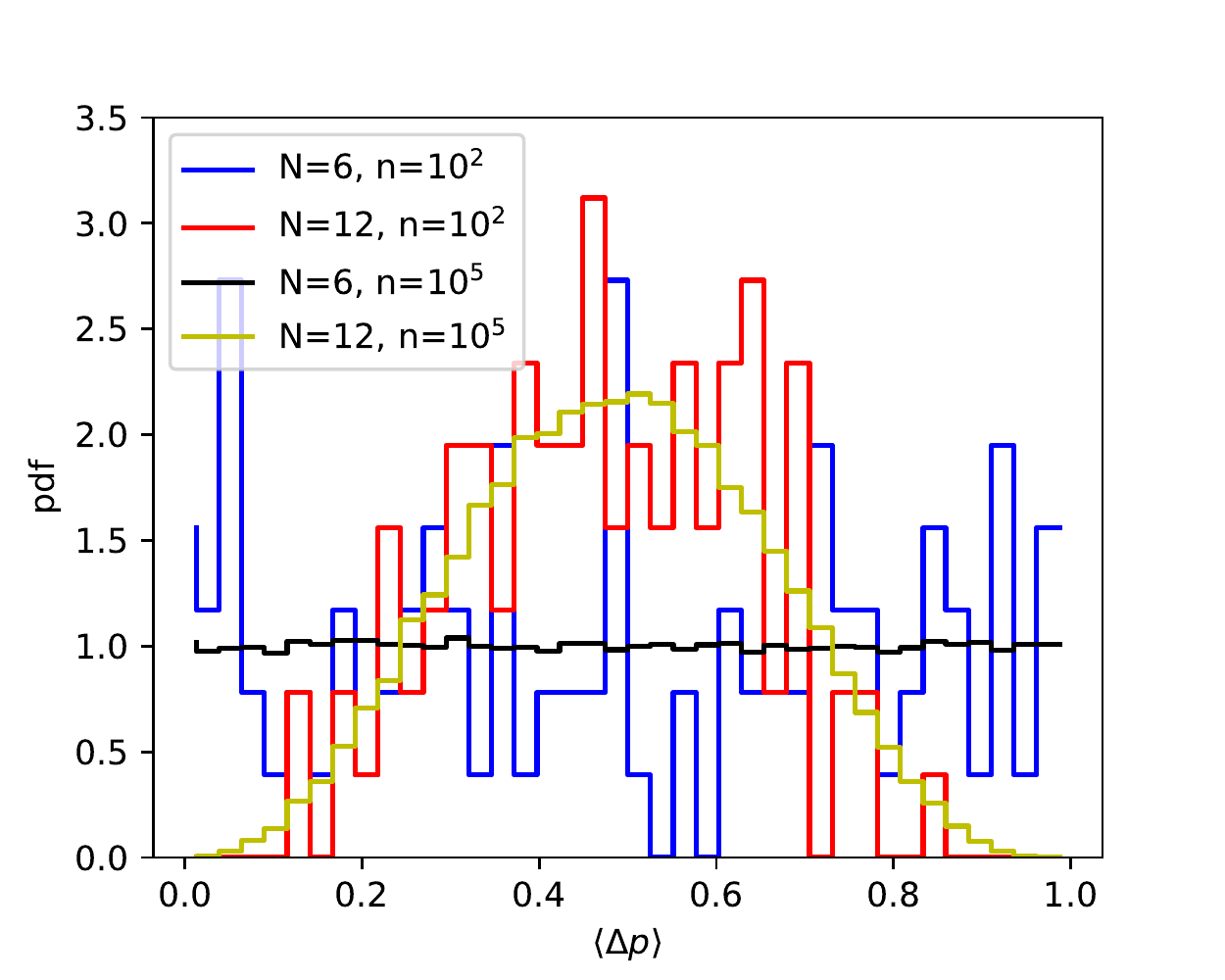} & 
\includegraphics[width=8cm]{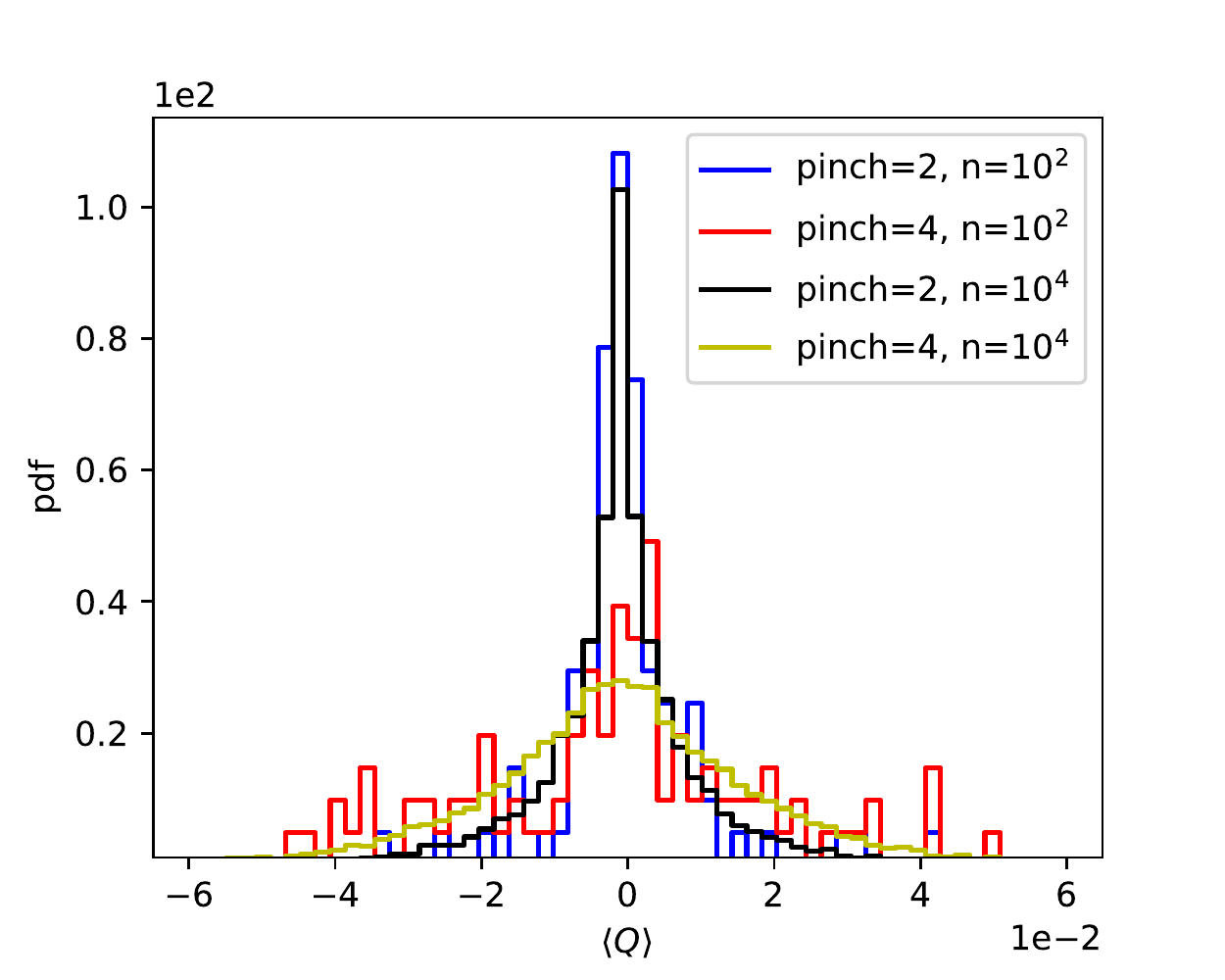}\tabularnewline
\end{tabular}\caption{\label{fig:distributions} \textbf{Kolmogorov-Smirnov test for distinguishing
Majorana from zero-energy Andreev bound states.} (a) Probability distribution
of the fermion parity signature for $N=6$ and $12$ Majorana modes
sampled with $n=10^{2}$ and $10^{5}$ realizations. (b) Probability
distribution of the charge signature obtained by suddenly decoupling
$2$ or $4$ Majorana modes from a quantum dot coupled to $N=8$ Majorana
modes sampled with $n=10^{2}$ and $10^{4}$ realizations.}
\end{figure}

Figure \ref{fig:distributions} shows sampled distributions of the
parity and charge signature with $n=100$ realizations and their comparison
to the reference distributions obtained by sampling $10^{4}$ and
$10^{5}$ realizations, respectively. We compare scenarios for distinguishing
Majorana zero modes from zero-energy Andreev bound states: For the
parity signature, we compare a quantum dot coupled to $N=6$ and $12$
Majorana zero modes, where the latter corresponds to six zero-energy
Andreev bound states. Evaluating the distance measure Eq. \ref{eq:KS_distance}
for the distributions shown in Fig. \ref{fig:distributions}(a),
\[
D_{n}\approx\left(\begin{array}{cc}
1.75 & 2.70\\
2.11 & 1.10
\end{array}\right)
\]
where the rows indicate the sampled distribution with 100 realizations
and $N=6$ and $12$ Majorana bound states and the columns indicate
the reference distribution with $N=6$ and $12$ obtained with $10^{5}$
realizations. 

For the charge signature, we consider a quantum dot connected to $N=8$
Majorana zero modes and consider pinching off the contacts to two
or four Majorana zero modes. The distributions are shown in Fig. \ref{fig:distributions}(b).
The distance measure, Eq. \ref{eq:KS_distance}, yields

\[
D_{n}\approx\left(\begin{array}{cc}
26 & 80\\
63 & 22
\end{array}\right)
\]
where the rows indicate the sampled distribution $100$ realizations
and pinching off two and four Majorana bound states and the columns
indicate the reference distribution with pinching off two and four
Majorana bound states sampled with $10^{4}$ realizations. 

In both cases, we find that the distance is minimal for configurations
with matching number of Majorana bound states. We conclude that a
sample size of ${\cal O}(100)$ statistically independent realizations
are sufficient to distinguish Majorana from zero-energy Andreev bound
states based on the probability distributions of their parity and
charge signatures.

Experimentally, the number of statistically independent realizations
depends on (i) the voltage range $\Delta V_{x,y}$ of the shape-defining
gates $x,y$ that is compatible with the system, (ii) the lever arm
$\partial_{V}x$, $\partial_{V}y$ that relates a change in gate voltage
with a change of the dimensionless parameter $x$, and (iii) the number
of quantum dot level $N_{\varepsilon}$ that can be tuned into resonance. 
The level arm $\partial_{V}x,\partial_{V}y$
can be determined experimentally by comparing current autocorrelation
measurements to predictions from random matrix theory. The number
of reachable statistically independent realizations is then $(\Delta V_{x}\partial_{V}x)(\Delta V_{y}\partial_{V}y)N_{\varepsilon}$.
To reach ${\cal O}(100$) statistically independent realizations,
the shape defining gates should each be tunable to reach around $\Delta V_{x}\partial_{V}x\approx\Delta V_{x}\partial_{V}x\approx4$
realizations and $N_{\varepsilon}\approx6$ different quantum dot
levels should be tunable into resonance. The number of statistically independent realizations can be increased by adding additional gates.

\subsection{Residual splitting due to coupling to non-resonant levels}

The residual splitting due to coupling to nonresonant quantum-dot levels has two effects (i) the eigenstates spanning the Majorana dark space have acquire a small support $\propto \frac{\eta}{\delta}$ on the quantum dot and (ii) the Majorana dark space evolves dynamically with frequency $\frac{\eta^2}{\delta}$ also during read-out and initialization. We expect that these to have no or only a weak effect $\propto \frac{\eta}{\delta}$ on the parity and charge signature.

To obtain the parity signature, the measured pair of MZMs is coupled to the chaotic quantum dot only when the loop is performed. Therefore, the evolution of the remaining MZMs coupled to the QD during initialization and readout does not affect the parity signature. 

The charge signature is produced by a quench at the end of the loop. In this case, the evolution in the Majorana dark space due to residual splittings leads to only a small, oscillating contribution $\propto \frac{\eta}{\delta}$ on top of the large charge signature $\overline{Q} \approx {\cal{O}}(1)e$ produced by the quench. This contribution is averaged over during the read-out with time $T_{\rm read} \gg \frac{h \delta}{\eta^2}$. 

\subsection{Conclusion}

We expect that the main energy and operation time scales lie within reach of the InAs/Al platform. 
The main challenge may be fabrication: The proposal requires to tune multiple external superconductors to host MZMs or zero-energy ABSs and couple them to a small quantum dot of size  $A \lesssim 0.1 \mu {\rm m}^2$ required to achieve a level splitting $\delta \gg \eta$ while having a readout setup nearby. 

The parity and charge signatures have individual advantages and challenges. The parity signature does not require initialization of the full Majorana dark space, but requires a parity readout of pairs of Majorana fermions that has not yet been demonstrated. In contrast, the charge signature requires an advanced initialization scheme or novel cooling technique. 

A simplest, first direction could be to pursue a verification of the proposal with zero-energy ABSs. A successful experiment with ABSs would already demonstrate the non-Abelian manipulation of fermionic degrees of freedom in superconductors.

\end{document}